\documentclass[floats,floatfix,
amssymb,prd,onecolumn,notitlepage,
superscriptaddress,nofootinbib]{revtex4-1}

\usepackage{amssymb,amsmath,verbatim,mathtools,
needspace,enumitem,etoolbox,graphicx,physics,microtype,afterpage,bm}

\usepackage[normalem]{ulem}

\usepackage[dvipsnames, usenames]{xcolor}

\definecolor{linkcolor}{rgb}{0.0,0.3,0.5}

\usepackage{orcidlink}

\usepackage{hyperref}

\usepackage{graphics,appendix,afterpage,makecell}

\usepackage{bbold}
\usepackage{tikz}
\usepackage{adjustbox}
\usepackage{tikz-feynman}
\usepackage{tcolorbox}

\newcommand{\MPl}{\bar{M}_{\textrm{\tiny{Pl}}}}

\definecolor{oucrimsonred}{rgb}{0.6, 0.0, 0.0}
\definecolor{persianblue}{rgb}{0.11, 0.22, 0.73}
\definecolor{forestgreen}{rgb}{0.13,0.35,0.13}
\definecolor{lightgray}{rgb}{0.83, 0.83, 0.83}
 \hypersetup{colorlinks, citecolor=oucrimsonred, linkcolor=persianblue, urlcolor=oucrimsonred}
\definecolor{cornellred}{rgb}{0.7, 0.11, 0.11}
\definecolor{navyblue}{rgb}{0.0, 0.0, 0.5}
\definecolor{amethyst}{rgb}{0.6, 0.4, 0.8}
\definecolor{yellow}{rgb}{1.0, 1.0, 0.0}
\definecolor{firebrick}{rgb}{0.7, 0.13, 0.13}
\definecolor{tangerineyellow}{rgb}{1.0, 0.8, 0.0}
\definecolor{deepfuchsia}{rgb}{0.76, 0.33, 0.76}
\definecolor{amber}{rgb}{1.0, 0.75, 0.0}
\definecolor{VioletRed4}{rgb}{0.55, 0.13, .32}
\definecolor{indiagreen}{rgb}{0.07, 0.53, 0.03}
\definecolor{VioletRed4}{rgb}{0.55, 0.13, .32}
\newcommand{\be}{\begin{equation}}
\newcommand{\ee}{\end{equation}}
\newcommand{\bea}{\begin{eqnarray}}
\newcommand{\eea}{\end{eqnarray}}
\newcommand{\nn}{\nonumber}

\definecolor{oucrimsonred}{rgb}{0.6, 0.0, 0.0}
\newcommand\vertarrowbox[3][6ex]{%
  \begin{array}[t]{@{}c@{}} #2 \\
  \left\uparrow\vcenter{\hrule height #1}\right.\kern-\nulldelimiterspace\\
  \makebox[0pt]{\scriptsize#3}
  \end{array}%
}

\definecolor{mtcolor}{rgb}{.8,.3,.1}

\definecolor{violachiaro}{rgb}{1,0.6,1}

\definecolor{gbcolor}{rgb}{.43,.22,.12}
 
\definecolor{gbcolor2}{rgb}{.9,.2,.6}
\definecolor{gbcolor3}{rgb}{.3,.2,.6}

\definecolor{verdechiaro}{rgb}{0.6,1,0.6}
\definecolor{giallochiaro}{rgb}{1,1,0.6}
\definecolor{bluscuro}{rgb}{0.15, 0.2, 0.9}
\definecolor{verdes}{rgb}{0.1, 0.5, 0.1}%
\definecolor{tangerineyellow}{rgb}{1.0, 0.8, 0.0}
\definecolor{smokyblack}{rgb}{0.06, 0.05, 0.03}

\definecolor{americanrose}{rgb}{1.0, 0.01, 0.24}
\definecolor{cobalt}{rgb}{0.0, 0.28, 0.67}
\definecolor{brandeisblue}{rgb}{0.0, 0.44, 1.0}
\definecolor{mycolor}{rgb}{0.0, 0.0, 0.5}
\definecolor{oxfordblue}{rgb}{0.0, 0.13, 0.28}
\definecolor{azure}{rgb}{0.0, 0.5, 1.0}
\definecolor{turquoiseblue}{rgb}{0.0, 1.0, 0.94}
\newtcolorbox{mynewbox}[1]{colback=white!5!white,colframe=azure!75!black,fonttitle=\bfseries,title=#1}
\newtcolorbox{mybox}{colback=mycolor!5!white,colframe=azure!75!black}
\newtcolorbox{mynamedbox}[1]{colback=mycolor!5!white,colframe=azure!75!black,title=#1}
\definecolor{venetianred}{rgb}{0.78, 0.03, 0.08}
\newtcolorbox{mynamedbox1}[1]{colback=venetianred!5!white,colframe=venetianred!80!black,title=#1}
\newtcolorbox{mynamedbox2}[1]{colback=azure!5!white,colframe=azure!80!black,title=#1}

\definecolor{rossocorsa}{rgb}{0.83, 0.0, 0.0}

\newcommand{\reh}{{\textrm{reh}}}

\tikzset{->-/.style={decoration={
  markings,
  mark=at position #1 with {\arrow{>}}},postaction={decorate}}}
\tikzset{-<-/.style={decoration={
  markings,
  mark=at position #1 with {\arrow{<}}},postaction={decorate}}} 

\newcommand{\uniroma}{Dipartimento di Fisica, Sapienza Università 
	di Roma, Piazzale Aldo Moro 5, 00185, Roma, Italy}
\newcommand{\infn}{INFN, Sezione di Roma, Piazzale Aldo Moro 2, 00185, Roma, Italy}
\newcommand{\jhu}{William H.\ Miller III Department of Physics and Astronomy, Johns Hopkins University, \\ 3400 N. Charles Street, Baltimore, Maryland, 21218, USA}

\begin{document}

\title[]{
ACT stands for Awkward Cosmology Theories
}

\date{\today}

\author{Loris Del Grosso\orcidlink{0000-0002-6722-4629}}
\email{ldelgro1@jh.edu}
\affiliation{\jhu}

\author{Alfredo Urbano\orcidlink{0000-0002-0488-3256}}
\email{alfredo.urbano@uniroma1.it}
\affiliation{\uniroma}
\affiliation{\infn}

\author{Marcello Ziccolella\orcidlink{0009-0009-6230-971X}}
\email{marcello.ziccolella@uniroma1.it}
\affiliation{\uniroma}
\affiliation{\infn}

\begin{abstract}
\noindent 
Recent Atacama Cosmology Telescope (ACT) observations, combined with Planck, favor a scalar spectral index in tension with Starobinsky inflation. Two main strategies have been proposed to reconcile Starobinsky inflation with these data: (i) including higher-curvature operators in the gravitational action, and (ii) modifying the post-inflationary reheating dynamics. We critically re-examine both approaches. When higher-curvature corrections are included, dimensional-analysis arguments allow introducing a second mass scale suppressing higher-dimensional operators, resulting in a one-coupling, two-scale effective theory. We show that fitting the Planck+ACT data drives this class of effective theories toward a regime of stronger sensitivity to UV physics, potentially implying fine-tuning of the inflationary observables.
We illustrate our general point with two concrete models, namely no-scale supergravity and metric-affine gravity, showing that they provide different realizations of the same underlying $f(R)$ theory.
On the reheating side, requiring the CMB modes to re-enter the horizon before recombination leads to stringent constraints on the reheating equation of state, and fitting the Planck+ACT data favors an exotic reheating phase with $\omega_{\rm reh} > 1/3$, incompatible with inflaton oscillations about the minimum of the standard Starobinsky potential. Finally, we explore the possibility that a phase of cosmological stasis, driven by a tower of decaying massive states, could play the role of the post-inflationary epoch. We find that accommodating ACT-preferred $n_s$ in this setup requires a tower of states with negative energy density.
Taken together, our results suggest that reconciling Starobinsky inflation with the Planck+ACT data demands ingredients that are challenging to motivate from the perspective of UV-complete theories, undermining the minimality of the original Starobinsky model.
\end{abstract}
\maketitle

%
\tableofcontents

\newpage

\section{Introduction}

Starobinsky inflation is one of the earliest and most successful models of cosmic inflation, originally proposed by Alexei Starobinsky in 1980\,\cite{Starobinsky:1983zz}. Unlike many models based on scalar fields with ad hoc potentials, Starobinsky inflation arises from a modification of the gravitational sector itself. Specifically, the action includes a quadratic correction to the Ricci scalar:
\begin{align}
\mathcal{S} = 
\frac{\MPl^2}{2} \int d^4x \sqrt{-g} \left( R + \frac{R^2}{6 M^2} \right),\label{eq:FirstStaro}
\end{align}
where $R$ is the Ricci scalar, $\MPl$ is the reduced Planck mass, and $M$ is a mass parameter that determines the scale of inflation. The $R^2$ term becomes significant at high curvatures and naturally leads to a period of accelerated expansion in the early universe.

This higher-curvature theory can be transformed into a standard Einstein-Hilbert action plus a scalar field through a conformal transformation. In the Einstein frame, the model is equivalent to a canonical scalar field $\phi$ with the following potential:
\begin{align}\label{eq:StaroPotPhi}
V(\phi) = \frac{3}{4} M^2 \MPl^2 \left[ 1 - \exp\left(-\sqrt{\frac{2}{3}} \frac{\phi}{\MPl} \right) \right]^2.
\end{align}
This potential is flat for large values of $\phi$, which ensures a prolonged period of slow-roll inflation and generates nearly scale-invariant, Gaussian, and adiabatic perturbations—features that are in excellent agreement with current observational data. One of the key predictions of the Starobinsky model is a very small tensor-to-scalar ratio, $r \sim 0.003$, and a scalar spectral index $n_s \sim 0.965$, both of which are well within the bounds measured by the Planck satellite\,\cite{Planck:2018jri}. Its minimal nature, strong predictive power, and geometric origin make Starobinsky inflation a benchmark model in modern cosmology.

However, recent combined data from the Planck satellite and the ACT have introduced potential tensions with the predictions of the Starobinsky model\,\cite{ACT:2025fju,ACT:2025tim,AtacamaCosmologyTelescope:2025blo}. These updated observations hint at a slightly higher tensor-to-scalar ratio and subtle shifts in the preferred values of the scalar spectral index, which may be difficult to reconcile within the strict predictions of Starobinsky inflation. While not conclusively ruling it out, these results invite a reevaluation of the model’s parameter space and have spurred renewed interest in alternative or extended inflationary scenarios. Indeed, several proposals have been put forward to reconcile the Starobinsky predictions with the new Planck+ACT constraints (see e.g. refs.\,\cite{Dioguardi:2025vci, Antoniadis:2025pfa, Salvio:2025izr,Shobcha:2026mpc,Odintsov:2026dss,Odintsov:2025eiv,Odintsov:2025jky,Yogesh:2025wak,Mohammadi:2025gbu,Yogesh:2026esn,Mohammadi:2026iod,Gialamas:2025ofz,Katsoulas:2025srh,Bezerra-Sobrinho:2025gfg}). One approach involves modifying the original action by including higher-order curvature corrections. These additional geometric terms can alter the prediction of the original Starobinsky model, cf. ref.\,\cite{Addazi:2025qra,Ketov:2025cqg,Kim:2025dyi}. A different route explores the role of reheating: by considering a prolonged and colder reheating phase, the number of $e$-folds $\Delta N_\star$ associated with CMB scales increases. This shifts the model predictions in the $(n_s, r)$ plane toward slightly higher values of $n_s$, thus alleviating the tension with Planck+ACT, cf. refs.\,\cite{Drees:2025ngb,Liu:2025qca,Zharov:2025evb,Dorsch:2026fgb} (see also refs.\,\cite{Gialamas:2026pjo, Gialamas:2022xtt}).

The purpose of this paper is to critically examine these proposals from the standpoint of effective field theory and ultraviolet completions. \textbf{Rather than simply demonstrating that certain parameter choices can fit the data, we ask what the implications are from the perspective of ultraviolet-complete theories. Our analysis reveals that in each case, the required modifications introduce fine-tuned hierarchies, exotic ingredients, or deformations that undermine the minimality of the original Starobinsky model.}

The paper is organized as follows. In sec.\,\ref{sec:lightning_review}, we present a self-contained review of Starobinsky inflation, collecting the key analytical results in the slow-roll approximation and comparing them with the latest observational constraints. In sec.\,\ref{sec:higher_curvature_operators}, we analyze the effect of higher-curvature operators on the inflationary predictions. We first carry out a phenomenological analysis (sec.\,\ref{sec:phenoanalyis}), then we use dimensional-analysis arguments to discuss multi-particle unitarity bounds and interpret the results through the lens of effective field theory, thereby identifying the fundamental couplings and mass scales governing the effective description (sec.\,\ref{sec:EFTperspective}).
We next focus on concrete realizations of higher-curvature corrections to Starobinsky inflation, exploring the embedding within no-scale supergravity in Sec.~\ref{sec:SUGRA}. We then consider the higher-curvature structure arising from metric-affine gravity (sec.\,\ref{sec:MetricAffineGravity}), showing that pseudoscalar inflation can be mapped onto an higher-curvature extension of Starobinsky inflation. We then discuss the naturalness of the two effective theories introduced above, showing that Planck+ACT-compatible data inevitably drive this class of effective theories toward a regime of increased fine-tuning (Sec.~\ref{sec:DiscussingNaturalness}).
In sec.\,\ref{sec:Reh}, we examine the reheating route, deriving the consistency conditions linking the equation-of-state parameter $\omega_{\textrm{reh}}$ and the reheating temperature $T_{\textrm{reh}}$ to the spectral index $n_s$, and we discuss the physical implications of requiring a stiff post-inflationary phase. In sec.\,\ref{sec:stasis}, we explore the possibility that a phase of cosmological stasis---driven by a tower of decaying massive states---could serve as the post-inflationary epoch, and we derive the corresponding phenomenological constraints. We present our conclusions in sec.\,\ref{sec:conclusions}. Appendix\,\ref{app:Conve} contains a detailed discussion of the unitarity bounds for multi-particle scattering amplitudes, including the careful tracking of coupling dimensions. Appendix~\ref{app:MetricAffine} contains a brief, self-contained review of metric-affine gravity, including the explicit derivation of the pseudoscalar potential driving inflation discussed in this work, as well as a discussion of the torsionless limit. In Appendix~\ref{app:C}, we present a metric-affine modification of Starobinsky inflation which reproduces the same pseudoscalar potential and manifestly reduces to Starobinsky inflation in the torsionless limit.

\section{A lightning review of Starobinsky inflation}\label{sec:lightning_review}

In this section, we present the key analytic results for the Starobinsky model in the slow-roll approximation, derive the dependence of observable quantities on the number of $e$-folds, and assess their alignment with the latest cosmological constraints.

In the slow-roll approximation, the inflationary dynamics can be solved semi-analytically. 
For ease of reading, we introduce the function
\begin{align}
f_x \equiv -\frac{1}{3}
(3+2\sqrt{3})e^{-1-2/\sqrt{3} - 4x/3}\,.\label{eq:ShortHandfx}
\end{align}
We indicate with $\Delta N_{\star}$ the number of $e$-folds between the horizon crossing of the CMB pivot scale $k_{\star} \equiv 0.05$ Mpc$^{-1}$ and the end of inflation.
The field value at the end of inflation is
\begin{align}
\phi_{\textrm{end}} = 
\sqrt{\frac{3}{2}}
\log\left(1+\frac{2}{\sqrt{3}}\right)\MPl\,.
\end{align}
The field value at the CMB pivot  scale is
\begin{align}
\phi_{\textrm{CMB}} = 
-\frac{1}{\sqrt{6}}\big[&
3+2\sqrt{3}+4
\Delta N_{\star} 
+ 
\log(-135+78\sqrt{3}) + 3W_{-1}(f_{\Delta N_{\star}})\big]
\MPl\,,
\end{align}
where $W_{-1}(z)$ is the branch with 
$k=-1$ of the Lambert W function $W_k(z)$. 
The scalar spectral index at the CMB pivot scale reads
\begin{align}
n_s = 1-\frac{16}{3[1+W_{-1}(f_{\Delta N_{\star}})]^2} 
+ \frac{8}{
3[1+W_{-1}(f_{\Delta N_{\star}})]
}\,,
\end{align}
while for the tensor-to-scalar ratio we find
\begin{align}
r = 
\frac{64}{
3[1+W_{-1}(f_{\Delta N_{\star}})]^2
}\,.
\end{align}
The amplitude of the scalar power spectrum at the CMB pivot scale is
\begin{align}
A_s = \frac{3M^2[1+W_{-1}(f_{\Delta N_{\star}})]^4}{128\pi^2 \MPl^2 
W_{-1}(f_{\Delta N_{\star}})^2}\,.
\end{align}
Finally, the square of the Hubble rate at the end of inflation is 
\begin{align}
H(N_{\textrm{end}})^2\equiv H_{\textrm{end}}^2 =  
3\left(\frac{7}{2}-2\sqrt{3}\right)M^2\,,
\end{align}
while its value at the time of horizon crossing for the CMB pivot scale  $k_{\star}$ is
\begin{align}
H(N_{\star})^2\equiv
H_{\star}^2 = \frac{M^2}{4}
\left[
1+\frac{1}{W_{-1}(f_{\Delta N_{\star}})}
\right]^2\,.
\end{align}
The value of the Hubble rate at the $e$-fold time $N_k$ at which the generic comoving wavenumber $k$ crosses the inverse comoving Hubble radius (that is, the time $N_k$ defined by the condition $k=a(N_k)H(N_k)$) can be obtained from the scaling
\begin{align}
\frac{k}{k_{\star}}=\frac{
a(N_k)H(N_k)
}{
a(N_{\star})H(N_{\star})
}\to
H(N_k)\equiv H_k = 
\left(\frac{k}{k_{\star}}\right)e^{N_* - N_k}H_{\star}\,.
\end{align}

\begin{figure}[h]
\begin{center}
$$\includegraphics[width=.495\textwidth]{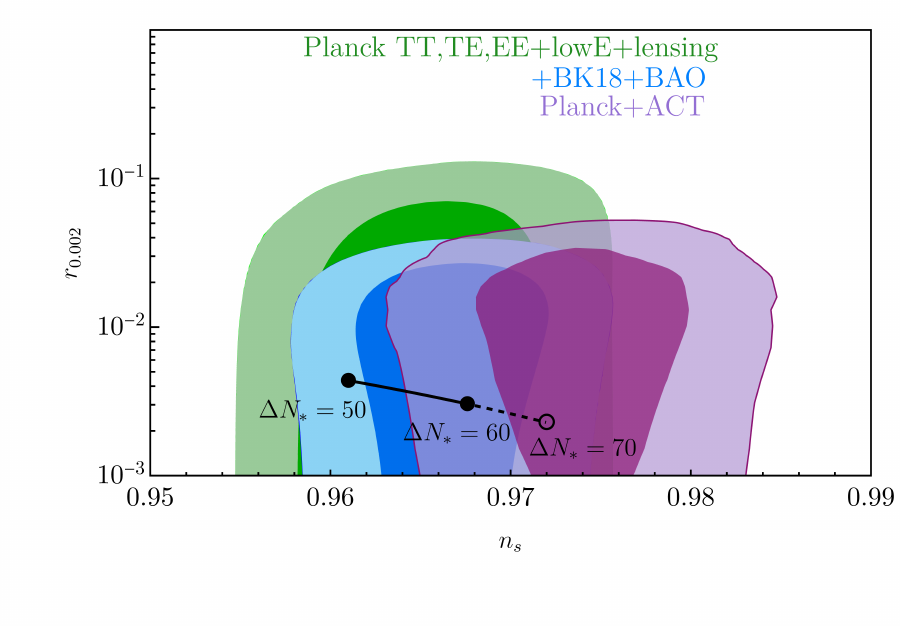}~~
$$\vspace{-0.5cm}
\caption{\em 
Comparison between the predictions of the Starobinsky inflation model (black line from $\Delta N_* =50$ to $\Delta N_* = 70$) and observational constraints on the scalar spectral index $n_s$ and tensor-to-scalar ratio $r_{0.002}$.
The shaded regions correspond to data from Planck (green), BICEP/Keck + BAO (blue), and Planck + ACT (purple), showing the 68\% and 95\% confidence contours. Here, 
$r_{0.002}$ denotes the tensor-to-scalar ratio evaluated at the pivot scale 
$k=0.002$ Mpc$^{-1}$, relevant for large angular scales in the CMB.
 }\label{fig:StaroBench}  
\end{center}
\end{figure}

These analytical predictions can be directly compared with current observational bounds. In fig.\,\ref{fig:StaroBench},we show the Starobinsky model predictions for two representative benchmark values of the number of $e$-folds: $\Delta N_\star = 50$ and $\Delta N_\star = 60$, shown as black filled dots. As expected, the model predicts a very low tensor-to-scalar ratio $r$ and a scalar spectral index $n_s$ slightly below unity, both consistent with Planck 2018 constraints. However, the combined Planck + ACT data introduce some tension with the Starobinsky predictions. In particular, the Planck+ACT data favor slightly higher values of $r$ and a spectral tilt closer to scale invariance. While Starobinsky inflation is still within the allowed region, it no longer sits at the center of the preferred contours. Notice that considering a prolonged period of inflation  $\Delta N_\star > 50$ (see the open circle in fig.\,\ref{fig:StaroBench}) comes with its own problems~\cite{Liddle:2003as} that will be explored in detail in sec.\,\ref{sec:Reh}.



\section{Starobinsky inflation and higher curvature operators}
\label{sec:higher_curvature_operators}

\subsection{Phenomenological analysis}\label{sec:phenoanalyis}

From the viewpoint of quantum gravity, 
it is natural to expect the presence of additional higher-curvature corrections in the effective action~\eqref{eq:FirstStaro}, such as terms proportional to $R^3$, $R^4$, and so on. Notably, even small corrections beyond the quadratic term can spoil the agreement with current CMB data, unless their coefficients are highly suppressed. Indeed, while the pure $R^2$ term leads to a flat plateau potential suitable for slow-roll inflation, higher-order corrections typically deform this shape, making the potential steeper or introducing new features~\cite{Sebastiani:2013eqa, Myrzakulov:2014hca}. 
%
%
Therefore, understanding and controlling higher-curvature terms is essential when embedding Starobinsky inflation into a UV-complete theory like supergravity or string theory.

In this section, we elaborate further on this point. Working in the general context of $f(R)$ theories of gravity~\cite{Sotiriou:2008rp,Nojiri:2010wj,Nojiri:2017ncd}, we show that, when rewritten in the Einstein frame, the $f(R)$ theory becomes equivalent to Einstein gravity minimally coupled to a scalaron field $\phi$ with a non-trivial potential.

We consider, in the spirit of ref.\,\cite{Brinkmann:2023eph}, the following gravitational action
\begin{align}\label{eq:StartingAction}
\mathcal{S} = \frac{\bar{M}_{\mathrm{Pl}}^2}{2} \int d^4x \sqrt{-g}\,f(R)
\end{align}
where
\begin{align}
f(R) = R + F(R)\,,~~~~~~~
F(R) \equiv \frac{R^2}{6M^2} 
+ \lambda_3 \frac{R^3}{M^4} + 
\lambda_4\frac{R^4}{M^6} + \dots 
= \frac{R^2}{6M^2} + \sum_{k = 3}^{k_{\textrm{max}}}
\lambda_k \frac{R^k}{M^{2k-2}}
\label{eq:fR}
\end{align}
In natural units, the Ricci scalar has dimension of a mass squared, $[R] = [\textrm{M}]^2$; 
the coefficients $\lambda_k$, therefore, are dimensionless numbers.
In eq.\,(\ref{eq:fR}), 
the lowest term of this type compatible with supersymmetry is $R^4$, cf. refs.\,\cite{Farakos:2013cqa,Ferrara:2013kca}. 
On the other hand, all powers of $R$, even and odd, starting from $R^4$ with arbitrary coefficients have
a supersymmetric generalization\,\cite{Ferrara:2013kca}. 
Classical and quantum
stability of the theory requires that 
$f^{\prime}(R) >0 $ and $f^{\prime\prime}(R) = F^{\prime\prime}(R) >0$, 
cf. ref.\,\cite{Appleby:2009uf}.

\begin{figure}[h]
\begin{center}
$$\includegraphics[width=.495\textwidth]{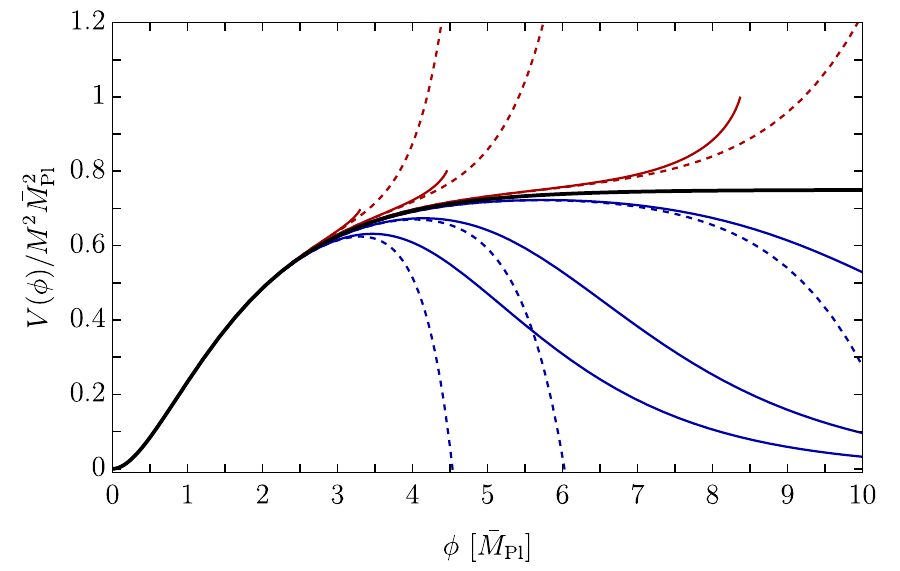}~~
\includegraphics[width=.495\textwidth]{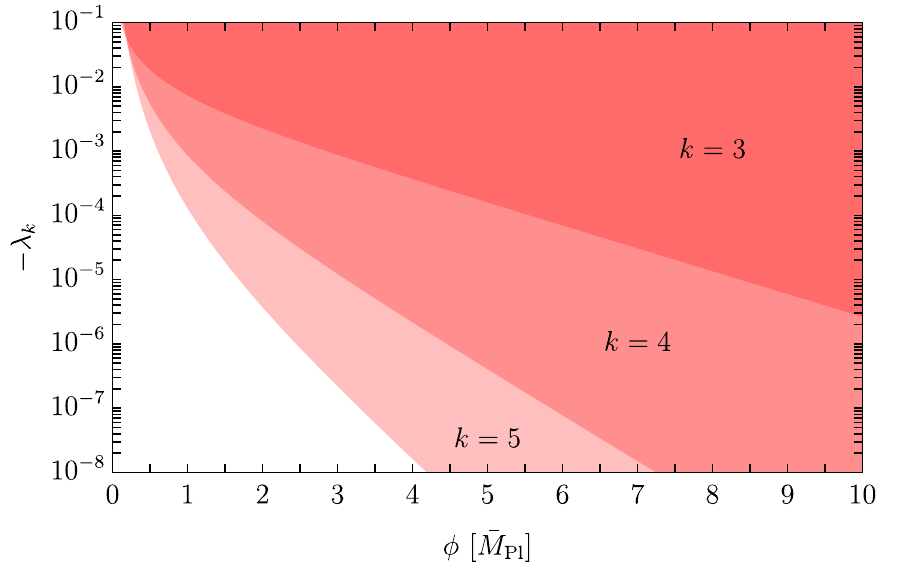}$$\vspace{-0.5cm}
\caption{\em
\textit{\textbf{Left panel:}} Einstein-frame potential $V^{(1)}_{k_{\textrm{max}}}(\phi)$ for the Starobinsky model supplemented by higher-curvature operators, shown for representative values of the coefficients $\lambda_k$. The solid black line corresponds to the baseline Starobinsky potential ($\lambda_k = 0$), while the colored curves illustrate the deformation induced by the leading corrections.
A negative sign of $\lambda_k$ steepens the potential at large field values, reducing the duration of slow-roll inflation and shifting the spectral index toward larger values.
\textit{\textbf{Right panel:}} 
Value of the inflaton field during slow-roll inflation as a function of $\lambda_k$, shown over a representative range of $\lambda_k$ values. Each $\lambda_k$ denotes the coefficient controlling the corresponding higher-curvature $R^k$ correction to the Starobinsky model, with $k=3,4,5.$
}\label{fig:modpotential}  
\end{center}
\end{figure}

A Legendre transformation allows us to recast the action in eq.\,(\ref{eq:StartingAction}) into a dynamically equivalent form in which the Lagrangian is linear in the scalar curvature, with the addition of an auxiliary scalar degree of freedom. 
We write
\begin{align}
  \mathcal{S} & = \frac{\bar{M}_{\mathrm{Pl}}^2}{2} \int d^4x \sqrt{-g}\,f(R) =   
  \frac{\bar{M}_{\mathrm{Pl}}^2}{2} \int d^4x \sqrt{-g}
  \left[
  f(\sigma) + f^{\prime}(\sigma)(R-\sigma)
  \right] 
 \nn\\
 &= 
  \bar{M}_{\mathrm{Pl}}^2
  \int d^4x \sqrt{-g}
  \left\{
  \frac{f^{\prime}(\sigma)}{2}R 
  - \frac{1}{2}
  \left[
  \sigma f^{\prime}(\sigma) - f(\sigma)
  \right]
  \right\}\,.
\end{align}
The action in the Einstein frame can be obtained via a Weyl transformation
\begin{align}
\bar{g}_{\mu\nu}\equiv \Omega^2 g_{\mu\nu}\,,~~~~~~~
\textrm{with}~~~\Omega^2 \equiv f^{\prime}(\sigma)\,. \label{eq:confTr}
\end{align}
Under such rescaling, in four space-time dimensions, we have
\begin{align}
 \sqrt{-g} &=  \Omega^{-4}\sqrt{-\bar{g}}\,,\\
R &= \Omega^2 \bar{R} 
+2(n-1)\bar{g}^{\mu\nu}
\Omega\big(
\bar{\nabla}_{\mu}
\bar{\nabla}_{\nu}\Omega
\big) - n(n-1)
\bar{g}^{\mu\nu}
\big(
\bar{\nabla}_{\mu}\Omega
\big)\big(
\bar{\nabla}_{\nu}\Omega
\big)\,,
\end{align}
with $n=4$. We thus have
\begin{align}
  \sqrt{-g}\,\Omega^2 R & =  
  \sqrt{-\bar{g}}
\left[
  \bar{R} 
+ 6\bar{g}^{\mu\nu}\Omega^{-1}\big(
\bar{\nabla}_{\mu}
\bar{\nabla}_{\nu}\Omega
\big) - 12
\bar{g}^{\mu\nu}
\Omega^{-2}
\big(
\bar{\nabla}_{\mu}\Omega
\big)\big(
\bar{\nabla}_{\nu}\Omega
\big)
  \right]\nn\\
  & = 
  \sqrt{-\bar{g}}
\left[
  \bar{R} 
  - 6
\bar{g}^{\mu\nu}
\Omega^{-2}
\big(
\bar{\nabla}_{\mu}\Omega
\big)\big(
\bar{\nabla}_{\nu}\Omega
\big)
  \right]\,,
\end{align}
where in the second line we performed an integration by parts (dropping the total derivative term). 
Moreover, we write 
\begin{align}
\bar{\nabla}_{\mu}\Omega = 
\bar{\nabla}_{\mu}\sqrt{f^{\prime}(\sigma)}= \frac{f^{\prime\prime}(\sigma)}{2\sqrt{f^{\prime}(\sigma)}}(\partial_{\mu}\sigma)\,.
\end{align}
We thus have
\begin{align}
  \sqrt{-g}\,\Omega^2 R = 
    \sqrt{-\bar{g}}
\left[
  \bar{R} 
  - 
\frac{3f^{\prime\prime}(\sigma)^2}{2}f^{\prime}(\sigma)^2
\bar{g}^{\mu\nu}(\partial_{\mu}\sigma)
(\partial_{\nu}\sigma)
  \right]
\end{align}
All in all, the action in the Einstein frame, $\mathcal{S}_{\textrm{E}}$, takes the form (we drop the symbol $\,\bar{}\,$ from all metric-dependent functions) 
\begin{align}\label{eq:EinsteinFrameFofR}
\mathcal{S}_{\textrm{E}} = \int d^4 x\sqrt{-g}\left[
\frac{\MPl^2}{2}R - \frac{h(\sigma)}{2}
g^{\mu\nu}
(\partial_{\mu}\sigma)(\partial_{\nu}\sigma) 
- V(\sigma)
\right]\,,
\end{align}
where
\begin{align}
h(\sigma) \equiv \frac{3\MPl^2
F^{\prime\prime}(\sigma)^2
}{2[1+ F^{\prime}(\sigma)]^2}\,,~~~~~~~~~~~~~
V(\sigma) \equiv 
\frac{\MPl^2[
\sigma F^{\prime}(\sigma) - F(\sigma)
]}{2[1+ F^{\prime}(\sigma)]^2}\,.\label{eq:potsigma}
\end{align}
In the above form, the scalar field $\sigma$ is not canonically normalized, and it has mass-dimension $[\sigma] = [\textrm{M}]^2$. 
If we define
\begin{align}
\varphi \equiv 1+ F^{\prime}(\sigma)\,,\label{eq:varphi} 
\end{align}
and introduce the field transformation 
\begin{align}
\varphi = \exp\left(
\sqrt{\frac{2}{3}}\frac{\phi}{\MPl}
\right)\,,\label{eq:FieldTra}
\end{align}
the action in the Einstein frame finally becomes
\begin{align}
\mathcal{S}_{\textrm{E}} = 
\int d^4 x\sqrt{-g}\left[
\frac{\MPl^2}{2}R - \frac{g^{\mu\nu}}{2}
(\partial_{\mu}\phi)(\partial_{\nu}\phi) 
- V(\phi)
\right]\,,~~~~\textrm{where}~~
V(\phi) = \frac{\MPl^2}{2\varphi^2}
\left\{
(\varphi - 1)\sigma(\varphi) - F[\sigma(\varphi)]
\right\}\,,\label{eq:EinsteinFrameAction}
\end{align}
where $\sigma(\varphi)$ can be obtained by solving eq.\,(\ref{eq:varphi}). 
More concretely, we write 
\begin{align}
1-\varphi + 
\frac{1}{3}\left(\frac{\sigma}{M^2}\right)
+
\sum_{k = 3}^{k_{\textrm{max}}}
k\lambda_k
\left(\frac{\sigma}{M^2}\right)^{k-1} = 0\,.\label{eq:Master}
\end{align}
In the limit $\lambda_k \to 0$, the previous equation gives 
$\sigma = 3M^2(\varphi - 1)$. 
Eq.\,(\ref{eq:Master}) admits $k_{\textrm{max}} - 1$ solutions, among which we select the one that continuously recovers the Starobinsky solution in the limit $\lambda_k \to 0$.
We are able to find exact analytic solutions up to $k_{\textrm{max}} = 5$. For larger values of $k_{\textrm{max}}$, or alternatively, we can construct perturbative solutions by expanding around small values of $\lambda_k$. In the left panel of fig.\,\ref{fig:modpotential} we show the corrected Starobinsky potential for different values of $k_{\textrm{max}}$ and $\lambda_k$, while in the right panel we show the constraint on $\lambda_k$ derived from the requirement that the higher-curvature corrections remain perturbative.
At the first order in $\lambda_k$, the potential we find takes the form 
\begin{align}
V^{(1)}_{k_{\textrm{max}}}(\phi) = 
\frac{3}{4} M^2 \MPl^2 \left[ 1 - \exp\left(-\sqrt{\frac{2}{3}} \frac{\phi}{\MPl} \right) \right]^2
\left\{
1-\sum_{k=3}^{k_{\textrm{max}}}
6\lambda_k \left[
-3+3\exp\left(\sqrt{\frac{2}{3}} \frac{\phi}{\MPl} \right) 
\right]^{k-2}
\right\}\,.
\end{align}

In fact, apart from the purpose of explicitly visualising the shape of the potential, there is no real need to go through this complicated inversion procedure, which is moreover applicable only in a limited set of cases.
The action in eq.\,(\ref{eq:EinsteinFrameAction}) delivers the EoM 
\begin{align}
\frac{d^2\phi}{dN^2} + 
\left[
3 - \frac{1}{2}\left(
\frac{d\phi}{dN}
\right)^2
\right]\left[\frac{d\phi}{dN} 
+ \frac{d\log V(\phi)}{d\phi}
\right] = 0\,,
\end{align}
where we wrote implicitly $\phi$ in units of the reduced Planck mass. 
Equivalently, we write 
\begin{align}
g(\sigma)
\frac{d^2\sigma}{dN^2}
+ g^{\prime}(\sigma)
\left(
\frac{d\sigma}{dN}
\right)^2 + 
\left\{
3 - \frac{1}{2}\left[
g(\sigma)
\frac{d\sigma}{dN}
\right]^2
\right\}
\left[g(\sigma)\frac{d\sigma}{dN} 
+ \frac{1}{g(\sigma)}\frac{d\log V(\sigma)}{d\sigma}
\right] = 0\,,\label{eq:EOM1}
\end{align}
with
\begin{align}
 g(\sigma) \equiv \frac{d\phi}{d\sigma}
 = \sqrt{\frac{3}{2}}\left[\frac{F^{\prime\prime}(\sigma)}{1+F^{\prime}(\sigma)}\right]
 \,,
\end{align}
and 
\begin{align}
\phi =    \sqrt{\frac{3}{2}}
\log\left[
1+ F^{\prime}(\sigma)
\right]\,.
\end{align}
\begin{figure}[h]
\begin{center}
$$\includegraphics[width=.495\textwidth]{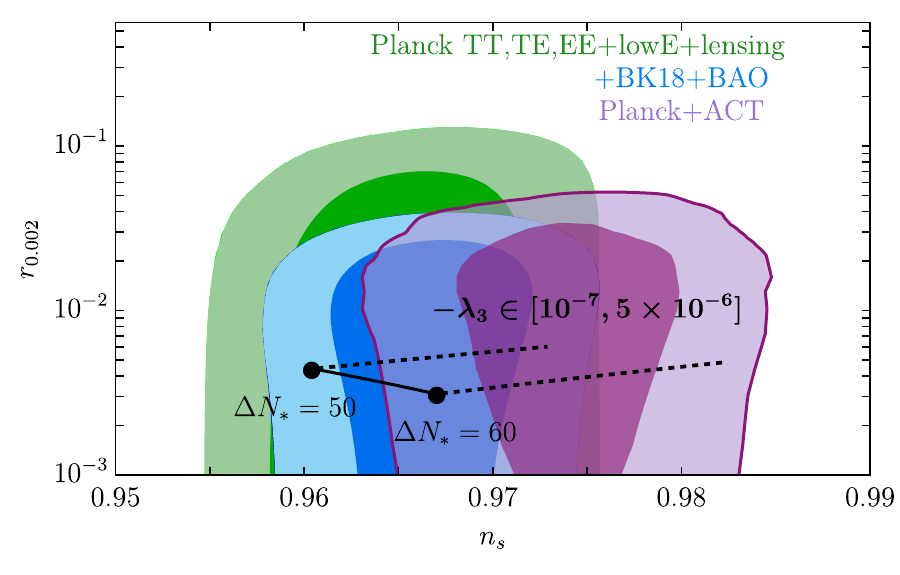}~~
\includegraphics[width=.495\textwidth]{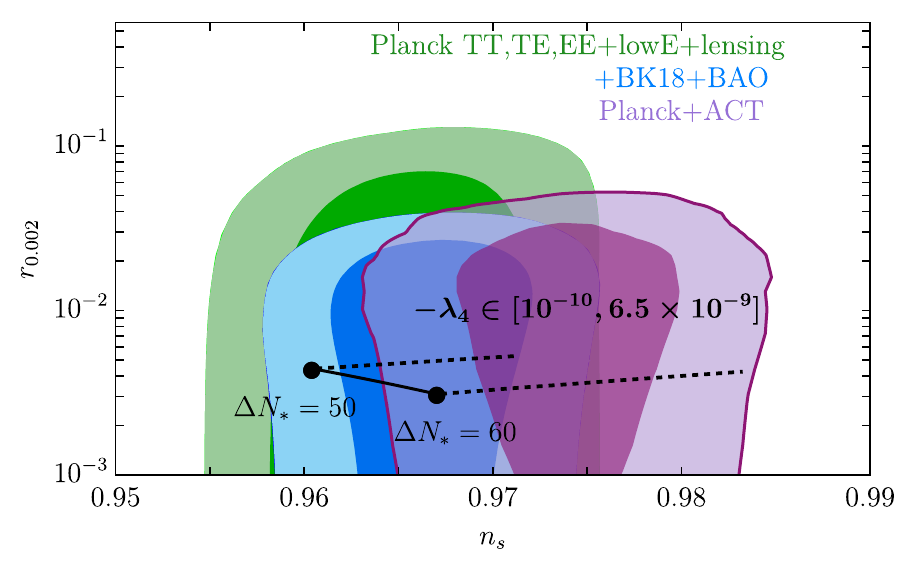}$$\vspace{-0.5cm}
\caption{\em
\textit{\textbf{Left panel:}} Predictions in the $(n_s, r)$ plane for the Starobinsky model extended by an $R^3$ operator with coefficient $\lambda_3 < 0$. The solid black line denotes the baseline Starobinsky prediction, while the dashed lines correspond to different values of $|\lambda_3|$, increasing in the direction of larger $n_s$.
\textit{\textbf{Right panel:}} Same as the left panel, but for the $R^4$ operator with coefficient $\lambda_4 < 0$ (and $\lambda_3 = 0$).
In both cases, a negative sign of the higher-curvature coefficient is required to shift the predictions toward the region favored by the Planck+ACT data combination (purple contours). The shaded regions correspond to the observational constraints from Planck 2018 (green), BICEP/Keck + BAO (blue), and Planck + ACT (purple), showing the 68\% and 95\% confidence contours. We require $50 \leq \Delta N_\star \leq 60$ $e$-folds of inflation after CMB modes exit the horizon.
}\label{fig:StaroFit}  
\end{center}
\end{figure}
The slow-roll parameters $\epsilon_V$ and $\eta_V$ takes the form
\begin{align}
\epsilon_V &\equiv \frac{\MPl^2}{2}
\left[\frac{V^{\prime}(\phi)}{V(\phi)}\right]^2 = 
\frac{\MPl^2}{2}\left[
\frac{1}{g(\sigma)}
\frac{V^{\prime}(\sigma)}{V(\sigma)}
\right]^2\,,\\
\eta_V &\equiv
\MPl^2
\frac{V^{\prime\prime}(\phi)}{V(\phi)}
= 
\frac{\MPl^2}{V(\sigma)}
\left[
\frac{V^{\prime\prime}(\sigma)}{g(\sigma)^2}  - 
\frac{g^{\prime}(\sigma)}{g(\sigma)^3}
V^{\prime}(\sigma)
\right]\,.
\end{align}
eq.\,(\ref{eq:EOM1}) describes  
the inflationary dynamics, and can be solved without any approximation using the 
potential $V(\sigma)$ in eq.\,(\ref{eq:potsigma}). 
We show our result in fig.\,\ref{fig:StaroFit}, where we show the effect of the $R^3$ and $R^4$ operators in the left and right panels, respectively, when each is switched on individually.
From this analysis, two important observations emerge.
\begin{itemize}
  \item[$\circ$] First, we find that a negative sign of the coefficient is required in order to shift the cosmological predictions toward the confidence region favored by the Planck+ACT data combination.
  \item[$\circ$] Second, we observe that an extremely small value of the coefficients \( \lambda_k \) is sufficient to alter the predictions of the baseline Starobinsky model.
\end{itemize}
These conclusions are in full agreement with previous studies (cf. refs.\,\cite{Addazi:2025qra,Ketov:2025cqg,Cheong:2025vmz,Qiu:2025uot}). However, in the following, we aim to provide a physical interpretation of these phenomenological results.

\subsection{The effective field theory perspective}\label{sec:EFTperspective}

In the MKS system, there are three fundamental units: length [L], mass [M], and time [T]. 
From these three fundamental units, all other units (such as force, energy, pressure, etc.) can be derived. Specifically, the energy [E] is given by the combination $[\textrm{E}]=[\textrm{M}][\textrm{L}]^2/[\textrm{T}]^2$. 
Let us now consider a system of units in which $c=\epsilon_0=1$ (where $\epsilon_0$ 
is the vacuum permittivity). In this case, we have [L]$=$[T] and, consequently, only two fundamental units remain, namely [L] and [M]. Since [E]$=$[M], we can work with 
[L]$=$[T] and [E].
The action $\mathcal{S}$ has the dimension of energy multiplied by time, 
$[\mathcal{S}] = [\textrm{E}][\textrm{L}]$, which corresponds to the dimensionality of the Planck constant. Therefore, we can write $[\mathcal{S}] = [\hbar] = [\textrm{E}][\textrm{L}]$.
It follows that the dimension of a Lagrangian density is given by 
$[\mathcal{L}] = [\hbar]/[\textrm{L}]^4$. 
From the corresponding canonical kinetic terms, we therefore obtain that the dimensionalities of a bosonic field $\phi$ and a fermionic field $\Psi$ are, respectively, 
$[\phi] = [\hbar]^{1/2}/[\textrm{L}]$ and 
$[\Psi] = [\hbar]^{1/2}/[\textrm{L}]^{3/2}$. 
It follows that a mass parameter $m_*$, whether bosonic or fermionic, appearing in the Lagrangian has the dimension $[m_*]=[\textrm{L}]^{-1}$ that is, it corresponds to the inverse of a length scale (consequently, note that $[m_*]=[\textrm{M}]/[\hbar]$). 
A typical coupling $g_*$ in the Lagrangian, such as a Yukawa coupling or a gauge coupling, has the dimension $[g_*]=[\hbar]^{-1/2}$.
In natural units, the dimensionality of the coupling is obscured, and one might be tempted to treat $g_{*}$ as if it were a dimensionless numerical constant. However, by reinstating $\hbar$, the fundamental distinction between 
$g_{*}$ and a pure number becomes evident.
In gravity, the Ricci scalar has the dimension $[R]=[\textrm{L}]^{-2}$ (since it is quadratic in spatial derivatives). Consequently, from the Einstein-Hilbert Lagrangian, we observe that the (reduced) Planck mass has units 
$[\MPl] = [\hbar]^{1/2}/[\textrm{L}]$.
Motivated by the previous relation, it is 
useful to introduce the coupling dimension $[\textrm{C}]$ such that 
$[\textrm{C}] \equiv [\hbar]^{-1/2} = [\textrm{E}]^{-1/2}[\textrm{L}]^{-1/2}$. 
The Lagrangian density has dimensions
\begin{align}
   [\mathcal{L}] = \frac{1}{[\textrm{C}]^2[\textrm{L}]^4}\,,
   \label{eq:LagrangianScaling}
\end{align}
and we also have 
\begin{align}
[\textrm{C}][\MPl] = [\textrm{L}]^{-1}\,,\label{eq:PlanckScaling}
\end{align}
that is, 
the product of the Planck mass times 
a fundamental coupling is the inverse of a length scale. 
Assuming the existence of one fundamental coupling $g_*$ and one fundamental scale $m_*$, 
we write the Lagrangian density in the form
\begin{align}
\mathcal{L} = 
\frac{m_*^4}{g_*^2}
\hat{\mathcal{L}}\left(
\frac{\partial}{m_*},
\frac{R}{m_*^2},
\frac{g_*\phi}{m_*},\frac{g_*\Psi}{m_*^{3/2}}
\right)\,,
\end{align}
where, assuming all fields canonically normalized, $\hat{\mathcal{L}}$ 
is a dimensionless polynomial functional with arbitrary order-one
numerical coefficients.

We now apply these considerations to the Starobinsky model in eq.\,(\ref{eq:FirstStaro}).
This theory can be considered of the one-coupling–one-scale type, since a fundamental coupling $g_*$ can be defined through the relation $g_*\MPl \equiv M$, which satisfies the scaling in eq.\,(\ref{eq:PlanckScaling}), with $M$ being a mass parameter (which plays the role of $m_*$ in the previous discussion).
Based on dimensional analysis, therefore,
we rewrite the action in the form
\begin{align}
\mathcal{S} = \int d^4x\sqrt{-g}\left(
\frac{\MPl^2}{2}R
+ \frac{R^2}{12 g_{*}^2}
\right)\,. \label{eq:StaroDim}
\end{align}
In terms of the canonically normalized scalar field $\phi$, at large field values, where the CMB observables are reproduced, the potential flattens, and dimensional analysis gives the scaling 
$V_{\textrm{CMB}} = O(M^4/g_*^2) = O(g_*^2 \MPl^4) = O(\MPl^2 M^2)$, 
in agreement with eq.\,(\ref{eq:StaroPotPhi}).
Using the slow-roll approximation and the requirement $r \lesssim 0.036$, we obtain an upper bound on the underlying coupling
\begin{align}
g_* \lesssim O(10^{-5})\,.\label{eq:WeakCoupling}
\end{align}
Consequently, $M \lesssim O(10^{-5})\MPl 
\ll \MPl$.

A common concern in inflationary effective field theories is the scale at which
perturbative unitarity is violated, i.e.~the energy $\Lambda_{\rm UV}$ at which
tree-level scattering amplitudes exceed partial-wave unitarity bounds and new
physics must enter, cf. ref.\,\cite{Burgess:2009ea,Hertzberg:2010dc,Burgess:2010zq}.
The consensus is that pure $R+R^{2}$ inflation is free of
sub-Planckian unitarity problems, with $\Lambda_{\rm UV}\sim M_{\rm Pl}$, cf. ref.\,\cite{Kehagias:2013mya,Cheong:2020rao}, and ref.\,\cite{He:2023fko} for a geometrical formulation. 

We here provide a simple argument based on dimensional analysis. The following discussion is motivated by the fact that ref.\,\cite{Burgess:2010zq} raised the concern that multi–particle final–state scattering processes, such as $\phi\phi \to \phi\phi\phi\phi$, could lead to a violation of tree–level unitarity at the scale $M$. 
However, even simply in light of the previous discussion, it is clear that it is not possible to identify the energy scale $\Lambda_{\rm UV}$ with the scale $M$ alone, as this would be incorrect already at the dimensional level, since $[M]=[\textrm{L}]^{-1}$ while $[\textrm{E}]=[\hbar][\textrm{L}]^{-1}$. In other words, it is a combination of mass parameters, i.e., inverse length scales, and couplings that defines the energy cutoff of the effective field theory, and not $M$ alone.
Let us further corroborate this point of view with the help of few examples. 

We start from some generic considerations. 
The scattering amplitude for a $n\to m$ process has dimension (cf. appendix\,\ref{app:Conve})
\begin{align}
[\mathcal{M}_{n\to m}] = [\textrm{C}]^{-2+n+m}[\textrm{L}]^{n+m-4}\,.\label{eq:MasterScaling}
\end{align}
This scaling is completely general and holds independently of the perturbative order at which the amplitude is computed. 
Consequently, we have $[\mathcal{M}_{2\to 2}] = [\textrm{C}]^{2}$ while $[\mathcal{M}_{2\to N}] = [\textrm{C}]^{N}[\textrm{L}]^{N-2}$. 

We are interested in the bounds placed on 
tree-level scattering amplitudes by unitarity. 
We extract the s-wave contribution to the scattering amplitude by evaluating the phase-space averaged scattering matrix
element that we denote with $\hat{\mathcal{M}}_{n\to m}$. We find (cf. appendix\,\ref{app:Conve})
\begin{align}
[\hat{\mathcal{M}}_{n\to m}] =  
[\textrm{C}]^{-2+n+m}\,.\label{eq:scalingAverage}
\end{align}
The phase-space averaged scattering matrix
element, therefore, verifies a precise scaling rule in terms of powers of quantities with dimension of a coupling. For example, in a one-coupling/one-scale EFT, the scaling of the scattering matrix will be saturated by an appropriate power of $g_*$, while the mass 
$M$
 can only enter through the dimensionless ratio $E/M$.

We consider the interactions contained in eq.\,(\ref{eq:StaroPotPhi}) after expanding around 
$\langle\phi\rangle$. We find the Lagrangian density 
\begin{align}
\mathcal{L} = \frac{1}{2}(\partial_{\mu}\phi)(\partial^{\mu}\phi) - \frac{1}{2}M^2\phi^2   
+ \frac{g_* M\phi^3 }{\sqrt{6}}
- \frac{7g_*^2\phi^4}{36} + 
\frac{g_*^3\phi^5}{6\sqrt{6}M} +\dots\,.\label{eq:StaroExp}
\end{align}
The generic structure of the contact interaction term with $N$ fields is, therefore, given by 
\begin{align}
\mathcal{L}_{\textrm{int}}^{(N)} =  
\frac{c_N g_*^{N-2}\phi^N}{M^{N-4}}\,,
\end{align}
in agreement with eq.\,(\ref{eq:LagrangianScaling}). 
In the above expression, $c_N$ are pure dimensionless numbers that can be extracted from the expansion in eq.\,(\ref{eq:StaroExp}).
The tree-level scattering amplitude for the process $n\to m$ is given by 
\begin{align}\label{eq:ScalingScatteringAMplipt}
\mathcal{M}_{n\to m} = c_{n+m}(n+m)!
\frac{g_*^{n+m-2}}{M^{n+m-4}}\left[
1+O\left(\frac{M^2}{E^2}\right)
\right]\,.
\end{align}
This expression  has already the correct coupling dimension to saturate the one in eq.\,(\ref{eq:scalingAverage}); the first term originates from the contact interaction $\mathcal{L}_{\textrm{int}}^{(n+m)}$, whereas the correction arises from tree-level diagrams in which a propagator is exchanged between interaction vertices of the form $\mathcal{L}_{\textrm{int}}^{(n+1)}$ and $\mathcal{L}_{\textrm{int}}^{(m+1)}$. 
In the high-energy limit $E\gg M$, the contact interaction dominates, and we henceforth focus on this contribution.

We now average over the initial and final phase space. 
We consider the limit in which $\phi$ is massless. 
Since we are considering contact interactions without any momentum dependence, we are allowed to replace the phase space integrals with the corresponding volumes. 
eq.\,(\ref{eq:ScalingScatteringAMplipt}) already contains the right power of coupling $g_*$ to reproduce the scaling in eq.\,(\ref{eq:scalingAverage}). 
Based solely on dimensional analysis, therefore, 
we expect 
\begin{align}
 \hat{\mathcal{M}}_{n\to m} \sim    
 g_*^{n+m-2}\left(\frac{E}{M}\right)^{n+m-4}\,.
\end{align}
A more precise computation gives
\begin{align}
\hat{\mathcal{M}}_{n\to m} 
= \left(\frac{\textrm{Vol}_n\textrm{Vol}_m}{n!m!}\right)^{1/2}\mathcal{M}_{n\to m} 
= \frac{
c_{n+m}(n+m)!
}{
8\pi[
n!m!(n-1)!(n-2)!(m-1)!(m-2)!
]^{1/2}
}\,g_*^{n+m-2}\,
\left(\frac{E}{4\pi M}
\right)^{n+m-4}\,,
\end{align}
which has the correct scaling in terms of powers of coupling as dictated by eq.\,(\ref{eq:scalingAverage}).
The matrix elements $|\hat{\mathcal{M}}_{n\to m}|$ are directly constrained by unitarity to be $\leq 1$. Consider the case $2\to m$. 
We find 
\begin{align}
\hat{\mathcal{M}}_{2\to m} = 
 \frac{
c_{2+m}(m+2)!
}{
8\pi[
2 m!(m-1)!(m-2)!
]^{1/2}
}\,g_*^{m}\,
\left(\frac{E}{4\pi M}
\right)^{m-2}\,.
\end{align}
In the limit of large $m$, the previous expression simplifies to 
\begin{align}
   \hat{\mathcal{M}}_{2\to m} \overset{m\gg 2}{=} 
 \frac{
c_{2+m}
}{
8\pi(
2 m!)^{1/2}
}
\left(\frac{g_* E}{4\pi M}
\right)^{m} = 
 \frac{
c_{2+m}
}{
8\pi(
2 m!)^{1/2}
}
\left(\frac{E}{4\pi \MPl}
\right)^{m}\,,
\end{align}
where powers of $g_*$ and $M$ combine into $\MPl$.
We thus find the unitarity bound for each $m$
\begin{align}
\frac{E}{4\pi \MPl} \leq \left[
\frac{8\pi(2m!)^{1/2}}{|c_{2+m}|}
\right]^{1/m} = \hspace{-0.3cm}
\vcenter{\hbox{\includegraphics[width=0.45\linewidth]{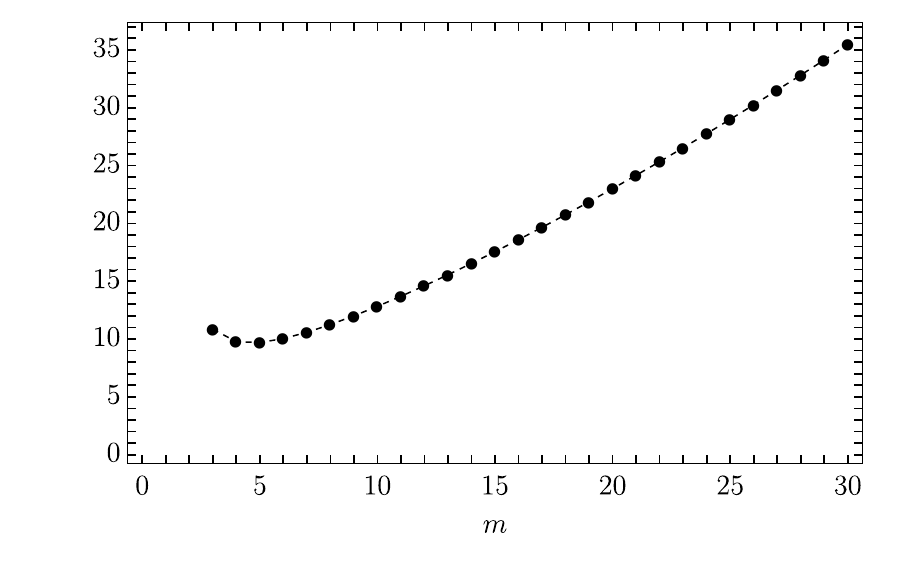}}}
\end{align}
with the numerical value of the right-hand side shown in the plot as function of $m$. 
This analysis shows that imposing tree-level unitarity on multi-particle processes does not lower the 
cut-off of the Starobinsky model below $\MPl$.

We now consider the presence of higher-curvature operators. 
For definiteness, we consider the case in which we include quadratic and cubic curvature corrections. 
Based solely on dimensional analysis, 
let us write the action in the form
\begin{align}
\boxed{
\mathcal{S} = \int d^4x\sqrt{-g}\left(
\frac{\MPl^2}{2}R 
+ \frac{R^2}{12 g_{*}^2} - 
\frac{R^3}{g_{*}^2\Lambda^2}
\right)\,
}  \label{eq:RewriteL}
\end{align}
where $g_{*}$ is a fundamental coupling while $\Lambda$ is a mass scale. 
The coefficient of the cubic term is negative, in accordance with the results of the phenomenological analysis carried out earlier, and should be understood as multiplied by a factor of order unity. For simplicity, we set this factor exactly to one. This choice is motivated by the fact that in the following sections we analyze distinct concrete realizations of the effective model in \eqref{eq:RewriteL}, arising from different underlying theories. 

We remark the following points.
\begin{itemize}
\item[$\circ$] 
Under the global scale transformation 
$g_{\mu\nu}(x) \to \lambda^2 g_{\mu\nu}(x)$
in four spacetime dimensions, one finds 
$\sqrt{-g} \to \lambda^4 \sqrt{-g}$ and $R \to \lambda^{-2} R$.
Consequently, the combination $\sqrt{-g}\,R^2$ is scale invariant, whereas 
$\sqrt{-g}\,R$ and $\sqrt{-g}\,R^3$ are not. 
The lack of invariance of the latter two operators explains why they must be accompanied by appropriate dimensionful scales in the action.
\item[$\circ$] The theory in eq.\,(\ref{eq:RewriteL}) has two, in principle different, mass scales namely $\Lambda$ and the combination $g_{*}\MPl$. 
It is therefore an effective theory of the one-coupling, two-scale type.
\item[$\circ$] 
The inflation observed at CMB scales takes place in a well-defined, fine-tuned regime, namely one in which the curvature $R$ is such that the quadratic term dominates. This corresponds to the condition
\begin{align}
  g_{*}^2 \MPl^2 \ll R \ll \Lambda^2\,.\label{eq:SeparationOfScales}  
\end{align}
\end{itemize}
In this perspective, the dimensionless coefficient $\lambda_3$ in eq.\,\eqref{eq:fR} reads 
\begin{align}
 \lambda_3 \equiv -\frac{2g_{*}^2 \MPl^2}{\Lambda^2}\,,
\end{align}
and the phenomenological condition 
$\lambda_3 \ll 1$ reflects the 
separation of scales implied by 
eq.\,(\ref{eq:SeparationOfScales}). 
Furthermore, we have the identification 
$M\equiv g_{*}\MPl$.
On the contrary, writing 
\begin{align}
  \mathcal{S} = \int d^4x\sqrt{-g}\left(
\frac{\MPl^2}{2}R 
+ \frac{R^2}{12 g_{*}^2} +\frac{\lambda_3}{2} 
\frac{R^3}{g_{*}^2 M^2}
\right)\,,    
\end{align}
makes the effective theory of the one-coupling, one-scale type. 
Consequently, requiring $\lambda_3 \ll 1$ appears unnatural. 
Although this may seem like a matter of semantics, distinguishing between one-scale and two-scale scenarios makes a substantial difference in the context of concrete realisations, as the underlying theoretical origin and the degree of naturalness of the required parameter hierarchies can be markedly different.

Starobinsky inflation thus operates on the delicate balance between two sources of explicit breaking of scale invariance.
The fragility of this solution becomes particularly apparent in light of the fact that there is no compelling reason for scale invariance to be a high-quality symmetry: in the infrared, it must be broken in order to recover Einstein gravity, while in the ultraviolet it is part of the common lore that a quantum theory of gravity does not accommodate global symmetries.

Undoubtedly, while corrections from higher-curvature operators are to be expected in the spirit of an effective field theory, it is equally undeniable that such sensitivity to ultraviolet physics undermines, to some extent, the original motivation behind the Starobinsky model, as {\it \textbf{inflation becomes a particular solution within a broader landscape of possible scenarios}}. 
This type of situation is referred to as accidental inflation, especially in the context of string-derived models\,\cite{Linde:2007jn}. The term highlights the idea that inflation is not a generic or inevitable outcome of the underlying theory, but rather the result of a specific - and possibly rare - choice of parameters within a high-dimensional theory landscape.

Moreover, achieving the hierarchy in eq.\,(\ref{eq:SeparationOfScales}) proves to be far from straightforward when attempting to embed the effective theory into the framework of a more fundamental underlying theory. 
With a possible embedding into string theory in mind, it would be desirable, for instance, if the coupling $g_*$ 
could be identified with the string coupling $g_s$, that controls the perturbative expansion in the string genus, and the mass scale $\Lambda$ with a mass scale associated to string theory. In the latter framework, two characteristic mass scales typically arise: the string mass scale $M_s = (\alpha^{\prime})^{-1/2} = \ell_s^{-1}$, which sets the energy threshold for the excitation of fundamental string modes, and the Kaluza–Klein (KK) mass scale $M_{\textrm{\tiny{KK}}}$, which corresponds to the inverse of the typical compactification radius and governs the excitation of the lowest modes associated with the compact extra dimensions. 
The dimensionless volume in string units 
is defined by $\mathcal{V} \equiv V_6/\ell_s^{6} \equiv (R/\ell_s)^6$ under the assumption of isotropic $R$ (that is, all compactified six extra dimensions have the same size $R$). 
Given that $M_{\textrm{\tiny{KK}}} \sim 1/R$, from the previous definition we find 
$M_{\textrm{\tiny{KK}}} \sim 1/R   = M_s/\mathcal{V}^{1/6}$. 
In string compactifications, the large-volume, small-coupling limit refers to the region of moduli space where the dimensionless compactification volume $\mathcal{V}\gg 1$ and the string coupling $g_s \ll 1$. In this regime, perturbative control is ensured: higher-genus worldsheet corrections are suppressed by powers of $g_s$, while $\alpha^{\prime}$ corrections - higher-derivative terms in the low-energy effective action arising from the finite length of the string -  are suppressed by inverse powers of $\mathcal{V}$. 
In this regime, we thus typically have
$M_s \gg M_{\textrm{\tiny{KK}}}$. 
The four-dimensional Planck scale becomes a derived scale in string theory. It is related to the string scale, the Kaluza-Klein scale and the string
coupling via
\begin{align}
g_s \MPl \approx \left(\frac{M_s}{M_{\textrm{\tiny{KK}}}}\right)^3 M_s \gg M_s\,. 
\end{align}
In most controlled treatments of string compactifications, therefore, we expect the hierarchy of scales
\begin{align}
M_{\textrm{\tiny{KK}}} \ll M_s 
\ll g_s \MPl\,,
\end{align}
which is the opposite compared to what eq.\,(\ref{eq:SeparationOfScales}) seems to suggest.

The preceding scalings should be regarded as order-of-magnitude estimates that hold in unwarped, isotropic compactifications and at tree level. They can be quantitatively modified by several effects: warping (which redshifts local mass scales and can lower the relevant KK thresholds), anisotropy (different radii produce multiple KK scale), as well as string-loop corrections (which can be non-negligible if curvatures are not parametrically small or if $g_s$ 
is not extremely weak). 
Consequently, while the hierarchy $M_{\textrm{\tiny{KK}}} \ll M_s 
\ll g_s \MPl$ 
is typical in controlled regimes, specific constructions can display deviations from these naive estimates~\cite{Chakraborty:2025wqn}.

In conclusion, if one insists on interpreting the shift toward larger values of $n_s$, as suggested by the Planck+ACT data, as being due to the presence of higher-curvature operators within the framework of the Starobinsky model, then one is inevitably led to introduce two distinct mass scales and to assume a parametric separation between them. Such a separation, however, does not admit an immediate interpretation in terms of models in which the Starobinsky action arises from controlled string-theoretic setups.

A direct identification of the scales entering the inflationary effective action with the fundamental string scales, such as $M_{\textrm{\tiny{KK}}}$ or $M_s$, can be overly simplistic. In realistic string compactifications, the path from the ten-dimensional string theory to four-dimensional inflationary dynamics is typically mediated by an intermediate stage in which the relevant degrees of freedom and interactions are described by a supergravity theory. This low-energy, four-dimensional supergravity arises after integrating out the massive string and Kaluza–Klein excitations, and it encodes the dynamics of the massless spectrum, including the moduli, gauge fields, and fermions. The inflationary sector is then embedded within this supergravity framework, and its parameters are determined not only by the underlying string scales, but also by the details of moduli stabilisation, supersymmetry breaking, and potential couplings to the rest of the light spectrum. As a result, the effective scales governing inflation may be related to, but need not coincide with, the naive values of $M_{\textrm{\tiny{KK}}}$ or $M_s$ obtained from simple compactification arguments.


\subsection{No-scale supergravity embedding of Starobinsky inflation with higher-curvature operators}\label{sec:SUGRA}
When string theory is examined at energies much lower than both the string scale $M_s$ and the Kaluza--Klein scale $M_{\textrm{\tiny{KK}}}$, the massive excitations of the string and the heavy modes associated with the compact extra dimensions decouple from the dynamics. The only degrees of freedom that remain relevant are the massless modes of the theory, including the graviton, its supersymmetric partners such as the gravitino, gauge bosons, and various scalar and fermionic fields. The resulting theory that describes the interactions of these fields is a supergravity theory (that is, a theory with local supersymmetry), which can therefore be viewed as the low-energy effective field theory of string theory.

In the context of cosmology, and in particular inflation, the supergravity theory obtained from string compactification provides the field content and interactions from which inflationary dynamics can emerge. The inflaton is identified with one of the scalar degrees of freedom present in the 4D effective theory. The specific choice of inflaton depends on the details of the compactification and on the mechanism used to stabilise the moduli, i.e.\ the scalar fields associated with the geometry and couplings of the extra dimensions. Moduli stabilisation is designed to fix the vacuum expectation values of these fields and to give them large masses, so as to remove unwanted light degrees of freedom from the low-energy spectrum. In an inflationary setup, however, one deliberately preserves a single scalar direction---identified with the inflaton---that remains sufficiently light after stabilisation to sustain a period of slow-roll inflation. The remaining scalar modes of the theory---namely, the moduli other than the inflaton---are therefore assumed to be stabilised with masses above the Hubble scale during inflation. This ensures that their quantum fluctuations are exponentially suppressed and that they can be consistently integrated out of the low-energy dynamics. As for the non-scalar fields in the spectrum, such as gauge bosons, fermions, and the gravitino, their masses and couplings depend on the details of supersymmetry breaking, and in realistic scenarios they are typically required either to be sufficiently heavy to avoid affecting the inflationary background (with the possible exception of the gravitino), or to interact so weakly that they remain spectators to the inflationary dynamics.

A particularly interesting class of supergravity theories that often emerges from string compactifications is provided by no-scale supergravity\,\cite{Cremmer:1983bf,Ellis:1983sf,Ellis:1983ei,Ellis:1984bm,Ellis:1984bs,Lahanas:1986uc} (cf. ref.\,\cite{Ellis:2020lnc} for a review). These models are characterised by a specific form of the Kähler potential that leads, at tree level, to a vanishing scalar potential along certain directions in field space, corresponding to flat moduli directions. This property makes no-scale supergravity especially appealing for cosmology: the flatness of the potential can naturally accommodate an inflaton candidate, while the remaining moduli can be stabilised through subleading corrections. 
Moreover, no-scale models often arise in controlled limits of string theory, particularly in large-volume compactifications, and thus provide a natural bridge between the UV structure of the theory and the low-energy inflationary dynamics. In what follows, we focus on this class of models as a concrete realisation of the effective supergravity framework introduced above.

The minimal
no-scale model appropriate for inflation involves two complex fields, cf. ref.\,\cite{Antoniadis:2025pfa,Ellis:2013xoa}.
The scalar potential for the canonical field $\phi$ takes the form
\begin{align}\label{eq:SUGRAPOT}
V(\phi) =  3M^2\MPl^2 \sinh^2\left(\frac{\phi}{\sqrt{6}\MPl}\right)\left[
\cosh\left(\frac{\phi}{\sqrt{6}\MPl}\right)
-\lambda\sinh\left(\frac{\phi}{\sqrt{6}\MPl}\right)
\right]^2\,.
\end{align}
Remarkably, for $\lambda = 1$ eq.\,\eqref{eq:SUGRAPOT} reduces to the Starobinsky potential in eq.\,\eqref{eq:StaroPotPhi}. In this section, we analyze the corrections arising when $\lambda \equiv 1-\epsilon$, with $\epsilon$ a small parameter. To this end, it is convenient to recast eq.\,\eqref{eq:SUGRAPOT} into an $f(R)$ theory proceeding as follows.

From eq.\,\eqref{eq:potsigma}, 
recalling the identification $\sigma = R$, we can then write the system (cf. also ref.\,\cite{Ivanov:2021chn})
\begin{align}
R(\varphi) = \frac{2\varphi}{\MPl^2}\left[
2V + \varphi\frac{dV(\varphi)}{d\varphi}
\right]\,,~~~~~~~
f(\varphi) = 
\frac{2\varphi^2}{\MPl^2}\left[
V + \varphi\frac{dV(\varphi)}{d\varphi}
\right]\,.
\end{align}
An inverse-engineering strategy is thus possible. If the potential $V(\phi)$ is known, the potential $V(\varphi)$ can be obtained through eq.\,(\ref{eq:FieldTra}). Consequently, the above system can be solved by eliminating the parametric dependence on $\varphi$, thereby allowing one to write $f(R)$ (and, consequently, $F(R)$).

We apply this procedure to the potential in eq.\,(\ref{eq:SUGRAPOT}), finding
\begin{align}
F(R) = \frac{M^2}{8\epsilon^2}\left\{
-1 + {}_{2}F_{1}\bigg(-\frac{2}{3},-\frac{1}{3},\frac{1}{2},\frac{3R^2\epsilon^2}{M^4}\bigg)
+ \frac{8R\epsilon}{M^2}
\left[
-1 + {}_{2}F_{1}\bigg(-\frac{1}{6},\frac{1}{6},\frac{3}{2},\frac{3R^2\epsilon^2}{M^4}\bigg)
\right]\right\}\,,
\end{align}
which admits the expansion 
\begin{align}\label{eq:F(R)noscale}
F(R) = \frac{R^2}{6M^2}
\left[
1 - \frac{1}{3}
\left(\frac{\epsilon R}{M^2}\right)+ 
\frac{2}{9}
\left(\frac{\epsilon R}{M^2}\right)^2 
- 
\frac{7}{36}
\left(\frac{\epsilon R}{M^2}\right)^3 
+ 
O\left(\frac{\epsilon^4 R^4}{M^8}\right)
\right]\,, 
\end{align}
under the assumption $\epsilon R/M^2 \ll 1$. The above expansion is consistent with the analysis of ref.\,\cite{Gialamas:2025ofz}, where the connection between the no-scale supergravity potential in eq.\,\eqref{eq:SUGRAPOT} and the higher-curvature corrections to the Starobinsky model was noted up to cubic order in the curvature. Following the analysis carried out in sec.\,\ref{sec:phenoanalyis}, we require the leading cubic correction to carry a negative sign, which translates into the condition $\epsilon \ge 0$ (or $\lambda \leq 1$).

\begin{figure}[h]
\begin{center}
$$\includegraphics[width=.495\textwidth]{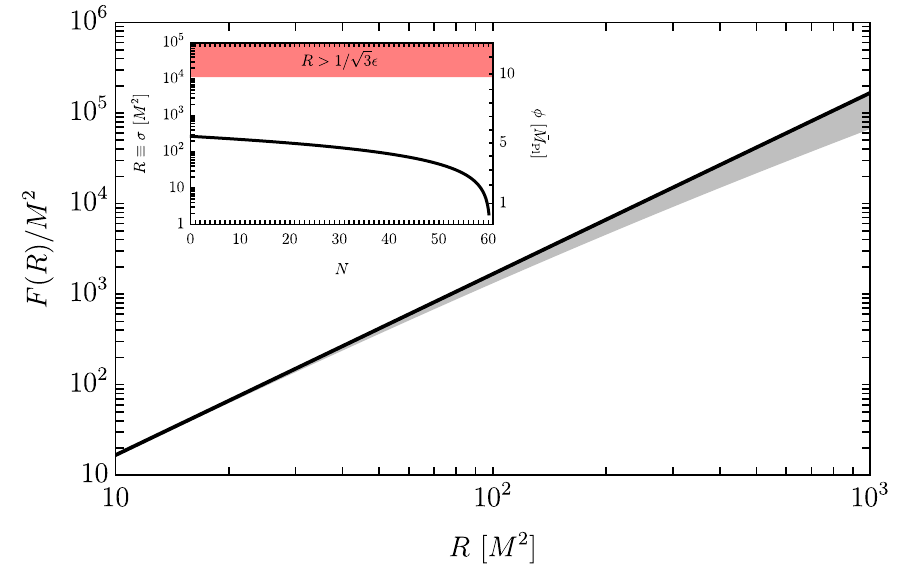}~~
\includegraphics[width=.495\textwidth]{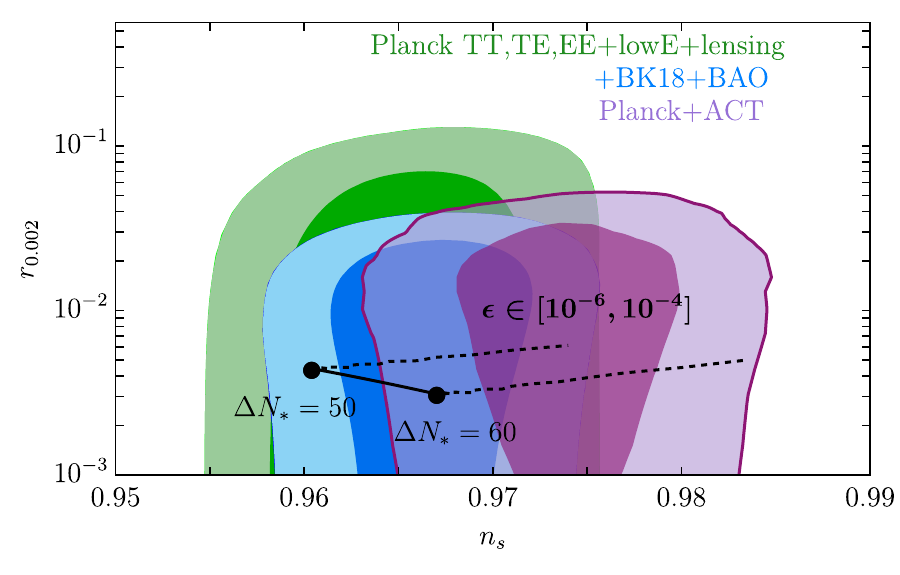}$$\vspace{-0.5cm}
\caption{\textit{
		\textit{\textbf{Left panel:}} $F(R)$ function given in eq.\,\eqref{eq:F(R)noscale} for $\epsilon\in(10^{-6},10^{-4})$. The solid black line denotes the Starobinsky limit (recovered for $\epsilon\lesssim 10^{-6}$), while the gray band corresponds to increasing values of $\epsilon$. The inset shows a representative numerical integration of the background inflaton field yielding $N=60$ $e$-folds of inflation after CMB modes exit the horizon. The red contour marks the region where the perturbativity condition $\epsilon R/M^2 \ll 1$ is violated. As shown, for the values of $R$ relevant to the inflationary dynamics, perturbativity is always satisfied.
		\textit{\textbf{Right panel:}} Prediction in the $(n_s,r)$ plane for the model given
		in eq.\,\eqref{eq:F(R)noscale} for different values of $\epsilon$. The solid black line is the Starobinsky prediction (recovered for $\epsilon\lesssim 10^{-6}$), while the dashed black lines denote increasing values of $\epsilon$, up to $\epsilon = 10^{-4}$.  We require $50 \leq N \leq 60$ e-folds of inflation after
		CMB modes exited the horizon. Experimental constraints are shown using the
		Planck 2018 baseline analysis (green regions) and including BICEP/Keck and baryon acoustic oscillation
		(BAO) data (blue regions), cf. ref.\,\cite{BICEP:2021xfz}. 
		We also show (purple regions) the results from the Atacama Cosmology Telescope (ACT), cf. ref.\,\cite{ACT:2025tim}.
		We show the 68\% and 95\% confidence contours.
}} 
\label{fig:HOV}  
\end{center}
\end{figure}

In the right panel of fig.~\ref{fig:HOV}, we show the predictions for the spectral index and tensor-to-scalar ratio in the model of eq.\,\eqref{eq:F(R)noscale}. As expected, for very small values of $\epsilon \lesssim 10^{-6}$ we recover the Starobinsky predictions (solid black line), whereas larger $\epsilon$ shifts the spectral index to higher values (along the dashed black lines), bringing the model into agreement with the Planck+ACT data. As complementary information, the left panel of fig.~\ref{fig:HOV} shows explicitly the function in eq.\,\eqref{eq:F(R)noscale} for the range $\epsilon\in(10^{-6},10^{-4})$, together with a representative inflationary solution (see inset).

In the spirit of sec.\,\ref{sec:EFTperspective}, we now analyze the structure of the higher-curvature operators appearing in the expansion \eqref{eq:F(R)noscale} from an effective field theory perspective. Based solely on dimensional analysis, we can write the action associated with \eqref{eq:F(R)noscale} as follows:
\begin{align}
    S=\int d^{4} x \sqrt{-g}\Bigg\{ &\frac{\MPl^2}{2}R+\frac{R^2}{12g_{*}^2}-\frac{R^3}{g_{*}^2\Lambda^2}+24\frac{R^4}{g_{*}^2\Lambda^4} +\dots\Bigg\} \label{eq:suEFT},
\end{align}
where we introduced the fundamental coupling 
$g_{*}$ such that $M\equiv g_{*}\MPl$ and the (UV) scale $\Lambda\equiv\frac{6{g_{*}}\MPl}{\sqrt{\epsilon}}$
by setting to $1$ the dimensionless coefficient multiplying the cubic term, in accordance with the convention adopted in \eqref{eq:RewriteL}.\\
Therefore, the theory in \eqref{eq:suEFT} can be interpreted as a one-coupling two-scale effective theory, as two distinct mass scales, $M$ and $\Lambda$, can in principle be identified as characterizing the effective description. From the fit to the cosmological observables (see fig.\,\ref{fig:HOV}), we deduce $\Lambda/M \sim 1/\sqrt{\epsilon} \gg 1$, in line with the general considerations discussed above (see in particular eq.\,\eqref{eq:SeparationOfScales}). However, while the limit $\epsilon \to 0$ gives $\Lambda \to \infty$, implying that higher-curvature operators completely decouple from the Starobinsky potential, recovering the Planck+ACT data sets a lower bound on $\epsilon$ and hence keeps $\Lambda$ finite. In other words, the scale $\Lambda$ controlling the higher-order operators cannot be too far from $M$. This introduces a sensitivity to the UV physics and ultimately a fine-tuning, as we will show in Sec.\,\ref{sec:DiscussingNaturalness}.

\subsection{Metric-affine gravity}
\label{sec:MetricAffineGravity}
Metric-affine gravity is a gravitational theory in which the affine connection and the metric are treated as independent dynamical variables, unlike in standard General Relativity. This approach introduces additional degrees of freedom, among which a pseudo-scalar component can be identified with the inflaton field, cf. ref.\,\cite{Pradisi:2022nmh}. 
This theoretical setup provides a geometric origin for the scalar field driving inflation, which appears as an alternative to that offered by the Starobinsky model. 
Building on the same theoretical setup, ref.\,\cite{Salvio:2025izr} demonstrates that metric-affine gravity can successfully reproduce the recent observational results reported by the ACT collaboration, offering an explanation that cannot be traced back to the predictions of the Starobinsky model. 

In this section, we critically reexamine the results of ref.\,\cite{Salvio:2025izr} showing that {\it \textbf{pseudo-scalar inflation from metric-affine gravity can actually be mapped onto an 
$f(R)$ theory that exactly reproduces the predictions of Starobinsky inflation, with the addition of higher-curvature operators}}.\\ 
In the metric-affine model studied in ref.\,\cite{Salvio:2025izr}, a pseudo-scalar degree of freedom arising from the independent affine connection can be promoted, after a field redefinition, to a canonical field~$\phi$ 
(see app.\,\ref{app:MetricAffine} for details).
In terms of $\phi$, the inflationary potential takes the form
\begin{align}
V(\phi) 
= \frac{3}{4}\MPl^2\bar{M}^2\left\{
1 - 
\frac{1}{4\tilde{\beta}}\sinh\Bigg[\frac{\phi}{\MPl\sqrt{3/2}}
+\tanh^{-1}\left(
\frac{4\tilde{\beta}}{
\sqrt{16\tilde{\beta}^2 + 1}
}
\right)\Bigg]
\right\}^2
\label{eq:Vphi}
\end{align}
where~$c$ and $\tilde{\beta}\equiv \beta/\MPl^2$ are dimensionless parameters inherited from the underlying metric--affine action, and we introduced the scale $\bar{M}^2\equiv \MPl^2\tilde{\beta}^2/3c$\footnote{A physical interpretation of the scale $\bar{M}$ is given in app.\ref{app:C}.}.\\
The potential in eq.\,(\ref{eq:Vphi}) exhibits a Starobinsky-like plateau for $|\tilde{\beta}|\gg1$, thus providing a viable slow-roll inflationary dynamics driven by the geometrically induced scalar field~$\phi$. More specifically, as noted in ref.\,\cite{Racioppi:2024pno}, the potential in eq.\,(\ref{eq:Vphi}) asymptotically converges to the Starobinksy potential for $|\tilde{\beta}|\to\infty$ (see also refs.\,\cite{He:2024wqv,Gialamas:2025thp,Gialamas:2026pjo} for further related examples of asymptotic convergence to the Starobinsky potential in metric-affine models). 
\\
Following the same reverse-engineering approach discussed in section\,\ref{sec:SUGRA}, it is possible to write a metric $f(R)$ which reproduce in the Einstein frame the scalaron potential \eqref{eq:Vphi}. Applying this procedure, we then obtain
\begin{align}
f(R) =& R + \frac{8cR^2}{\MPl^2(1+16\tilde{\beta}^2)} -\frac{R^3}{\bar{M}^4}\left[\left(\frac{1}{24\tilde{\beta}}\right)^2+O(\tilde{\beta}^{-4}) \right]+\frac{R^4}{\bar{M}^6}\left[3\left(\frac{\sqrt{2}}{24\tilde{\beta}}\right)^4+O(\tilde{\beta}^{-6}) \right]+\dots \label{eq:resummedfR}
\end{align}
By construction, the coefficients of the terms quadratic in $R$ determine the inverse squared mass of the pseudo-scalaron introduced in ref.\,\cite{Salvio:2025izr}\footnote{In $f(R)$ theories of gravity, for a Minkowski background, the scalaron mass is given by $M_{\phi}^{2} = \frac{1}{3 f''(0)}$, cf. ref.\,\cite{Sotiriou:2008rp}. 
}.
Comparing with eq.\,\eqref{eq:fR}, we indeed recognize the following squared mass
\begin{align}
M^2 \equiv   \frac{\MPl^2(1+16\tilde{\beta}^2)}{48c}\,.
\label{eq:SaMass}
\end{align}
The conceptual connection is significant: demonstrating that pseudo-scalar inflation in metric-affine gravity can be mapped onto an $f(R)$ model with a ``Starobinsky + higher-curvature'' structure effectively unifies two seemingly distinct geometric approaches—the metric-affine and purely metric formulations—within a single effective framework. 
The phenomenological implication is also immediate: the metric $f(R)$ in \eqref{eq:resummedfR} exactly reproduces the fit to the Planck+ACT data reported in ref.\,\cite{Salvio:2025izr}, thus proving that, in this case, the metric-affine geometry does not generate new inflationary dynamics.

\begin{figure}[h]
\begin{center}
$$\includegraphics[width=.495\textwidth]{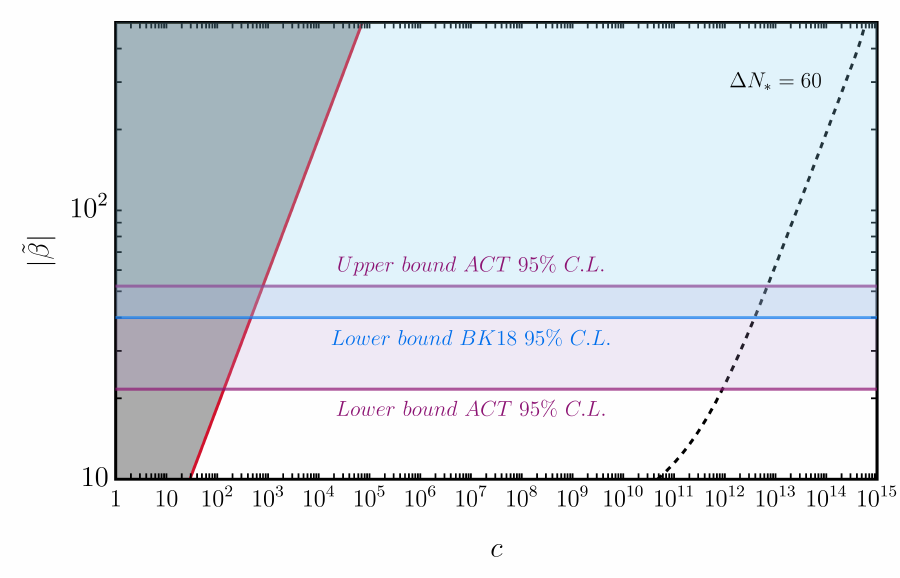}~~
\includegraphics[width=.495\textwidth]{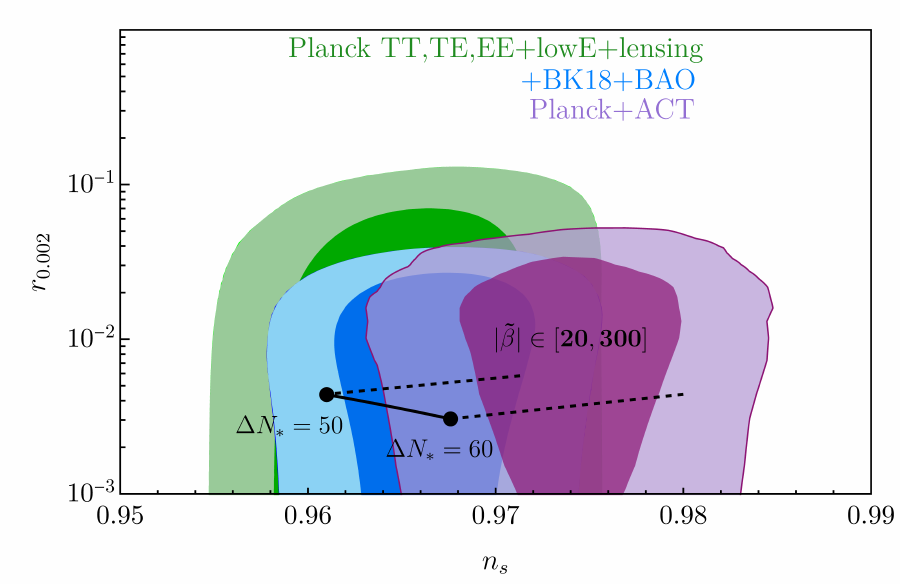}$$\vspace{-0.5cm}
\caption{\textit{\textit{\textbf{Left panel:}} The red line denotes the upper bound on $|\tilde{\beta}|$ in \eqref{eq:evenUpperBound} derived imposing tree-level unitarity. The gray shaded region indicates the portion of parameter space where tree-level unitarity up to scales parametrically larger than $\MPl$ is violated. The black dashed curve corresponds to pairs $(|\tilde{\beta}|, c)$ reproducing the observed scalar power spectrum for $\Delta N_* = 60$. The purple band corresponds to the range of $|\tilde{\beta}|$ for which the predictions of the potential \eqref{eq:Vphi} in the $(n_s, r)$ plane at $\Delta N_* = 60$ are consistent with Planck+ACT data at the $95\%$ confidence level. The blue line marks the lower bound on $|\tilde{\beta}|$ for which the same predictions are consistent with BICEP/Keck $+$ BAO data at the $95\%$ confidence level. \textit{\textbf{Right panel:}} Prediction in the $(n_s,r)$ plane for the scalar potential in eq.\,\eqref{eq:Vphi} for distinct values of $|\tilde{\beta}|$. The solid black line is the Starobinsky prediction (recovered for $|\tilde{\beta}|\gtrsim 10^{2}$), while the dashed black lines denote decreasing values of $|\tilde{\beta}|$, up to $|\tilde{\beta}| = 20$.  We require $50 \leq N \leq 60$ e-folds of inflation after CMB modes exited the horizon. Experimental constraints are shown using the Planck 2018 baseline analysis (green regions) and including BICEP/Keck and baryon acoustic oscillation (BAO) data (blue regions), cf. ref.\,\cite{BICEP:2021xfz}. We also show (purple regions) the results from the Atacama Cosmology Telescope (ACT), cf. ref.\,\cite{ACT:2025tim}.
We show the 65\% and 95\% confidence contours.
}} 
\label{fig:MetricAffine}  
\end{center}
\end{figure}

In fig.~\ref{fig:MetricAffine}, we show the inflationary predictions of the theory in eq.\,\eqref{eq:resummedfR}. In the left panel, we display the values of the parameters $(|\tilde{\beta}|,c)$ (dashed black line) that reproduce the observed scalar power spectrum. In the right panel, we show the predictions for the spectral index and tensor-to-scalar ratio for different values of $|\tilde{\beta}|$.
We see that the predictions converge to those of Starobinsky in the large-$|\tilde{\beta}|$ limit, in agreement with the $f(R)$ expansion in \eqref{eq:resummedfR}, which reduces to Starobinsky plus vanishing corrections as $|\tilde{\beta}|\to\infty$. For decreasing values of $|\tilde{\beta}|$, the spectral index shifts to higher values, bringing the model into agreement with the Planck+ACT data. 



We analyze the structure of the higher-curvature operators appearing in the expansion \eqref{eq:resummedfR} in the spirit of sec.~\ref{sec:EFTperspective}. Unlike the metric $f(R)$ model derived from no-scale supergravity in \eqref{eq:F(R)noscale}, the series of coefficients multiplying the higher-curvature operators in \eqref{eq:resummedfR} cannot be resummed into a closed form. Therefore, in order to allow a comparison with the analysis performed in sec.~\ref{sec:SUGRA}, we restrict our attention to the leading contribution associated with each higher-curvature operator. 
Based solely on dimensional analysis, we can write the action associated with \eqref{eq:resummedfR} as follows
\begin{align}
    S=\int d^{4} x \sqrt{-g}\Bigg\{ &\frac{\MPl^2}{2}R+\frac{R^2}{12g_{*}^2}-\frac{R^3}{g_{*}^2\Lambda^2}\left[1+O(\tilde{\beta}^{-2}) \right]+\frac{R^4}{g_{*}^2\Lambda^4}\left[24\frac{16\tilde{\beta}^2}{1+16\tilde{\beta}^2}+O(\tilde{\beta}^{-2}) \right] +\dots\Bigg\} \label{eq:MAEFT},
\end{align}
where we introduced the fundamental coupling 
$g_{*}^2\equiv\frac{1+16\tilde{\beta}^2}{48c}$ such that $M\equiv g_{*}\MPl$ and the (UV) scale 
\begin{equation}\label{eq:Lambdametricaffine}
    \Lambda\equiv\sqrt{2}\cdot 24\,|\tilde{\beta}|\left(\frac{16\tilde{\beta}^2}{1+16\tilde{\beta}^2}\right)g_{*}\,\MPl \, ,
\end{equation}
by setting to unity the dimensionless coefficient multiplying the cubic term, in accordance with the convention adopted in \eqref{eq:RewriteL} \footnote{Note that, alternatively, by defining $\Lambda\equiv24|\tilde{\beta}|\sqrt{\frac{16\tilde{\beta}^2}{1+16\tilde{\beta}^2}}\MPl g_{*}$, one obtains the following action $S=\int d^{4} x \sqrt{-g}\Bigg\{ \frac{\MPl^2}{2}R+\frac{R^2}{12g_{*}^2}-\frac{1}{2}\frac{R^3}{g_{*}^2\Lambda^2}\left[\frac{1+16\tilde{\beta}^2}{16\tilde{\beta}^2}+O(\tilde{\beta}^{-2}) \right] +6\frac{R^4}{g_{*}^2\Lambda^4}\left[\frac{1+16\tilde{\beta}^2}{16\tilde{\beta}^2}+O(\tilde{\beta}^{-2}) \right] +\dots\Bigg\} \label{eq:maEFT},$ which presents the same dimensionless dependence on $\tilde{\beta}$ in the coefficients multiplying each leading higher curvature operator.}.
Therefore, the theory in \eqref{eq:MAEFT} can also be interpreted as a one-coupling, two-scale effective theory.\\
The fit to the cosmological observables (see fig.~\ref{fig:MetricAffine}) yields $\Lambda/M\gg1$, in agreement with the general considerations on the scale hierarchy discussed previously (see eq.~\eqref{eq:SeparationOfScales}). As follows from eq.~\eqref{eq:Lambdametricaffine}, in the Starobinsky limit $c$ is chosen such that $g_{*}$ (and thus $M$) remains finite, while $\Lambda\to\infty$, so that higher-curvature operators completely decouple.
In analogy with what was done in sec.~\ref{sec:SUGRA}, agreement with the Planck+ACT data sets an upper bound on $|\tilde{\beta}|$, which in turn constrains the scale separation between $M$ and $\Lambda$ to be finite. Once again, this introduces a sensitivity to the UV physics and ultimately a fine-tuning (see sec.\,\ref{sec:DiscussingNaturalness}).


\subsubsection{Metric-Affine vs $f(R)$: (In)equivalence and convergence to Starobinsky predictions}

It is important to clarify that, although the metric $f(R)$ in eq.\,\eqref{eq:resummedfR} and the metric-affine model of ref.\,\cite{Salvio:2025izr} reproduce the same inflationary potential, they are not equivalent theories in general. To point out this in-equivalence, we stress two key points. 
\begin{itemize}
\item[$\circ$] 
In the Einstein frame, defined by the metric ${\bar{g}}_{\mu\nu}$ in \eqref{eq:confTr}, the metric $f(R)$ theory is dynamically equivalent to the Einstein–Hilbert action supplemented by a scalar degree of freedom with potential in eq.~\eqref{eq:Vphi}. By contrast, the metric-affine model of ref.\,\cite{Salvio:2025izr}, after integrating out the non-dynamical components of the torsion tensor, reduces directly in the Jordan frame\footnote{We here call "Jordan frame" representation of the metric-affine model, the frame with metric $g_{\mu\nu}$, in order to distinguish it from the representations in which the metric has been conformally transformed, even if the gravitational sector in the $g_{\mu\nu}$ representation already takes the Einstein-Hilbert form.} (i.e., the frame with metric $g_{\mu\nu}$) to the Einstein–Hilbert action plus a pseudo-scalar degree of freedom with the same potential, without requiring any conformal transformation (see app~\ref{app:MetricAffine} for details).
\item[$\circ$] 
The two models feature inequivalent connections. In the Einstein frame, the connection of the metric $f(R)$ model is related to Levi-Civita of $g_{\mu\nu}$ through the conformal transformation
\begin{align}
    {{\{\}}_{\mathbf{\bar{g}}}}^{\lambda}{}_{\mu\nu}
    &= {{\{\}}_{\mathbf{g}}}^{\lambda}{}_{\mu\nu}+\frac{1}{\Omega}\Big(\delta^{\lambda}{}_{\nu}\partial_{\mu}\Omega+\delta^{\lambda}{}_{\mu}\partial_{\nu}\Omega-{g}^{\lambda\sigma}{g}_{\mu\nu}\partial_{\sigma}\Omega \Big), \label{eq:GammaEinstein}
\end{align}
where ${{\{\}}_{\mathbf{\bar{g}}}}^{\lambda}{}_{\mu\nu}$ and ${{\{\}}_{\mathbf{g}}}^{\lambda}{}_{\mu\nu}$ are respectively Levi-Civita of $\bar{g}_{\mu\nu}$ and $g_{\mu\nu}$.\\
On the other hand, in the Jordan frame, the connection of the metric-affine model deviates from Levi-Civita due to the presence of torsion
\begin{align}
    {\tilde{\Gamma}}^{\lambda}{}_{\mu\nu} \equiv {{\{\}}_{\mathbf{g}}}^{\lambda}{}_{\mu\nu}+ K^{\lambda}{}_{\mu\nu}, \label{eq:disto0}
\end{align}
where $K^{\lambda}{}_{\mu\nu}$ is the contorsion tensor constructed from the torsion tensor, whose explicit form for the metric-affine model of ref.\,\cite{Salvio:2025izr} is provided in app.~\ref{app:MetricAffine}.
\end{itemize}
The two points discussed above highlight that the two theories are in-equivalent. Since conformally related frames (such as Jordan and Einstein frame) of a given metric $f(R)$ theory are, at least classically, physically equivalent \footnote{As explained in ref.\,\cite{PhysRev.125.2163}, where conformal transformations in the context of scalar–tensor theories were originally introduced, a conformal transformation can be seen as a rescaling of units. In this view, the transformation \eqref{eq:confTr} connecting the two frames must be understood together with a corresponding and appropriate rescaling of the fundamental units in the conformally transformed frame. Since physical measurements can only probe dimensionless quantities, which are invariant under such rescalings, conformally related frames are physically equivalent. Ref.\,\cite{Postma_2014} further shows that metric $f(R)$ theories can indeed be cast into a manifestly frame-independent formulation.} (see, e.g., refs.\,\cite{PhysRev.125.2163,Faraoni_2007,Postma_2014}), it is convenient to consider the metric $f(R)$ in eq.\,\eqref{eq:resummedfR} in the Einstein frame for facilitate comparison with the metric-affine model of ref.\,\cite{Salvio:2025izr}. In this representation, as already emphasized in the first point above, it becomes manifest that, although the two theories are described by the same action, they are formulated in terms of different, albeit conformally related, metrics. Of course, one is always free to adopt a prescription that identifies the metric associated with the metric-affine action with $\bar{g}_{\mu\nu}$.
This amounts to adopting a prescription in which time, length and mass units are appropriately rescaled with the conformal factor $\Omega^2$ relating $\bar{g}_{\mu\nu}$ and $g_{\mu\nu}$ \cite{PhysRev.125.2163}. However, even with such a prescription, the two theories should not be regarded as equivalent.
The key distinction 
lies in the different geometrical nature of the field $\phi$: it is a pseudo-scalaron degree of freedom, a manifestation of torsion, in the metric-affine model, whereas it is a scalaron in the metric $f(R)$ theory. Nevertheless, although the parity-violating nature of the pseudo-scalaron leads to interaction terms with matter describing the post-inflationary phase (see refs.\,\cite{DiMarco:2023ncs,DiBenedetto:2026onj} for recent studies of the post-inflationary phase of pseudo-scalar inflation) that have no analogue for the scalaron,  in a purely inflationary setting the two theories are phenomenologically equivalent, in the sense that they generate the same inflationary dynamics.\\
The second point highlighted above identifies a further source of inequivalence, related to the distinct nature of $\phi$ in the two theories. The two theories feature inequivalent connection, meaning that ${{\{\}}_{\mathbf{\bar{g}}}}^{\lambda}{}_{\mu\nu}$ is conformally equivalent to ${{\{\}}_{\mathbf{g}}}^{\lambda}{}_{\mu\nu}$, whereas $\tilde{{\Gamma}}^{\lambda}{}_{\mu\nu}$ is not due to the presence of dynamical torsion. 

It is interesting to investigate the limit in which the two theories become completely equivalent.
In a framework where $\tilde{\beta}$ and $c$ are treated as independent parameters, an interesting feature emerges in the large-$|\tilde{\beta}|$ limit. In this regime, eq.\,\eqref{eq:SaMass} implies $M\to\infty$ so that the metric model \eqref{eq:resummedfR} reduces to General Relativity. Moreover, as one expects, the conformal transformation relating the Jordan and Einstein frames, $\Omega^2 = f'(R)$, approaches the identity
\begin{align}
\Omega^2 \to 1 \quad \text{for} \quad |\tilde{\beta}| \to \infty,
\label{eq:trivialconf}
\end{align}
meaning that, in this limit Einstein and Jordan frames coincide and the connection reduces to ${{\{\}}_{\textbf{g}}}^{\lambda}{}_{\mu\nu}$. 
On the other hand, in the metric-affine model of ref.~\cite{Salvio:2025izr}, the same large-$|\tilde{\beta}|$ limit drives the torsion tensor and hence the contorsion tensor to zero, so that the connection also reduces to ${{\{\}}_{\textbf{g}}}^{\lambda}{}_{\mu\nu}$, and the theory reduces to purely metric. In particular, one can readily show that it reduces to General Relativity (see app.~\ref{app:MetricAffine} for details).
Hence, the two theories become equivalent for $|\tilde{\beta}|\to\infty$, as both reduce to General Relativity.
In this regime, the two models share the same Einstein-Hilbert action, defined on a spacetime with identical metric $g_{\mu\nu}$ and Levi-Civita connection ${{\{\}}_{\textbf{g}}}^{\lambda}{}_{\mu\nu}$.\\
In an inflationary context, however,  $\tilde{\beta}$ and $c$ are not independent parameters: for each $|\tilde{\beta}|$, $c$ is fixed to reproduce the observed scalar power spectrum. This prevents $M$ from diverging in the large $|\tilde{\beta}|$-regime, giving
\begin{align}
M\sim\bar{M}\lesssim O(10^{-5})\MPl\Rightarrow \frac{\tilde{\beta}^2}{3c}\lesssim O(10^{-10}),\label{eq:cOkApprox}
\end{align}
so that the metric theory \eqref{eq:resummedfR} reduces to the Starobinsky model with $M\sim\bar{M}$. Eq.\,\eqref{eq:cOkApprox} implies that the ratio $c/\tilde{\beta}^2$ is fixed to a large value. As a result, the limit in \eqref{eq:trivialconf} does not lead to a conformal transformation approaching the identity
and the connection remains as in eq.\eqref{eq:GammaEinstein}.
On the metric-affine side, eq.~\eqref{eq:cOkApprox} implies that the torsion tensor does not vanish for $|\tilde{\beta}|\to\infty$ (see app.~\ref{app:MetricAffine} for details), so that the connection is given by eq.~\eqref{eq:disto0}, meaning that the theory does not reduce to a metric one. Thus, in an inflationary context, the large-$|\tilde{\beta}|$ limit does not reduce the two theories to be equivalent, leaving them as two inequivalent descriptions of the same inflationary dynamics.
\\
Remarkably, one can uncover the physical origin of the non-trivial exact convergence of the inflationary potential associated with the metric-affine model of ref.\,\cite{Salvio:2025izr} to the Starobinsky potential for $|\tilde{\beta}|\to\infty$. 
To make this point explicit, in app.~\ref{app:C} we present an alternative construction corresponding to a metric-affine counterpart of Starobinsky supplemented with a purely torsional operator. This model exhibits two interesting properties. First, it manifestly reduces to Starobinsky in the large-$|\tilde{\beta}|$ limit, where the torsion tensor is switched off and $M$ sets the inflationary scale as in eq.\,\eqref{eq:cOkApprox}, while it reduces to General Relativity in the torsionless limit with $M\to\infty$. Second, it reproduces, in the Einstein frame, the scalaron potential in eq.\,\eqref{eq:Vphi}. It follows that, in contrast with the metric-affine model of ref.\,\cite{Salvio:2025izr}, in the large-$|\tilde{\beta}|$ regime this metric-affine construction becomes equivalent to the metric $f(R)$ theory in eq.\,\eqref{eq:resummedfR} also in the inflationary context (see app.\,\ref{app:C} for details).\\
Although the metric-affine model of ref.\cite{Salvio:2025izr} is not a priori related to Starobinsky, its inflationary potential coincides with that of the construction in app.\,\ref{app:C}. Thus, the Starobinsky limit of eq.\,\eqref{eq:Vphi}, as well as the corresponding inflationary predictions shown in fig.\,\ref{fig:MetricAffine} (right panel) and the specific structure of the metric $f(R)$ theory in eq.\,\eqref{eq:resummedfR}, can be understood physically as a consequence of the torsionless limit of the metric-affine deformation of Starobinsky presented in app.\,\ref{app:C}.

\subsubsection{Tree-level unitarity bounds}

The phenomenological behavior in the large-$|\tilde{\beta}|$ regime motivates investigating upper bounds on $|\tilde{\beta}|$ by imposing tree-level unitarity, to determine whether the model can reproduce viable slow-roll inflationary dynamics while remaining free of sub-Planckian unitarity problem, as in the case of Starobinsky inflation.\\
Following the discussion in section \ref{sec:EFTperspective}, we impose tree-level unitarity by looking at phase-space averaged scattering amplitudes.\\
We consider the interactions contained in \eqref{eq:Vphi} after expanding around $\langle\phi\rangle=0$. We find the Lagrangian density
\begin{align}
    \mathcal{L} =& \frac{1}{2}(\partial_{\mu}\phi)(\partial^{\mu}\phi) - \frac{1}{2}{M^2}\phi^2
- \frac{\sqrt{2}}{6}\tilde{\beta}\left(\frac{48}{1+16\tilde{\beta}^2}\right)^{1/2}g_{*}M\phi^3
- \frac{1+28\tilde{\beta}^2}{432}\left(\frac{48}{1+16\tilde{\beta}^2}\right)g_{*}^2\phi^4\nn \\
&- 
\frac{\tilde{\beta}(1+16\tilde{\beta}^2)}{864\sqrt{2}}\left(\frac{48}{1+16\tilde{\beta}^2}\right)^{3/2}\frac{g_{*}^3}{M}\phi^5 +\dots\, \label{eq:SalvioExp}
\end{align}
The generic structure of the contact interaction term with $N$ fields is, therefore, given by
\begin{align}
\mathcal{L}_{\textrm{int}}^{(N)} =  c_N(\tilde{\beta})\frac{\MPl^{4-N}\phi^N}{c}=
c_N(\tilde{\beta})\left(\frac{48}{1+16\tilde{\beta}^2}\right)\frac{ g_*^{N-2}\phi^N}{M^{N-4}}\,. \label{eq:lSalvioint1}
\end{align}
in agreement with eq.\,(\ref{eq:LagrangianScaling}). In the above expression, 
$c_{N}(\tilde{\beta})$ are pure dimensionless numbers that can be extracted from the expansion of the potential in eq.\,\eqref{eq:Vphi}
\begin{align}
    c_{N}(\tilde{\beta})&=-\frac{1}{64}\frac{(2/3)^{N/2}}{N!}\Big[2^{N-1}+32\tilde{\beta}^2\big(2^{N-1}-1 \big) \Big]\quad \text{for $N$ even}, \label{eq:evencoef}\\
    c_{N}(\tilde{\beta})&=-\frac{1}{16}\frac{(2/3)^{N/2}}{N!}\tilde{\beta}\sqrt{1+16\tilde{\beta}^2}\big( 2^{N}-2\big)\quad \text{for $N$ odd} \label{eq:oddcoef}.
\end{align}
The tree-level scattering amplitude for the process $n\to m$ is given by
\begin{align}\label{eq:ScalingScatteringAMplipt2}
\mathcal{M}_{n\to m} = c_{n+m}(\tilde{\beta})(n+m)!\left(\frac{48}{1+16\tilde{\beta}^2}\right)
\frac{g_*^{n+m-2}}{M^{n+m-4}}\left[
1+O\left(\frac{M^2}{E^2}\right)
\right]\ ,
\end{align}
whose mass dimension is consistent with the scaling relation in eq.\,(\ref{eq:MasterScaling}).
In the ultrarelativistic limit $E\gg M$, $\phi$ can be approximated as masseless, and the contact interaction dominates. We therefore focus on this contribution, namely the first term in eq.\,\eqref{eq:ScalingScatteringAMplipt2}.
In this set-up, averaging over the initial and final phase space corresponds to replace the phase space integrals with the corresponding volumes, yielding
\begin{align}
\hat{\mathcal{M}}_{n\to m} 
= \frac{
c_{n+m}(\tilde{\beta})(n+m)!
}{
8\pi[
n!m!(n-1)!(n-2)!(m-1)!(m-2)!
]^{1/2}
}\,\left(\frac{48}{1+16\tilde{\beta}^2} \right)g_*^{n+m-2}\,
\left(\frac{E}{4\pi M }
\right)^{n+m-4},
\label{eq:Maveraged}
\end{align}
which has the expected dependence on the coupling as dictated by eq.\,(\ref{eq:scalingAverage}).
The matrix elements $|\hat{\mathcal{M}}_{n\to m}|$ are directly constrained by unitarity to be $\leq 1$, this automatically gives upper bounds on $|\tilde{\beta}|$. Note that $|\hat{\mathcal{M}}_{n\to m}|$ is a function of both $|\tilde{\beta}|$ and $c$, the resulting upper bounds on $|\tilde{\beta}|$ depend on $c$ for a given cutoff energy. In the following, in order to make this dependence explicit, and because matrix elements are independent of the specific definition of the coupling $g_{*}$, we express all results in terms of $c$ and $\tilde{\beta}$.\\
As a benchmark, we consider $n\to n$ scatterings and derive the corresponding analytic unitarity bound.
Inserting \eqref{eq:evencoef} into \eqref{eq:Maveraged} yields
\begin{align}
|\hat{\mathcal{M}}_{n\to m}|
&= \frac{1}{c}\frac{2^{n+m-1}+32\tilde{\beta}^2(2^{n+m-1}-1)}{(\frac{3}{2})^{\frac{n+m}{2}}}\cdot t(n,m) \left(\frac{E}{4\pi \MPl }
\right)^{n+m-4}, \label{eq:evenSM}
\end{align}
where, for ease of reading, we define the function $t(n,m)$ as follows
\begin{align}
    t(n,m) \equiv \frac{1}{512\pi \sqrt{n!m!(n-1)!(n-2)!(m-1)!(m-2)!}}.
\end{align}
For the case $n\to n$, we find
\begin{align}
|\hat{\mathcal{M}}_{n\to n}| = \frac{1}{c}\frac{2^{2n-1}+32\tilde{\beta}^2(2^{2n-1}-1)}{(\frac{3}{2})^{n}}\cdot t(n,n) \left(\frac{E}{4\pi \MPl }
\right)^{2n-4}. \label{eq:evenindep}
\end{align}
In the large-$n$ limit, $t(n,n)$ can be approximated using Stirling's formula, and consequently \eqref{eq:evenindep} can be written as
\begin{align}
    |\hat{\mathcal{M}}_{n\to n}| = \frac{1}{c}\frac{2^{2n-1}+32\tilde{\beta}^2(2^{2n-1}-1)}{1024\sqrt{2}\pi^{5/2}{(\frac{3}{2})}^{n}}\cdot  \sqrt{n}(n-1)\Big( \frac{e}{n}\Big)^{3n}\left(\frac{E}{4\pi \MPl }
\right)^{2n-4}.\label{eq:evenindepapp}
\end{align}
The maximum of \eqref{eq:evenindepapp} over n is obtained by solving $\partial_{n}|\hat{\mathcal{M}}_{n\to n}|=0$.
To derive an approximate analytical expression for $n_{\max}$ valid in the large-$n$ limit, we solve the previous equation by considering only the leading term in the large-$n$ regime, obtaining
\begin{align}
    n_{max}\simeq \Big(\sqrt{\frac{8}{3}}\frac{E}{4\pi\MPl} \Big)^{2/3}. \label{eq:nmax}
\end{align}
The unitarity bound $|\hat{\mathcal{M}}_{n_{max}\to n_{max}}|\leq1$,  in the large-$n$ limit, defines the cutoff energy scale
\begin{align}
&\frac{E_{\textrm{UV}}}{4\pi\MPl}\sim\sqrt{\frac{3}{8}}\left[\frac{1}{3}\log\left(\frac{288\sqrt{2}\pi^{5/2}c}{32\tilde{\beta}^2+1} \right) \right]^{3/2},
\end{align}
where $|\hat{\mathcal{M}}_{n_{max}\to n_{max}}|$ has been approximated by a majorizing function to obtain non-implicit expression for $E_{\textrm{UV}}$.\\
The cut-off scale $E_{\textrm{UV}}$ is parametrically larger than $4\pi\MPl$ if
\begin{align}
    |\tilde{\beta}|<\frac{\sqrt{B\cdot c-1}}{4\sqrt{2}},\label{eq:evenUpperBound}
\end{align}
with $B\equiv 288\sqrt{2}\pi^{5/2}e^{-2\cdot 3^{2/3}}\sim110$, and $c$ must satisfy $c>1/B$.
Eq.\,\eqref{eq:evenUpperBound} shows that $|\tilde{\beta}|$ can be large, i.e., $|\tilde{\beta}| \gg 1$, provided that $c$ is sufficiently large, $c\gg33/B$.\\
The unitarity bound is shown in the left panel of fig.\,\ref{fig:MetricAffine}.
It is evident from the figure that the portion of the dashed curve corresponding to values of $|\tilde{\beta}| \in (20,40)$ compatible with the Planck+ACT data lies safely within the region required for ensuring tree-level unitarity at scales parametrically larger than $\MPl$. In fact, the full range of $|\tilde{\beta}|$ values allowed by the BICEP-Keck + BAO data remains entirely within the region where tree-level unitarity is preserved. This is consistent with the fact that Starobinsky inflation, recovered in the large-$|\tilde{\beta}|$ regime, is free of sub-Planckian unitarity problems, as shown in sec.\,\ref{sec:EFTperspective}.\\ 

\section{Discussing fine-tuning}
\label{sec:DiscussingNaturalness}
Building on the considerations discussed in sections~\ref{sec:SUGRA} and~\ref{sec:MetricAffineGravity}, we now quantitatively analyze the degree of fine-tuning of the two inflationary EFTs considered in this work, namely the $f(R)$ model embedded in no-scale supergravity (SUGRA) and the $f(R)$ theory arising from metric-affine gravity. Both EFTs under consideration can be viewed as of the one-coupling, two-scales type: the combination $g_* \MPl$, which sets the scalaron mass scale, and an additional (UV) scale $\Lambda$, which controls the suppression of higher-dimensional operators. The scale $\Lambda$ is determined as a function of the dimensionless IR parameter controlling the EFT, namely $\epsilon$ in the no-scale SUGRA case and $|\tilde{\beta}|$ in the metric-affine case. 
As introduced in Sec.~\ref{sec:EFTperspective}, to facilitate a direct comparison between different EFTs, we define the UV scale $\Lambda$ by normalizing the dimensionless coefficient of the cubic curvature operator to unity, so that each $f(R)$ EFT can be written in the form of eq.~\eqref{eq:RewriteL}.


Heuristically speaking, the amount of fine-tuning is proportional to how sensitively a generic physical observable $O$ reacts to variations of $\Lambda$. 
The naive expectation is that in a \textit{natural} EFT the UV physics should be decoupled in the IR, namely the physical observables should not be sensitive to the UV scale $\Lambda$. The fine-tuning $\Delta_{O}$ of an observable $O$ is quantified following \cite{Anderson:1994dz} as
\begin{align}
\Delta_{O}(a)\equiv\frac{\tilde{\Delta}_{O}(a)}{\bar{c}},\quad \tilde{\Delta}_{O}\equiv\left|\frac{d\log O(a)}{d\log a}\right |\,,\label{eq:fineTuning}
\end{align}
where $\tilde{\Delta}_{O}$ is the "absolute" sensitivity first introduced in \cite{Barbieri:1987fn}, $a$ denotes a dimensionless IR parameter defining $\Lambda$ ($\epsilon$ or $|\tilde{\beta}|$ in the present case), and $\bar{c}$ denotes the average sensitivity defined through the following equation
\begin{align}
    \frac{1}{\bar{c}}\equiv \frac{\int da \,\tilde{\Delta}^{-1}_{O}(a)}{\int da}\,, \label{eq:InvAveSens}
\end{align} 
%
where we restrict the integration to be performed over a physically viable range of the IR parameter $a$. The integration domain entering the average is therefore not an arbitrary prior, but rather a precise specification of the physical question under consideration~\cite{Iovino:2025tcv}. By evaluating $\Delta_{n_s}$, we quantify the extent to which, over a physically relevant range of $n_s$, fitting the Planck+ACT data drives the theory toward a region of parameter space characterized by anomalous sensitivity. 
In the following, we define an EFT to be fine-tuned when $\Delta_{O}>1$.
%
%
Note that, the absolute sensitivity $\tilde{\Delta}_{O}(a)$ in principle depends on the specific relation between the IR parameters and the UV scale $\Lambda$ via the following chain rule
\begin{align}
    \tilde{\Delta}_{O}=\left|\frac{d\log O}{d\log M/\Lambda}\right|\cdot\left|\frac{d \log M/\Lambda}{d \log a}\right|= \left|\frac{d\log O}{d\log M/\Lambda}\right|\cdot \tilde{\Delta}_{M/\Lambda}.
    \label{eq:AbsoluteSens}
\end{align}
In one-coupling, two-scale EFTs, it is reasonable to assume a power-law parametrization of the scale hierarchy of the form $M/\Lambda\propto a^x$, with $|x|=\mathcal{O}(1)$, leading to $\tilde{\Delta}_{M/\Lambda}=|x|=\mathcal{O}(1)$. Thus, $\tilde{\Delta}_{M/\Lambda}$ does not introduce any parametric enhancement or suppression in \eqref{eq:AbsoluteSens}, showing that $\tilde{\Delta}_{O}$ is not particularly sensitive to the specific choice of parametrization $a(M/\Lambda)$ and is primarily controlled by the scale separation. We conclude that the fine-tuning ${\Delta}_{O}$ can be regarded as a measure of the direct sensitivity of the observables to variations of the UV scale $\Lambda$ rather than $a$.
We explicitly check the power-law parametrization for the two EFTs studied in this work, for which $\tilde{\Delta}_{M/\Lambda}$ takes the form
\begin{align}
\tilde{\Delta}_{M_{SU}/\Lambda_{SU}} = \frac{1}{2},\quad \tilde{\Delta}_{M_{MA}/\Lambda_{MA}}=1+\frac{2}{1+16\tilde{\beta}^2}\,,\label{eq:AbsSensParam}
\end{align}
in the no-scale SUGRA and metric-affine cases, respectively.
As expected, $\tilde{\Delta}_{M_{MA}/\Lambda_{MA}} \sim 1$ in the large-$|\tilde{\beta}|$ regime, where the parametrization reduces to the power law $M_{MA}/\Lambda_{MA} \propto |\tilde{\beta}|^{-1}$.

\begin{figure}[h]
\begin{center}
$$\includegraphics[width=.495\textwidth]{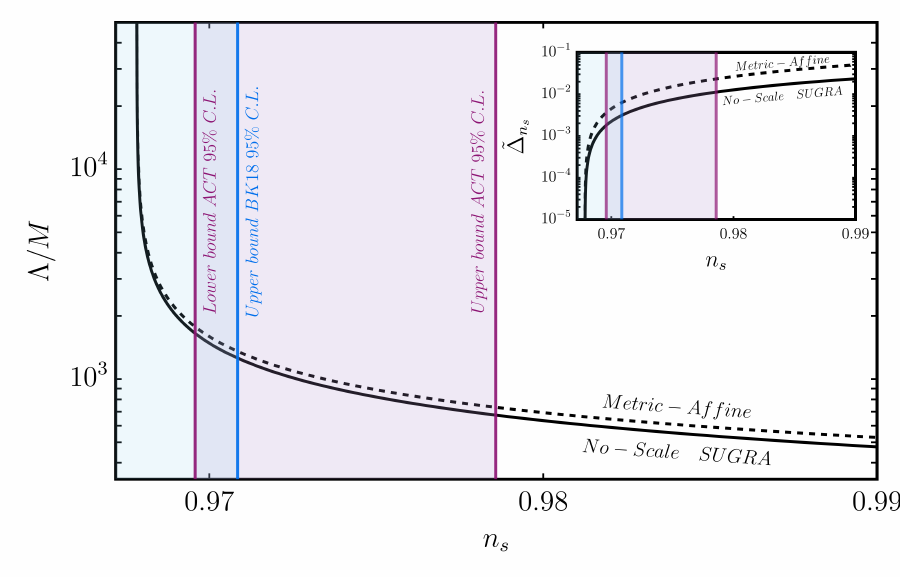}~~
\includegraphics[width=.495\textwidth]{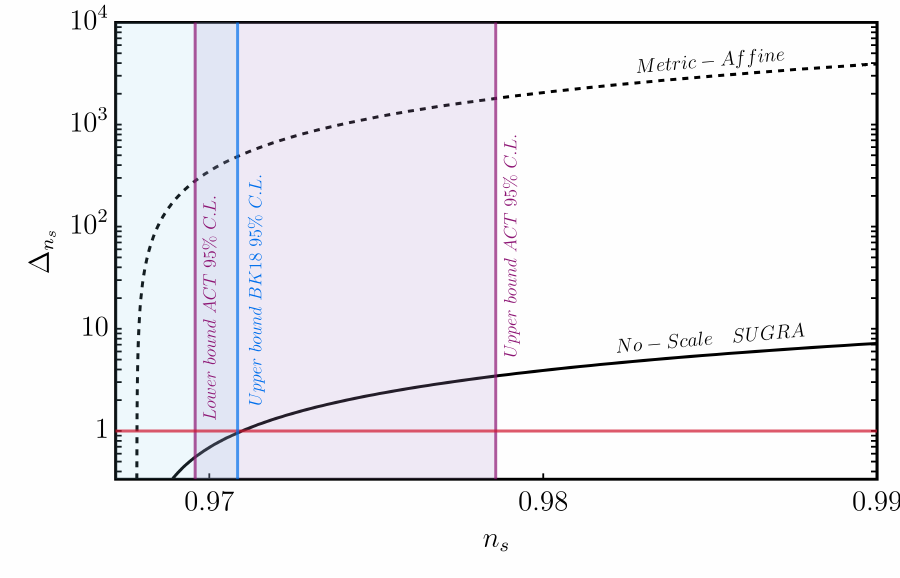}
$$\vspace{-0.5cm}
\caption{\textit{
\textit{\textbf{Left panel:}}
Scale hierarchy $\Lambda/M$ as a function of $n_s$ for the no-scale SUGRA and metric-affine and  EFTs, shown as solid and dashed lines, respectively. The purple band indicates the range of $n_s$ values for which the predictions of the potentials \eqref{eq:SUGRAPOT} and \eqref{eq:Vphi} in the $(n_s,r)$ plane, for $\Delta N_* = 60$, are consistent with Planck+ACT data at the $95\% $  confidence level. The blue line denotes the upper bound on $n_s$ for which the corresponding predictions remain consistent with BICEP/Keck$+$BAO data at the $95\% $ confidence level.
The inset shows the absolute sensitivity $\tilde{\Delta}_{n_s}$ defined in \eqref{eq:fineTuning} as a function of $n_s$ for $\Delta N_* = 60$ in the no-scale SUGRA EFTs and metric-affine cases, shown as solid and dashed lines, respectively. 
\textit{\textbf{Right panel:}} The amount of fine-tuning, computed according to \eqref{eq:fineTuning}, required for the no-scale SUGRA and metric-affine EFTs to reproduce a given value of $n_s$ at $\Delta N_* = 60$, shown as solid and dashed lines, respectively. The horizontal red line at $\Delta_{n_s}=1$ marks the threshold above which a theory is considered fine-tuned. The purple band indicates the range of $n_s$ values for which the predictions of the potentials \eqref{eq:SUGRAPOT} and \eqref{eq:Vphi} in the $(n_s,r)$ plane, for $\Delta N_* = 60$, are consistent with Planck+ACT data at the $95\% $  confidence level. The blue line denotes the upper bound on $n_s$ for which the corresponding predictions remain consistent with BICEP/Keck$+$BAO data at the $95\% $ confidence level.
}} 
 \label{fig:FineTuning}  
\end{center}
\end{figure}

In fig.~\eqref{fig:FineTuning} (left-panel), we show the scale separation between $\Lambda$ and $M$ as a function of the observable $n_s$ for the no-scale SUGRA and metric-affine EFTs, displayed as continuous and dashed lines, respectively. The figure shows that the hierarchy between $\Lambda$ and $M$ decreases with increasing $n_s$, indicating that $n_s$ is a monotonically increasing function of $M/\Lambda$. 
The inset displays the absolute sensitivity $\tilde{\Delta}_{n_s}$ for the two theories, which also increases with $n_s$, and therefore is a monotonically increasing function of $M/\Lambda$.
This means that ACT-compatible values of $n_s$ imply a smaller scale separation compared to that associated with values of $n_s$ compatible with the BICEP/Keck $+$ BAO data, thereby moving effective theories of the form \eqref{eq:RewriteL} towards a regime characterized by a larger absolute sensitivity to UV physics, as already noticed in sections~\ref{sec:SUGRA} and~\ref{sec:MetricAffineGravity}. From fig.~\eqref{fig:FineTuning} (left panel), it is also evident that the scale separation in the two theories is comparable for all values of $n_s$. Since fine-tuning is directly related to the absolute sensitivity as in \eqref{eq:fineTuning}, we expect that a shift towards ACT-compatible values of $n_s$ corresponds to driving the theory into a more fine-tuned regime. This is shown quantitatively in the right panel of fig.~\eqref{fig:FineTuning}, where we plot the amount of fine-tuning required by the SUGRA and metric-affine EFTs to reproduce a given value of $n_s$, as a function of $n_s$. To compute the average sensitivity $\bar{c}$ entering \eqref{eq:fineTuning}, we considered ranges of $\epsilon$ and $|\tilde{\beta}|$ yielding values $n_{s}\in[n^{\text{Starobinsky}}_{s}, 1]$ in the respective EFT. We observe that for both models $\Delta_{O} \gtrsim 1$ for values of $n_s$ consistent with the Planck+ACT data, implying that both EFTs are fine-tuned. The actual degree of fine-tuning depends on the model, with metric-affine gravity reaching fine-tuning up to $\Delta_{O} \sim 10^4$, while in no-scale SUGRA $\Delta_{O} \sim 10$ at most. This difference between the two models can be traced back to the fact that the metric-affine effective theory exhibits a smaller average sensitivity $\bar{c}$ compared to the no-scale SUGRA case. Notably, there is an intrinsic source of systematic error in the definition of fine-tuning in eq.~\eqref{eq:fineTuning}. Indeed, there is no unique way of defining the average sensitivity $\bar{c}$. 
We have numerically verified that the fine-tuning shown in the right panel of fig.~\eqref{fig:FineTuning} is reduced by approximately one order of magnitude in the no-scale SUGRA case and by approximately two orders of magnitude in the metric-affine case when other definitions of $\bar{c}$ are adopted, e.g.\ $\bar{c} = \int da \,\tilde{\Delta}_{O}(a)/\int da$. 
This implies that, while we can confidently conclude that metric-affine gravity is fine-tuned, the same conclusion is not definitive for no-scale SUGRA. Nevertheless, the central physical point—that higher values of the spectral index correspond to higher absolute sensitivity to the UV scale—does not depend on the choice of $\bar{c}$.

\section{Starobinsky inflation and reheating}\label{sec:Reh}

An alternative way to reconcile the Starobinsky model with Planck+ACT data is to consider a longer observed phase of inflation, namely the interval between the horizon exit of the CMB pivot scales and the end of inflation. Clearly, this effect must be properly compensated during the post-inflationary epoch by an appropriate reheating phase; otherwise, the CMB scales would be stretched to such an extent that they would not re-enter the horizon before recombination epoch, in contradiction with what is observed in the post-inflationary Universe.
In the left panel of fig.\,\ref{fig:StaroBench}, we display, using an open circle, 
the prediction of the Starobinsky model for $\Delta N_* = 70$ $e$-folds of observed inflation, and we note that such an extended inflationary period can be readily reconciled with the ACT-preferred values of $n_s$ and $r$.

Reheating is the non-adiabatic transition that converts the inflaton’s coherent energy into a thermal bath of relativistic particles, thereby initiating the standard radiation-dominated era. In a model-agnostic treatment it can be characterized by two effective parameters: the reheating temperature $T_{\textrm{reh}}$ and the (average) equation-of-state parameter $\omega_{\textrm{reh}}$. 
The temperature $T_{\textrm{reh}}$ marks the end of reheating when the plasma attains (near) thermal equilibrium, and thus sets the energy scale of the handoff to radiation domination. The parameter $\omega_{\textrm{reh}}$ controls how the background energy density redshifts during reheating and therefore how long it lasts.

Refs.\,\cite{Drees:2025ngb,Zharov:2025evb} show  that the predictions for the spectral index and tensor-to-scalar
ratio can lie within 1-$\sigma$ of the Planck+ACT constraints if the reheating temperature
satisfies $4\,\textrm{MeV} \lesssim T_{\textrm{reh}}  \lesssim 10\,\textrm{GeV}$ for $0.8 \lesssim \omega_{\textrm{reh}} \lesssim 1$ ( the lower bound on the reheating temperature, $T_{\textrm{reh}} \gtrsim  4\,\textrm{MeV}$, arises from the requirement not to spoil Big Bang nucleosynthesis).
\begin{figure}[h!]
\begin{center}
$$\includegraphics[width=.495\textwidth]{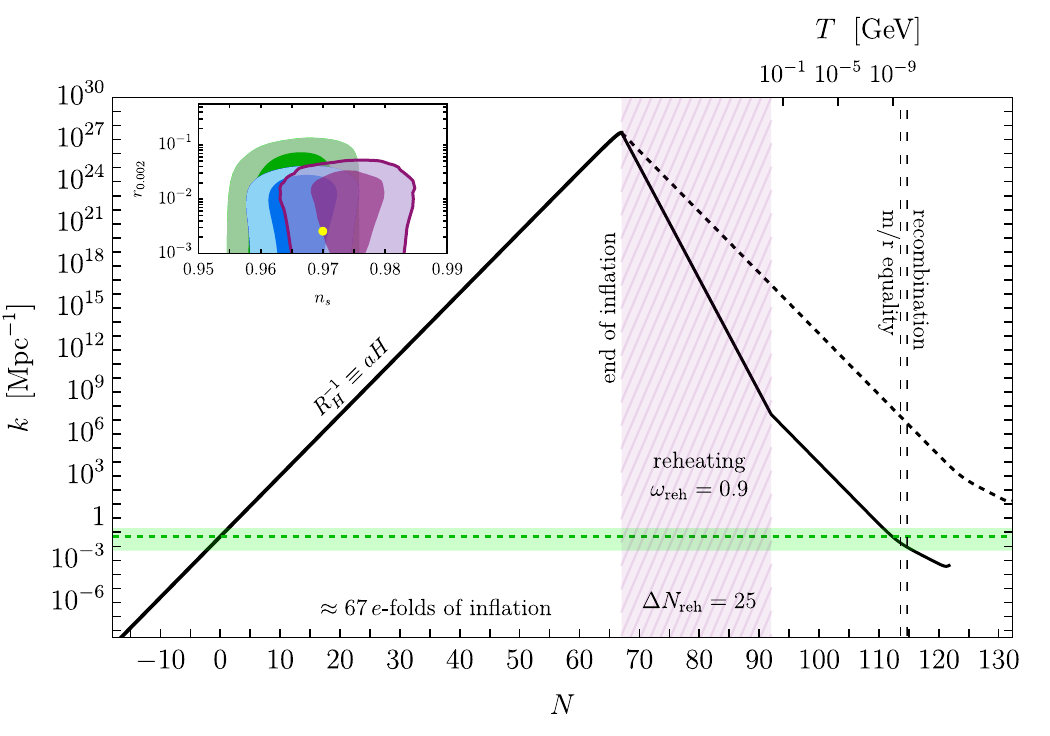}~~
\includegraphics[width=.495\textwidth]{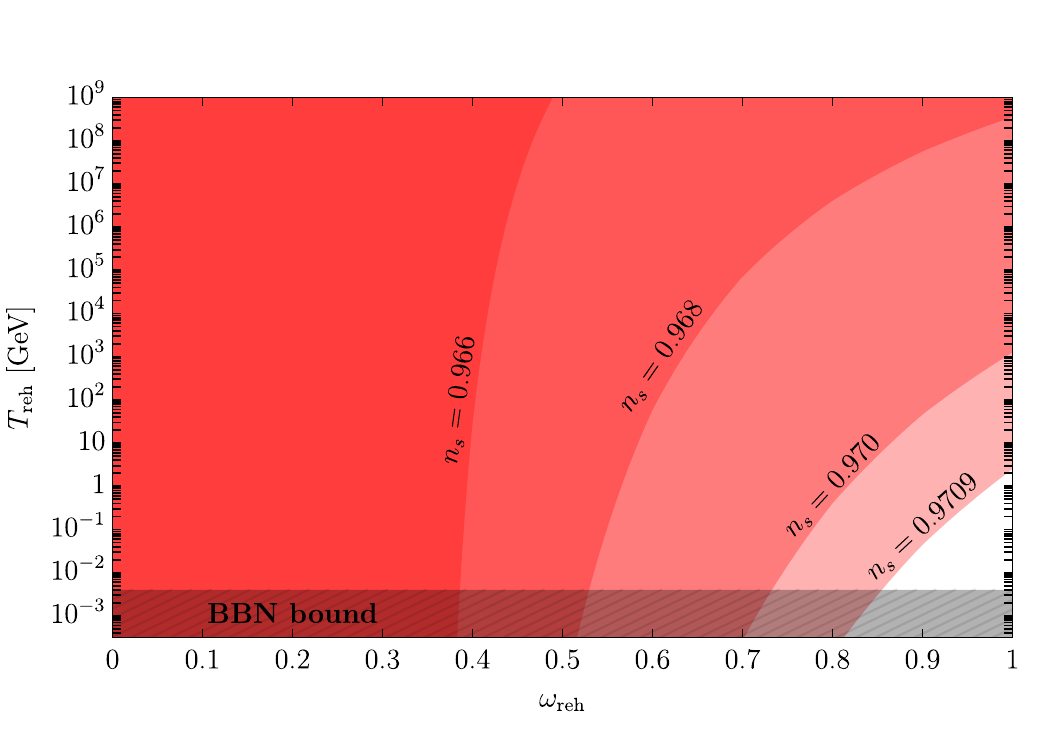}
$$\vspace{-0.5cm}
\caption{\textit{
\textit{\textbf{Left panel:}}
Evolution of the inverse comoving Hubble radius $R_H^{-1} = aH$  during inflation and in the subsequent phase. 
We consider $\Delta N_* = 67$ $e$-folds of inflation in the Starobinsky model which give 
$n_s = 0.97$ and $r=2.5\times 10^{-3}$ (yellow dot in the inset plot). 
After inflation, the solid line corresponds to a reheating phase with equation-of-state parameter 
$\omega_{\textrm{reh}} = 0.9$ and duration $\Delta N_{\textrm{reh}} = 25$, while the dashed line represents the case of instantaneous reheating. The green band highlights the cosmological pivot scale, the shaded purple region denotes the reheating epoch, and the vertical dashed lines indicate the times of matter–radiation equality and recombination. 
The reheating temperature is approximately $T_{\textrm{reh}} = 0.7$ GeV.
\textit{\textbf{Right panel:}} Excluded regions of the parameter space $(\omega_{\mathrm{reh}}, T_{\mathrm{reh}})$ resulting from the requirement that CMB modes remain sub-horizon at the time of recombination, shown for different values of the scalar spectral index $n_s$. The gray-shaded region indicates the lower bound $T_\reh \gtrsim 4\,\mathrm{MeV}$, imposed by the requirement of successful Big Bang nucleosynthesis.
}} 
 \label{fig:ReheatingStaro}  
\end{center}
\end{figure}
We illustrate this point in the left panel of fig.\,\ref{fig:ReheatingStaro}. 
We plot the evolution of the inverse comoving Hubble radius. The inflationary solution is chosen with $\Delta N_* = 67$
$e$-folds, which makes the Starobinsky model compatible with the recent ACT observations, yielding $n_s = 0.97$ and $r=2.5\times 10^{-3}$. Without loss of generality, we set $N = 0$ at the moment in which the CMB modes exit the horizon during inflation.
As shown, the CMB modes do not have time to re-enter the horizon in the post-inflationary evolution if one assumes instantaneous reheating (dashed line). On the other hand, introducing a prolonged and stiff reheating phase, here modeled with 
$\omega_{\textrm{reh}} = 0.9$ and 
$\Delta N_{\textrm{reh}} = 25$ (solid line), significantly modifies the evolution of the inverse Hubble radius after inflation, allowing the CMB scales to re-enter the horizon before matter–radiation equality and recombination. 

Let us reinforce this result through a more quantitative analysis. We consider some
value of comoving wavenumber $k$ and compare it with the
inverse comoving Hubble radius at the time of recombination $N_{\textrm{rec}}$, $k/a_{\textrm{rec}}H_{\textrm{rec}}$. We introduce the e-fold
time $N_k$ defined through the equation $k = a_k H_k$, where $N_k$ indicates
the moment in time during inflation when $k$ transitions
from being sub-horizon to super-horizon. We thus write (cf. refs.\,\cite{Munoz:2014eqa,Allegrini:2024ooy})
\begin{align}\label{eq:consistencyofscales}
    \log\left(
\frac{k}{a_{\textrm{rec}}H_{\textrm{rec}}}
\right) = -(\Delta N_{k} +\Delta N_{\textrm{reh}}+\Delta N_{\textrm{rec}}) + \log\left(
\frac{H_k}{H_{\textrm{rec}}}
\right) \,.
\end{align}
The term $\Delta N_{k}$ denotes the number of $e$-folds between horizon exit of the mode $k$ and the end of inflation, while $\Delta N_{\textrm{reh}}$ and the equation of state parameter $\omega_{\textrm{reh}}$ characterize the duration and dynamical properties of reheating. Finally, $\Delta N_{\textrm{rec}}$ is the number of $e$-folds from the end of the reheating phase to recombination epoch. 

The ratio of Hubble parameters $H_k / H_{\textrm{reh}}$ in the last term of eq.\,\eqref{eq:consistencyofscales} can be rewritten as $(\rho_k / \rho_{\textrm{rec}})^{1/2}$, where $\rho_k$ and $\rho_{\textrm{rec}}$ are, respectively, the energy density of the universe at $N_k$ and $N_{\textrm{rec}}$. Furthermore, since the Universe is radiation-dominated after reheating, the continuity equation allows us to express $\Delta N_{\textrm{rec}}$ as
\begin{equation}
    \Delta N_{\textrm{rec}} = \frac{1}{4} \log\left(
\frac{\rho_{\textrm{reh}}}{\rho_{\textrm{rec}}}
\right) \,.
\end{equation}
Plugging back into eq.\,\eqref{eq:consistencyofscales}, we finally obtain
\begin{equation}\label{eq:consistencyofscales2}
    (3 \omega_{\textrm{reh}} -1) \Delta N_{\textrm{reh}} =   4 \Delta N_k  -\log\left(
\frac{\rho_{\textrm{end}}}{\rho_{\textrm{rec}}}
\right)+4  \log\left(
\frac{k}{a_{\textrm{rec}}H_{\textrm{rec}}}
\right)\, ,
\end{equation}
where we used the fact that $ \rho_k \approx \rho_{\textrm{end}}$.  

The quantities $\rho_{\mathrm{rec}}$ and $a_{\mathrm{rec}} H_{\mathrm{rec}}$ are fixed by the standard $\Lambda$CDM cosmological model. In contrast, $\Delta N_k$ and $\rho_{\mathrm{end}}$ are determined by the inflationary dynamics. In the case of Starobinsky inflation, both are uniquely determined as functions of the scalar spectral index $n_s$. Requiring the CMB modes $k_{\rm CMB}$ to be sub-horizon at the epoch of recombination amounts to imposing $ \log(k_{\rm CMB} / a_{\textrm{rec}}H_{\textrm{rec}}) > 0$. Consequently, eq.\,\eqref{eq:consistencyofscales2} should be interpreted as a consistency relation between $\omega_{\mathrm{reh}}$ and $T_{\textrm{reh}}$ for a fixed value of $n_s$, where the relation between the reheating temperature $T_{\textrm{reh}}$ and the duration of the reheating stage $\Delta N_{\textrm{reh}} $ is given by 
\begin{align}\label{eq:connectionTrehDeltaNreh}
T_{\textrm{reh}}^4 = 
\frac{30\rho_{\textrm{end}}}{g_{\textrm{reh}}\pi^2}e^{-3\Delta N_{\textrm{reh}}(1+\omega_{\textrm{reh}})}\,,
\end{align}
and $g_{\textrm{reh}}$ denotes the effective number of relativistic degrees of freedom contributing to the energy density at thermalization, i.e. at the end of reheating. For simplicity, we fix the fiducial value $g_{\textrm{reh}} = 100$. 

Crucially, the left-hand side of eq.\,\eqref{eq:consistencyofscales2} changes sign at $\omega_\reh = 1/3$. As a consequence, for $\omega_\reh > 1/3$ a longer duration of the observed inflation $\Delta N_k$ can always be compensated by a longer reheating phase $\Delta N_\reh$, without spoiling the consistency of scales. On the other hand, for $\omega_\reh < 1/3$, higher values of $\Delta N_k$ require smaller values of $\Delta N_\reh$, which is however constrained to be non-negative. This observation explains the two different behaviors shown in fig.\,\ref{fig:ReheatingStaro} (left panel).

\begin{figure}[h]
\begin{center}
$$\includegraphics[width=.495\textwidth]{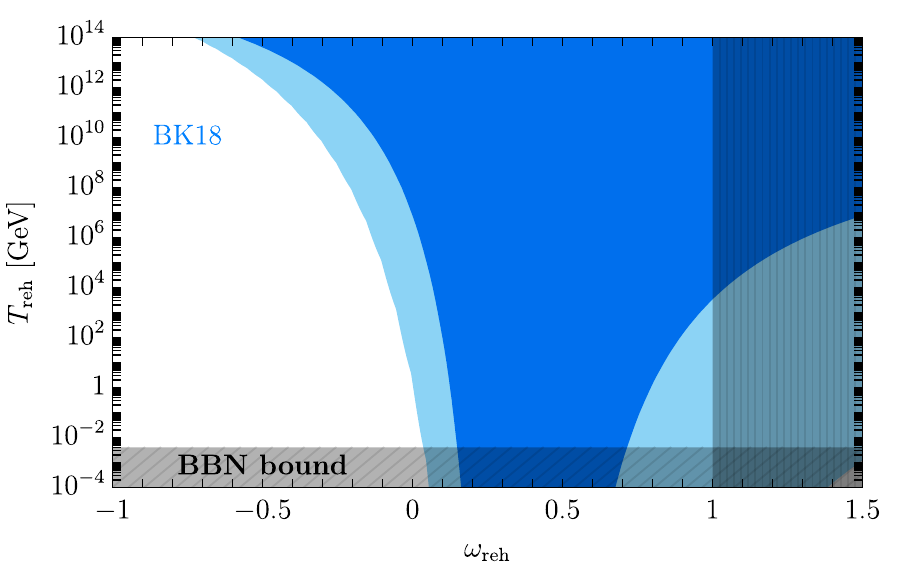}~~
\includegraphics[width=.495\textwidth]{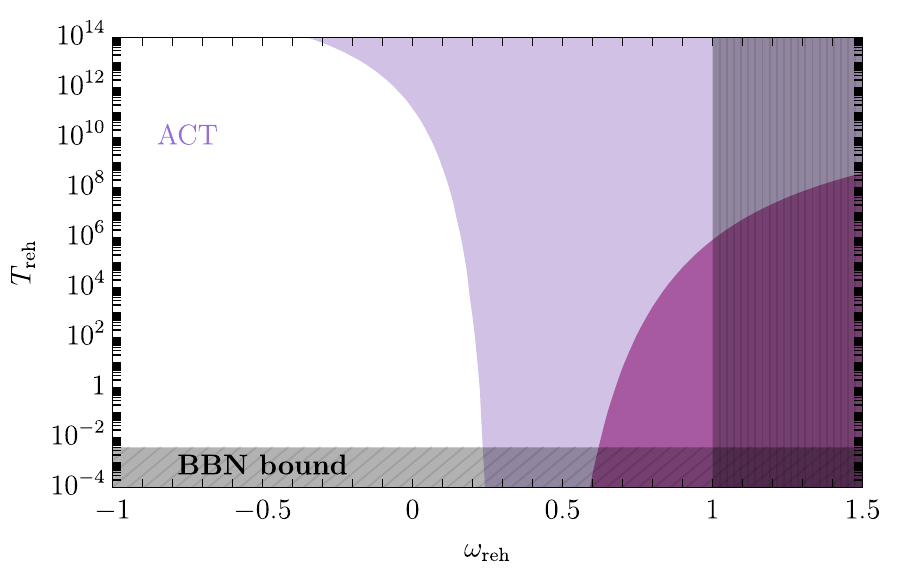}$$\vspace{-0.5cm}
\caption{\em Constraints in the $(\omega_\reh, T_\reh)$ plane using the joint analysis of Planck data with additional
BICEP/Keck and BAO data (left panel)~\cite{BICEP:2021xfz} and the recent results from the
Atacama Cosmology Telescope (right panel)~\cite{AtacamaCosmologyTelescope:2025nti}, for the 68\% and 95\% confidence contours. The vertical gray-shaded region denotes parameter values that violate the causality bound on the reheating equation of state, i.e. $\omega_{\rm reh} > 1$, while the horizontal gray-shaded region indicates the lower bound on the reheating temperature $T_{\rm reh} \gtrsim 4\,\mathrm{MeV}$, imposed by the requirement of successful Big Bang nucleosynthesis.}\label{fig:Reheating}  
\end{center}
\end{figure}

In the right panel of fig.\,\ref{fig:ReheatingStaro}, we present the constraints in the parameter space $(\omega_{\mathrm{reh}}, T_{\mathrm{reh}})$ derived from eq.\,\eqref{eq:consistencyofscales2}. Each shade of red denotes the region where $\log\left(k_{\rm CMB}/a_{\mathrm{rec}} H_{\mathrm{rec}}\right) < 0$ for a fixed value of $n_s$. We find that accommodating larger values of $n_s$, consistent with Planck+ACT data, requires a stiff reheating phase (namely $\omega_{\mathrm{reh}} > 1/3$) in agreement with Refs.\,\cite{Drees:2025ngb,Zharov:2025evb}. A complementary information is presented in fig.\,\ref{fig:Reheating}. In this case, we fix $k_{\rm CMB} = 0.05 \, {\rm Mpc^{-1}}$ and we find the region of parameter space $(\omega_{\mathrm{reh}}, T_{\mathrm{reh}})$ consistent with Planck+BICEP/Keck data~\cite{BICEP:2021xfz} (left panel) and ACT recent results~\cite{AtacamaCosmologyTelescope:2025nti} (right panel). In both cases, the darker shaded region corresponds to values of $n_s$ lying within one standard deviation of the respective central values ($\approx 0.965$ for Planck and $\approx 0.974$ for Planck+ACT), while the lighter shaded region corresponds to agreement at the two-sigma level. Crucially, the Planck+BICEP/Keck data are consistent at the one-sigma level with a standard reheating phase $\omega_\reh \simeq 0$~\cite{Saha:2020bis}, provided that the reheating temperature $T_\reh$ is sufficiently high (or, equivalently, that the reheating duration $\Delta N_\reh$ is sufficiently short). In contrast, one-sigma agreement with the Planck+ACT results implies a stiffer equation-of-state parameter $\omega_\reh$ and a lower reheating temperature $T_\reh$ (i.e., a longer duration of the reheating phase $\Delta N_\reh$). 

We now turn to the physical significance of our findings. Addressing the tension with the Planck+ACT data through a prolonged phase of inflation and an extended reheating stage dominated by a stiff equation of state, while phenomenologically viable, suffers from some conceptual shortcomings. Towards the end of inflation, the Starobinsky inflaton undergoes oscillations around the minimum of the potential, which is dominated by the quadratic term. 
These oscillations behave like a pressureless fluid, i.e. with an effective equation of state 
$\omega_{\textrm{reh}} = 0$ (matter-dominated)~\cite{Saha:2020bis, Allegrini:2024ooy}, rather than by a stiff phase with $\omega_{\textrm{reh}} \to 1$. Therefore, accommodating an exotic reheating stage requires deforming the Starobinsky potential in the vicinity of its minimum $\phi \approx 0$. This is opposite in spirit to the approach adopted in sec.\,\ref{sec:higher_curvature_operators}, where we considered deformations of the baseline Starobinsky potential at large field values, $\phi \gg 1$ (see the left panel of fig.\,\ref{fig:HOV}). It should be noted that the equation of state parameter of a homogeneous condensate oscillating in a potential with a minimum of the form
$V(\phi) \propto \phi^p$ is given by
$\omega_{\textrm{reh}} = (p-2)/(p+2)$, see ref.\,\cite{Turner:1983he}. 
Consequently, $\omega_{\textrm{reh}} > 1/3$ requires $p>4$, indicating a potential dominated near its minimum by higher-dimensional operators, a scenario that is certainly not natural. The condition $\omega_{\mathrm{reh}} > 1/3$ may also be achieved by replacing the minimum at $\phi \approx 0$ with a steep step in the potential, leading to a kination era~\cite{deHaro:2021swo}.

\section{Cosmological stasis: Theoretical considerations and phenomenological constraints}\label{sec:stasis}

In addition to the well-known conceptual issues of the standard cosmological model—such as the horizon and flatness problems—there is another, often underappreciated, feature: large portions of the universe's early evolution remain observationally unconstrained. While the CMB and BBN provide strong empirical anchors, 
the post-inflationary and pre-BBN epoch spans many orders of magnitude in time and is largely free from direct observational constraints. Within this window, the expansion rate, energy content, and thermal history of the universe could deviate significantly from standard assumptions without conflicting with known data. This observational gap provides ample room for nonstandard cosmological phases to occur.

The concept of \textit{stasis} deviates from the standard behavior. In a stasis era, two (or more) components maintain constant fractional energy densities over extended cosmological timescales, even as the universe continues to expand. This scenario is not achievable with traditional perfect fluids alone, due to their fixed redshifting behavior. Instead, it requires non-trivial dynamics—such as particle decays within a structured mass spectrum—that can inject energy into one component while simultaneously compensating the dilution of another.
%
%
%
In this section, we will adopt the mechanism proposed in \cite{Dienes:2021woi}. After the end of inflation, the scalar field undergoes an oscillatory phase around the minimum of the potential. During this stage, we assume that the inflaton decays into a tower of $N_{\rm max}$ massive states, denoted by $\phi_l$, with indices $l = 0, 2, \dots,N_{\rm max}-1$ assigned in order of increasing mass, and decay rate $\Gamma^{\phi}_l$. Each state is then assumed to subsequently decay into radiation with rate $\Gamma_l$, while direct interactions among the tower states are neglected. Using the fact that the kinetic energy of the inflaton is dominating over its potential energy ($\dot{\phi}^2 \gg V$), and denoting with $\Omega_\phi$, $\Omega_l$, and $\Omega_\gamma$, respectively, the abundances of the inflaton field, the $l$-th state in the tower, and of radiation, the evolution of the system is governed by the following set of equations
\begin{align}
 \frac{d\Omega_l}{dN} &= \Omega_l \left(-1 - \frac{\Gamma_l}{H} + \sum_{l = 1}^{N_{\rm max}} \Omega_l + 2 \Omega_\gamma + 4  \Omega_\phi \right)+\frac{\Gamma^{\phi}_l}{H} \Omega_\phi \, , \label{eq:continuity} \\
 \frac{d\Omega_\phi}{dN} &= \Omega_\phi \left(-4-\sum_{l = 1}^{N_{\rm max}} \frac{\Gamma^{\phi}_l}{H}+\sum_{l = 1}^{N_{\rm max}} \Omega_l + 2 \Omega_\gamma + 4  \Omega_\phi \right) \, ,\\
 \Omega_\gamma &= 1 -  \sum_{l = 1}^{N_{\rm max}} \Omega_l-\Omega_\phi  \, , \label{eq:1friedmann}\\
 \frac{dH}{dN} &= -\frac{1}{2}H \left(  2 +  \sum_{l = 1}^{N_{\rm max}} \Omega_l + 2 \Omega_\gamma +4\Omega_\phi  \right) \label{eq:2friedmann}\,,
\end{align}
where the evolutions is given in terms of $e$-folds, according to the relation $dN = H dt$.
The first two equations are the continuity equation for the cosmological components, while the last two are essentially the first and second Friedmann equations, respectively.  It is convenient to further introduce the total abundance of massive states
\begin{equation}\label{eq:OmegaMdef}
	\Omega_{\rm M} = \sum_{l = 1}^{N_{\rm max}} \Omega_l  \,. 
\end{equation}
Following ref.\,\cite{Dienes:2021woi}, we then parametrize the decay rate of the $l$-th state as 
\begin{equation}
		\Gamma_l = \Gamma_0 \left(\frac{m_l}{m_0}\right)^\gamma \,, 
\end{equation}
and $m_l = m_0 + (\Delta m) l^\delta$. Thus, $\Gamma_0$ and $m_0$ set, respectively, the decay rate and the mass of the lowest-energy state, while $\Delta m$, $\delta$, and $\gamma$ control the mass hierarchy within the tower. We introduce an analogous parametrization for the inflaton decay rates $\{\Gamma^{\phi}_l\}$, but with different parameters $\Gamma^\phi_0$ and $\gamma_\phi$, while $m_0$, $\Delta m$, and $\delta$ cannot change, as they depend only on the mass structure of the tower. In the context of Starobinsky inflation, the conformal transformation used to derive the theory in the Einstein frame universally fixes the decay rates $\{\Gamma_l^{\phi}\}$ in terms of the inflaton mass, the masses of the states in the tower, and the reduced Planck mass~\cite{Gorbunov:2010bn}. For instance, assuming that the tower is made of fermionic states, the decay rate into the heaviest state of mass $m_\psi$, reads~\cite{Gorbunov:2010bn}
\begin{equation}\label{eq:StaroGamma}
	\Gamma_{N_{\rm max}-1}^\phi = \frac{m_{\phi} m_{\psi}^2}{48\,\pi \, \MPl^2} \,,
\end{equation}
where $m_{\phi} \simeq 10^{13} \, {\rm GeV}$ is the inflaton mass.

\begin{figure}[h!]
\begin{center}
 $$\includegraphics[width=.495\textwidth]{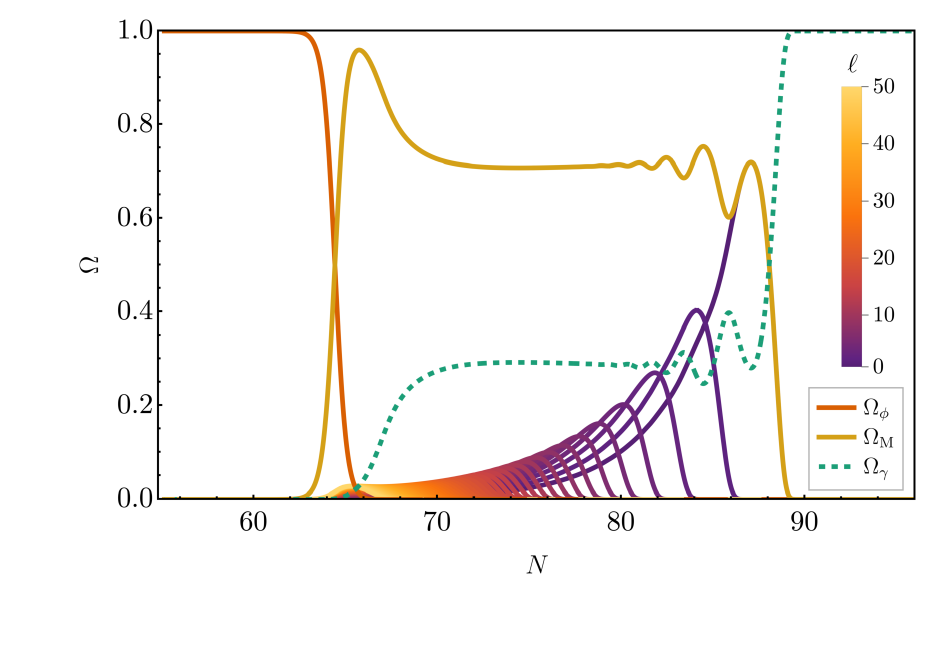}~~
\includegraphics[width=.495\textwidth]{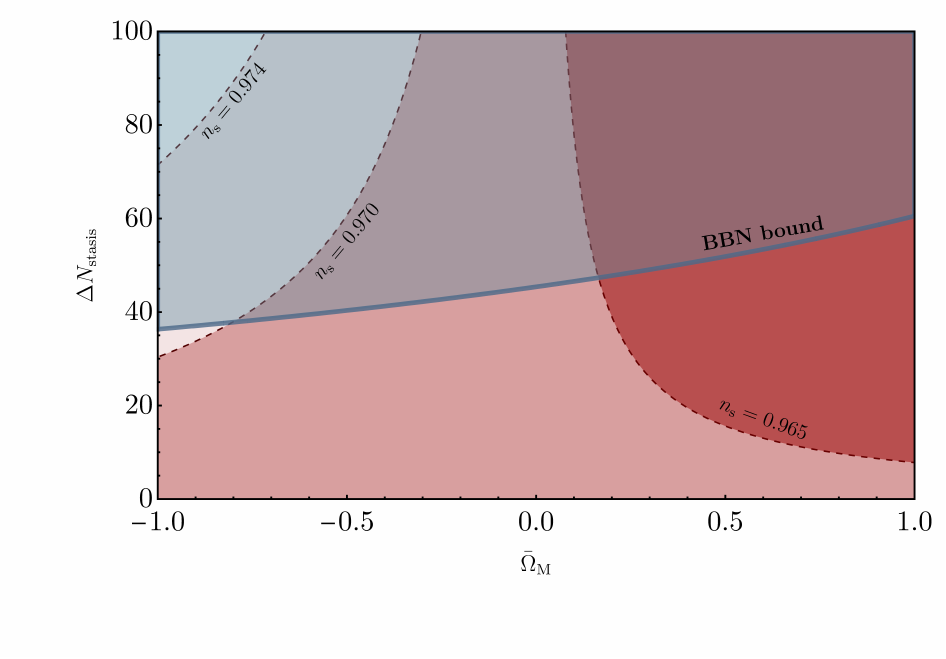}$$
\caption{\textit{\textit{\textbf{Left panel:}} A representative solution of eqs.\,\eqref{eq:continuity}-\eqref{eq:2friedmann} that leads to a cosmological stasis. We set $\Delta m=m_0$, $\delta=1.1$, $\gamma=6.3$, $\gamma_\phi=0.5$, $\Gamma^\phi_{N_{\rm max}-1}=10^{-2} \, {\rm GeV}$, $N_{\rm max}=50$, and $\Gamma_0=10^{-13}\,\Gamma^\phi_0$. We begin the numerical integration at $N=55$ with $\Omega_\phi=1$ and $\Omega_\gamma=\Omega_l=0$ for all $l$. The pile of colored curves at the bottom represents $\Omega_l \ \forall l$. \textit{\textbf{Right panel:}} Excluded regions of the parameter space $(\bar{\Omega}_{\rm M}, \Delta N_{\rm stasis})$ resulting from the requirement that CMB modes remain sub-horizon at the time of recombination, shown for different values of the scalar spectral index $n_s$. The blue-shaded region indicates the upper bound on $\Delta N_{\rm stasis}$, imposed by the requirement of successful Big Bang nucleosynthesis.
}}
\label{fig:Stasis}  
\end{center}
\end{figure}

In the left panel of fig.\,\ref{fig:Stasis}, we present a numerical integration of eqs.\,\eqref{eq:continuity}-\eqref{eq:2friedmann} for the parameter choice $\Delta m=m_0$, $\delta=1.1$, $\gamma=6.3$, $\gamma_\phi=0.5$, $\Gamma^\phi_{N_{\rm max}-1}=10^{-2} \, {\rm GeV}$, $N_{\rm max}=50$, and $\Gamma_0=10^{-13}\,\Gamma^\phi_0$. The large hierarchy between $\Gamma_0^\phi$ and $\Gamma_0$ ensures a sufficient temporal separation between inflaton decay and the onset of the stasis epoch. For the chosen initial conditions, eq.\,\eqref{eq:StaroGamma} implies $m_{\psi}\sim 10^{11} \, {\rm GeV}$.


We begin the numerical integration at $N=55$ with $\Omega_\phi=1$ and $\Omega_\gamma=\Omega_l=0$ for all $l$. In this representative solution, the inflaton decays completely into the tower of massive states after $\sim 10$ e-folds. The universe then undergoes a period of cosmological stasis lasting $\Delta N_{\rm stasis}\sim 15$ e-folds, after which it becomes radiation-dominated.

As explicitly shown in fig.~\ref{fig:Stasis}, during the cosmological stasis $\Omega_{\rm M}(N) \equiv \bar{\Omega}_{\rm M}$, i.e. the abundance of massive states is constant. This allows us to obtain an analytical solution for the evolution of the background geometry. Indeed, substituting eq.~\eqref{eq:OmegaMdef} and eq.~\eqref{eq:1friedmann} in eq.~\eqref{eq:2friedmann}, and using $\Omega_\phi \simeq 0$, yields
\begin{align}
	\frac{dH}{dN} &= -\frac{1}{2} H\left(  2 +  \Omega_{\rm M} + 2  - 2\Omega_{\rm M}  \right) =  -\frac{1}{2}H \left( 4 -  \Omega_{\rm M}\right) = -\frac{1}{2}H \left( 4 - \bar{ \Omega}_{\rm M}\right)  \, , 
\end{align}
which in turn gives
\begin{equation}\label{eq:Hubblesolution}
\frac{H(N)}{H_i} = e^{-\frac{1}{2} \left( 4 - \bar{ \Omega}_{\rm M}\right) (N - N_i)}  = \left(\frac{a}{a_i}\right)^{-\frac{1}{2}( 4 - \bar{ \Omega}_{\rm M}) }\, ,
\end{equation}
where  $N_i$ and $H_i$ are, respectively, the efold number and the Hubble parameter at the beginning of the stasis. We used the fact that $a/a_i = e^{N-N_i}$, in which $a$ is the scale factor. 

eq.~\eqref{eq:Hubblesolution} can be compared with the evolution of the Hubble parameter during a phase of reheating in which the universe is dominated by a single fluid with equation of state parameter $\omega_{\rm reh}$, that is
\begin{equation}
	\frac{H}{H_i} = \left(\frac{a}{a_i}\right)^{-\frac{3}{2}(1+\omega_{\rm reh})} \,.
\end{equation}
Thus, we see that during a stasis the universe geometry effectively evolves as if it was dominated by a fluid with equation of state parameter $\omega_{\rm reh}$, provided that
\begin{equation}\label{eq:barOmegaM}
	\bar{ \Omega}_{\rm M} = 1 - 3 \omega_{\rm reh} \,.
\end{equation}

In light of the preceding discussion, we now explore weather a cosmological stasis could reconcile the Starobinsky model with Planck+ACT data. Notice that, in contrast to previous approaches, we do not modify the baseline Starobinsky potential, neither in the large-field nor in the small-field regime. 
%
%
 %
%
%
It has been shown in ref.\,\cite{Dienes:2021woi} that $\Delta N_{\rm stasis} \propto \log N_{\rm max}$. Therefore, in the limit $N_{\rm max}\gg 1$ it is reasonable to neglect “edge effects” at the beginning and end of the decay process. We then impose the consistency relation eq.\,~\eqref{eq:consistencyofscales2} to the present scenario, where $\Delta N_\reh \approx \Delta N_{\rm stasis}$ and $\omega_\reh$ given by eq.~\eqref{eq:barOmegaM}. We present the resulting constraints in fig.\,\ref{fig:Stasis} (right panel). Each shade of red denotes the region where $\log\left(k_{\rm CMB}/a_{\mathrm{rec}} H_{\mathrm{rec}}\right) < 0$ for a fixed value of $n_s$. The blue-shaded region denotes the upper bound on the duration of the stasis phase, imposed by the requirement that the Universe be radiation dominated by the time its energy density reaches $\rho_{\rm BBN}$ at the epoch of Big Bang nucleosynthesis. This condition is equivalent to imposing $T_\reh \gtrsim 4\,\mathrm{MeV}$ in eq.\,\eqref{eq:connectionTrehDeltaNreh} and solving for $\Delta N_\reh \simeq \Delta N_{\rm stasis}$ as a function of $\bar{\Omega}_{\rm M}$ using eq.~\eqref{eq:barOmegaM}.

We notice that accommodating larger values of $n_s$, consistent with Planck+ACT data, requires negative values of $\bar{\Omega}_{\rm M}$, which in turn implies the existence of a tower of negative-energy density states. Remarkably, the condition $\bar{\Omega}_{\rm M} \leq 0$ translates into $\omega_\reh \geq 1/3$ via eq.~\eqref{eq:barOmegaM}. Therefore, once again, we find that resolving the tension with the Planck+ACT data appears to require introducing ingredients that are, from a fundamental physics perspective, somewhat pathological. Note that allowing for a non-zero duration of the edge effects can only strengthen this conclusion, since they would have to be compensated by an even longer period of stasis in eq.\,\eqref{eq:consistencyofscales2}.

\newpage

\section{Conclusions}\label{sec:conclusions}

The release of the ACT DR6 data\,\cite{ACT:2025fju,ACT:2025tim}, reporting a scalar spectral index $n_s \simeq 0.974 \pm 0.003$ when combined with previous Planck data~\cite{AtacamaCosmologyTelescope:2025blo}, has raised the question of whether the Starobinsky model of inflation\,\cite{Starobinsky:1980te} can remain a viable benchmark in the landscape of inflationary scenarios. In this work, we have examined the two principal strategies that have been proposed to reconcile the model with the Planck+ACT observations---higher-curvature corrections and a modified post-inflationary phase---and we have critically assessed their viability.

Concerning the inclusion of higher-curvature operators in the gravitational action, we have shown that while corrections of the form $R^3$, $R^4$, etc.\ can indeed shift the predictions of the Starobinsky model toward the ACT-favored region in the $(n_s,r)$ plane, this implies a reduced hierarchy between $\Lambda$, the mass scale controlling the size of the higher-order terms, and the inflaton mass $M$. As a consequence, Planck+ACT data inevitably drive higher-curvature extensions of Starobinsky inflation toward a regime characterized by a stronger sensitivity to ultraviolet physics, ultimately leading to fine-tuning in the spectral index $n_s$. We have substantiated our argument using concrete realizations of the higher-curvature correction scenario within no-scale supergravity and metric-affine gravity. In the latter case, proving that the metric-affine framework does not generate genuinely new inflationary dynamics but rather provides an alternative geometric realization of the same effective theory is in itself a non-trivial and original result. It would be interesting to generalize eq.~\eqref{eq:RewriteL} so that, rather than considering only powers of the scalar curvature, one includes all curvature operators at a given order in the effective action, since these are unavoidably produced by loop corrections~\cite{Birrell:1982ix}. We leave this for future work.

Turning to the reheating route, we have shown that extending the observed inflationary phase to $\Delta N_\star \simeq 67$--$70$ e-folds, accompanied by a prolonged reheating epoch with a stiff equation of state $\omega_{\textrm{reh}} \gtrsim 0.8$--$1$, brings the Starobinsky predictions within the $1\sigma$ Planck+ACT contours, in agreement with previous findings\,\cite{Drees:2025ngb,Zharov:2025evb}. However, such a scenario requires deforming the Starobinsky potential near its minimum with operators of the form $V(\phi) \propto \phi^p$ with $p > 4$, or alternatively replacing the minimum altogether with a steep runaway direction giving rise to a kination epoch\,\cite{deHaro:2021swo}. Moreover, we have explored the possibility that a phase of cosmological stasis\,\cite{Dienes:2021woi}, during which the fractional energy densities of matter and radiation remain constant thanks to the sequential decay of a tower of massive states, could serve as the post-inflationary epoch. We have derived the consistency conditions linking the stasis parameters $(\bar{\Omega}_{\textrm{M}},\Delta N_{\textrm{stasis}})$ to the observed value of $n_s$, imposing that the CMB modes re-enter the horizon before recombination and that the universe is radiation-dominated by the time of Big Bang nucleosynthesis. We find that accommodating the ACT-preferred spectral index requires $\bar{\Omega}_{\textrm{M}} \leq 0$. In other words, consistency with the Planck+ACT data would demand a tower of negative-energy-density states.

In summary, our analysis reveals a common thread: every attempt to reconcile the Starobinsky model with the ACT observations inevitably introduces ingredients---whether a fine-tuned separation of scales, an ad hoc deformation of the potential near the minimum, or a tower of negative-energy states---that undermine the minimality and geometric elegance constituting the original appeal of the Starobinsky model. Along these lines, ref.~\cite{Lust:2023zql} has questioned even the original Starobinsky model on effective-field-theory grounds, arguing that the theory lies in the Swampland. Although these results point toward a genuinely different inflationary model rather than a deformation of the Starobinsky paradigm (see also~\cite{Catinari:2025itc} for a broader discussion of naturalness in the context of inflationary models), future measurements by CMB-S4~\cite{CMB-S4:2016ple} and LiteBIRD~ \cite{LiteBIRD:2022cnt} will be decisive in sharpening or relaxing the present tension~\cite{Ferreira:2025lrd}.

\section{Acknowledgements}

L.D.G. is supported by NSF Grants No. PHY-2207502, AST-2307146, PHY-090003 and PHY-20043, by NASA Grant No. 21-ATP21-0010, by the John Templeton Foundation Grant 62840, and by the Simons Investigator Grant No. 144924. This work is partially supported by ICSC - Centro
Nazionale di Ricerca in High Performance Computing,
Big Data and Quantum Computing, funded by European
Union-NextGenerationEU and by the research grant number 20227S3M3B
under the program PRIN 2022 of
the Italian Ministero dell’Università e Ricerca (MUR). 

\newpage

\appendix 

\section{Unitarity of the $S$-matrix}\label{app:Conve}

Consider the case of a free, massive scalar field with mass $M$. 
Keeping track of powers of $\hbar$, we write its decomposition in terms of creation and annihilation operators in the form 
\begin{align}
\phi(x) = \sqrt{\hbar}\int 
\frac{d^3\boldsymbol{p}}{(2\pi\hbar)^3}\sqrt{\frac{\hbar}{2E_p}}\left[
a(\boldsymbol{p})e^{-ip\cdot x/\hbar} +
a^{\dagger}(\boldsymbol{p})e^{+ip\cdot x/\hbar}
\right]
 = \phi_+(x) + \phi_-(x)\,,\label{eq:ScalarFieldH}
\end{align}
with $E_p^2 = |\boldsymbol{p}|^2 + M^2\hbar^2$.
Equal-time commutation relations read
$[\phi(\boldsymbol{x},t),\dot{\phi}(\boldsymbol{y},t] = i\hbar\delta(\boldsymbol{x}-\boldsymbol{y})$ with $[a(\boldsymbol{p}),
a^{\dagger}(\boldsymbol{k})] = (2\pi\hbar)^3
\delta(\boldsymbol{p}-\boldsymbol{k})$. 
Creation and annihilation operators, therefore, have dimension $[a(\boldsymbol{p})]
= [a^{\dagger}(\boldsymbol{p})] = 
[\textrm{L}]^{3/2}$. 
Relativistically normalized single-particle states 
$|\boldsymbol{p}\rangle \equiv \sqrt{2E_p/\hbar}a^{\dagger}(\boldsymbol{p})|0\rangle$ 
have dimension $[|\boldsymbol{p}\rangle] = 
[\textrm{L}]$. 
Consequently, the $S$-matrix element 
$S_{fi} \equiv \langle f |S| i \rangle$ 
between some initial $| i \rangle \equiv |\boldsymbol{p_1},\dots , \boldsymbol{p_n}\rangle$ and final $| f \rangle \equiv |\boldsymbol{q_1},\dots,\boldsymbol{q_m}\rangle$ multi-particle states has dimension 
$[S_{fi}] = [\textrm{L}]^{n+m}$. 
Consider now the computation of the connected part of $S_{fi}$ with Feynman rules. 
For each external particle, we have 
$\phi_+(x)|\boldsymbol{p}\rangle 
= 
\hbar^{1/2}e^{-ip\cdot x/\hbar}
|0\rangle$, 
so that we have a factor $\hbar^{1/2}$ for each external leg while the phases $e^{-ip\cdot x/\hbar}$ eventually reconstruct a delta function of total four-momentum conservation, $(2\pi\hbar)^4\delta(p_1+\dots p_n - q_1 - \dots q_m)$, which carries the dimension $[\textrm{L}]^4$. We thus write
\begin{align}
S_{fi} =  
(2\pi\hbar)^4\delta(p_1+\dots p_n - q_1 - \dots q_m)\hbar^{(n+m)/2}i\tilde{\mathcal{M}}_{n\to m}\,,
\label{eq:SfiScaling}
\end{align}
with the scattering amplitude $i\tilde{\mathcal{M}}_{n\to m}$ computed with ordinary Feynman rules. 
Consider the contribution to $i\tilde{\mathcal{M}}_{n\to m}$ from a generic connected diagram with $P$ propagators and $V$ vertices. 
Each propagator contributes with a power of $\hbar$ while each vertex with a power $1/\hbar$.
Consequently, 
a total power of $\hbar^{P-V} = \hbar^{L-1}$ is associated to each diagram, 
where $L=P-(V-1)$ is the corresponding loop order. 
Schematically, we have
\begin{align}
i\tilde{\mathcal{M}}_{n\to m} = 
\frac{1}{\hbar}\,\textrm{(tree-level)} + 
\hbar^0\,\textrm{(one-loop)} +
 \hbar^1\,\textrm{(two-loops)} + \dots.
\end{align}
If we incorporate a power of $\hbar$ into 
$\tilde{\mathcal{M}}_{n\to m}$,  eq.\,(\ref{eq:SfiScaling}) takes the form $S_{fi} =  
(2\pi\hbar)^4\delta(p_1+\dots p_n - q_1 - \dots q_m)\hbar^{(n+m)/2 - 1}i(\hbar\tilde{\mathcal{M}}_{n\to m})$. 
Following from the previous argument, 
the quantity ${\mathcal{M}}_{n\to m} \equiv \hbar\tilde{\mathcal{M}}_{n\to m}$ 
 does not have any power of $\hbar$  at the tree-level while a factor of $\hbar^{L}$ must be included working 
at $L$ loops.
With these conventions, we write
\begin{align}
S_{fi} =  
(2\pi\hbar)^4\delta(p_1+\dots p_n - q_1 - \dots q_m)\hbar^{(n+m)/2 - 1}i\mathcal{M}_{n\to m}\,.
\label{eq:SfiScaling2}
\end{align}
Solving for the dimension of $\mathcal{M}_{n\to m}$, we find eq.\,(\ref{eq:MasterScaling})
\begin{align}
\boxed{
[\mathcal{M}_{n\to m}] = [\textrm{C}]^{-2+n+m}[\textrm{L}]^{n+m-4}
}\label{eq:MasterScaling1}
\end{align}

We are interested in the bounds placed on tree-level scattering amplitudes by unitarity. 
We refer to refs.\,\cite{Chang:2019vez,Abu-Ajamieh:2020yqi,Falkowski:2019tft,Cohen:2021ucp} for some original literature in the context of Higgs physics, and to ref.\,\cite{Steingasser:2025txd} for a recent application in the context of Higgs inflation. 
We write the $S$ matrix as
$S = \mathbb{1} + iT$, where the identity contribution represents the free propagation of particles without interactions, and the transition matrix $T$ describes interactions. 
 Unitarity of the $S$ matrix implies
that if $|i\rangle$ and $|f\rangle$ are unit normalized states we have $|\langle f|S|i\rangle| \leq 1$. 
For $|i\rangle \neq |f\rangle$, this condition implies 
$|\langle f|T|i\rangle| \leq 1$. 
Plane-wave states are not unit normalized, but we can 
 put forward the following construction. 
We consider the scattering of $n = \{n_1,\dots, n_r\}$ ingoing scalar
particles into $m = \{m_1,\dots, m_s\}$ outgoing scalar particles, where $n_I$ and $m_I$ respectively
denote the number of particles of the species $I$. We have $\sum_{i=1}^{r}n_i = n$ and $\sum_{j=1}^{s}m_j = m$. 
We define the $n$-particle 
state with total four-momentum $P^{\mu}$
\begin{align}
|P;\{n_1,\dots, n_r\}\rangle 
\equiv 
C_{\{n_1,\dots, n_r\}}
\int d^4x e^{-iP\cdot x/\hbar}
\left[\phi_1^-(x)\right]^{n_1}
\dots
\left[\phi_r^-(x)^{n_r}\right]
|0\rangle\,,
\end{align}
where $\phi_I^-(x)$ is the part of the (interaction picture) field $\phi_I(x)$, normalized as in 
eq.\,(\ref{eq:ScalarFieldH}), 
that contains a creation operator for the scalar particle of type $I$.  
More precisely, let us write 
\begin{align}
\phi_I^-(x) = \sqrt{\hbar}\int 
\frac{d^3\boldsymbol{p}}{(2\pi\hbar)^3}\sqrt{\frac{\hbar}{2E_p}}
a^{\dagger}(I,\boldsymbol{p})e^{+ip\cdot x/\hbar}\,,
\end{align}
where we have added a species index $I$ such that 
$|I,\boldsymbol{p}\rangle = 
\sqrt{2E_p/\hbar}\,a^{\dagger}(I,\boldsymbol{p})|0\rangle
$ with $[a(I,\boldsymbol{p}),a^{\dagger}(J,\boldsymbol{p})] = 
(2\pi\hbar)^3\delta_{IJ}\delta(\boldsymbol{p}-\boldsymbol{q})$.
Consider a multi-particle state defined by the partition $n = \{n_1,\dots, n_r\}$. 
Let us group the momenta according to the notation
\begin{align}
|
\underbrace{\boldsymbol{p}^{(1)}_1,\dots,\boldsymbol{p}^{(1)}_{n_1}}_{\textrm{type}\,1};
\underbrace{\boldsymbol{p}^{(2)}_1,\dots,\boldsymbol{p}^{(2)}_{n_2}}_{\textrm{type}\,2};
\dots ;
\underbrace{\boldsymbol{p}^{(r)}_1,\dots,\boldsymbol{p}^{(r)}_{n_r}}_{\textrm{type}\,r}
\rangle \equiv |\{\boldsymbol{p}^{(1)}\};\dots;
\{\boldsymbol{p}^{(r)}\}
\rangle
\,,
\end{align}
such that we have
\begin{align}
 |\boldsymbol{p}^{(1)}_1,\dots,\boldsymbol{p}^{(1)}_{n_1};
\boldsymbol{p}^{(2)}_1,\dots,\boldsymbol{p}^{(2)}_{n_2};
\dots ;
\boldsymbol{p}^{(r)}_1,\dots,\boldsymbol{p}^{(r)}_{n_r}
\rangle = 
\prod_{I=1}^{r}\prod_{a=1}^{n_I}
\sqrt{\frac{2E_{p_{a}^{(I)}}}{\hbar}}
a^{\dagger}(I,\boldsymbol{p}^{(I)}_{a})|0\rangle\,.
\end{align}
The normalization of these states is given by 
\begin{align}
 \langle\{
 \boldsymbol{q}^{(1)}
 \};\dots;\{
  \boldsymbol{q}^{(r)}
 \}
    |\{\boldsymbol{p}^{(1)}\};\dots;
\{\boldsymbol{p}^{(r)}\}
\rangle = 
\prod_{I=1}^{r}
\sum_{\sigma_I \in S_{n_I}}
\prod_{a=1}^{n_I}
(2\pi\hbar)^3
\frac{2E_{p^{(I)}_{\sigma_I(a)}}}{\hbar}
\delta\big(
\boldsymbol{q}_a^{(I)} - \boldsymbol{p}_{\sigma_I(a)}^{(I)}
\big)\,,
\end{align}
where we sum 
 over all permutations of the $n_I$
 identical particles of species $I$.
We write 
\begin{align}
   |P;\{n_1,\dots, n_r\}\rangle 
&= 
C_{\{n_1,\dots, n_r\}}\hbar^{n/2}
\int d^4x\left[
\prod_{I=1}^{r}\prod_{a=1}^{n_I}
\frac{d^3\boldsymbol{p}^{(I)}_{a}}{
(2\pi \hbar)^3(2E_{p_a^{(I)}}/\hbar)
}
\right]e^{-i(P-\sum_{a,I} p_a^{(I)})\cdot x/\hbar}
|\{\boldsymbol{p}^{(1)}\};\dots;
\{\boldsymbol{p}^{(r)}\}
\rangle
\nonumber\\
& =
C_{\{n_1,\dots, n_r\}}\hbar^{n/2}
\int \left[
\prod_{I=1}^{r}\prod_{a=1}^{n_I}
\frac{d^3\boldsymbol{p}^{(I)}_{a}}{
(2\pi \hbar)^3(2E_{p_a^{(I)}}/\hbar)
}
\right]
(2\pi\hbar)^4\delta\big(P-\sum_{a,I} p_a^{(I)}\big)
|\{\boldsymbol{p}^{(1)}\};\dots;
\{\boldsymbol{p}^{(r)}\}
\rangle
\end{align}
Given that the n-body phase space volume is given by 
\begin{align}
\textrm{Vol}_n \equiv \int\textrm{dLIPS}_n  = 
\int\left[\prod_{I=1}^{r}\prod_{a=1}^{n_I}
\frac{d^3\boldsymbol{p}^{(I)}_{a}}{
(2\pi \hbar)^3(2E_{p_a^{(I)}}/\hbar)
}\right]
(2\pi\hbar)^4\delta\big(P-\sum_{a,I} p_a^{(I)}\big)\,,
~~~~~\textrm{with}~~~[\textrm{Vol}_n] = 
[\textrm{L}]^{4-2n}\,,
\end{align}
we write
\begin{align}
    |P;\{n_1,\dots, n_r\}\rangle =
C_{\{n_1,\dots, n_r\}}\hbar^{n/2}
\int \textrm{dLIPS}_n
|\{\boldsymbol{p}^{(1)}\};\dots;
\{\boldsymbol{p}^{(r)}\}
\rangle
\,. 
\end{align}
In the case of massless particles with center of mass energy $E=\sqrt{P^2}$, the total volume of phase space
is given by\,\cite{Kleiss:1985gy}
\begin{align}
\textrm{Vol}_n  = \int\textrm{dLIPS}_n 
= \frac{1}{8\pi(n-1)!(n-2)!}\left(\frac{E}{4\pi}\right)^{2n-4}\,.
\end{align}
The coefficients $C_{\{n_1,\dots, n_r\}}$ can be fixed by imposing the normalization condition
\begin{align}
\langle Q;\{m_1,\dots, m_s\}|P;\{n_1,\dots, n_r\}\rangle = (2\pi\hbar)^4\delta(Q-P)\delta_{m_1 n_1}\dots \delta_{m_s n_r}\,,
\end{align}
that gives
\begin{align}
C_{\{n_1,\dots,n_r\}}
=\frac{1}{\sqrt{\hbar^{\,n}\mathrm{Vol}_n
\big(\prod_{I=1}^{r} n_I!}\big)}
\end{align}
Equivalently, the normalized states $|P;\{n_1,\dots,n_r\}\rangle$ can be written as
\begin{align}
|P;\{n_1,\dots,n_r\}\rangle
&=\frac{1}{\sqrt{\mathrm{Vol}_n
\big(\prod_{I=1}^{r} n_I!}\big)}\;
\int d\mathrm{LIPS}_n\;
|\{\boldsymbol{p}^{(1)}\};\dots;
\{\boldsymbol{p}^{(r)}\}
\rangle\,.
\end{align}
We define the scattering matrix elements $\hat{\mathcal{M}}_{n\to m}$ according to
\begin{align}
\langle Q,\{m_1,\dots,m_s\}|iT| P;\{n_1,\dots,n_r\}\rangle = (2\pi\hbar)^4\delta(P-Q)i\hat{\mathcal{M}}_{n\to m}\,.
\end{align}
Unitarity of the S-matrix directly implies
\begin{align}
 |\hat{\mathcal{M}}_{n\to m}| \leq 1\,,   
\end{align}
for $n\neq m$.
The relation with the ordinary 
scattering matrix elements defined in 
eq.\,(\ref{eq:SfiScaling2}) is given by 
\begin{align}
\hat{\mathcal{M}}_{n\to m} \equiv  
\left[\frac{1}{
\big(\prod_{i=1}^r n_i!\big)
\big(\prod_{j=1}^s m_j!\big)
\textrm{Vol}_n\textrm{Vol}_m
}\right]^{1/2}
\int \textrm{dLIPS}_n 
\int \textrm{dLIPS}_m\,
\hbar^{(n+m)/2 - 1}
\mathcal{M}_{n\to m}\,.
\end{align}
Therefore, we find that
\begin{align}
\boxed{
[\hat{\mathcal{M}}_{n\to m}] =  \hbar^{(n+m)/2 - 1}[\textrm{C}]^{n+m-2}  
}
\end{align}
This formula is illustrative, as it shows that phase-space–averaged scattering matrix elements are indeed dimensionless quantities, yet they exhibit a well-defined dependence on parameters with the dimensions of a coupling. 
If we now set the reduced Planck constant to unity, we recover exactly the scaling in eq.\,(\ref{eq:scalingAverage}).

We now provide an example of the scaling above. 

\begin{itemize}
    \item[$\circ$] Ref.~\cite{Creminelli:2025tae} 
    considers the single-field potential
    \begin{align}
     V(\phi) = \Lambda^4\cos\left(\frac{\phi}{f}\right) \,.  
    \end{align}
On dimensional ground, we have
\begin{align}
 [\Lambda] =\frac{1}{[\textrm{C}]^{1/2}[\textrm{L}]}\,,~~~~~~~~~
 [f] =\frac{1}{[\textrm{C}][\textrm{L}]}\,.
\end{align}
Consequently, using these two quantities we can form the following combinations 
with dimension, respectively, 
of a coupling and the inverse of a length
\begin{align}
\bigg[\frac{\Lambda^2}{f^2}\bigg] = [\textrm{C}]\,,~~~~~~~~~
\bigg[\frac{\Lambda^2}{f}\bigg] = \frac{1}{[\textrm{L}]}\,.\label{eq:CremiScaling}
\end{align}
According to eq.\,(\ref{eq:MasterScaling1}), therefore, 
the scattering amplitude for the process $n\to n$ 
can only scale according to
\begin{align}
\mathcal{M}_{n\to n} = \left(\frac{\Lambda^2}{f^2}\right)^{2n-2}    
\left(\frac{f}{\Lambda^2}\right)^{2n-4}\,.
\end{align}
The phase-space–averaged scattering matrix elements, in the limit of massless $\phi$, 
are given by 
\begin{align}
\hat{\mathcal{M}}_{n\to n} 
= \frac{\textrm{Vol}_n}{n!}\mathcal{M}_{n\to n} 
=
\frac{1}{8\pi n!(n-1)!(n-2)!}
\left(\frac{\Lambda^2}{f^2}\right)^{2n-2}
\left(\frac{f E}{4\pi\Lambda^2}\right)^{2n-4}\,,
\end{align}
which agrees with the results of ref.~\cite{Creminelli:2025tae}. 
Let us rewrite the above amplitude as
\begin{align}
\hat{\mathcal{M}}_{n\to n} 
=
\frac{1}{8\pi n!(n-1)!(n-2)!}
\left(\frac{\Lambda^2}{f^2}\right)^{2}
\left(\frac{E}{4\pi f}\right)^{2n-4}\,.
\end{align}
The amplitude is maximized by considering 
\begin{align}
n_{\textrm{max}} = \left(\frac{E}{4\pi f}\right)^{2/3}    \,,
\end{align}
and the unitarity bound $|\hat{\mathcal{M}}_{n\to n}| \leq 1$ defines the cut-off energy scale 
\begin{align}
 \frac{\Lambda_{\textrm{UV}}}{4\pi f}
 \approx \frac{1}{3\sqrt{3}}
 \left[
 \log\left(
 \frac{2\sqrt{2}\pi^{3/2}f^4}{\Lambda^4}
 \right)
 \right]^{3/2}\,.
\end{align}
Following from eq.\,(\ref{eq:CremiScaling}), it is convenient to introduce the 
coupling $g_* \equiv \Lambda^2/f^2$. 
We thus write
\begin{align}
  \frac{\Lambda_{\textrm{UV}}}{4\pi f}
 \approx \frac{1}{3\sqrt{3}}
 \left[
 \log\left(
 \frac{2\sqrt{2}\pi^{3/2}}{g_*^2}
 \right)
 \right]^{3/2}\,.
\end{align}
The cut-off scale 
$\Lambda_{\textrm{UV}}$ is parametrically 
larger than $4\pi f$ if 
\begin{align}
 g_* < \frac{(2\pi)^{3/4}}{e^{3/2}}\approx 0.9\,.   
\end{align}
\end{itemize}

\section{Metric-affine gravity}
\label{app:MetricAffine}
In General Relativity, the connection is assumed to be symmetric in its lower indices and metric-compatible. These two conditions uniquely determine the connection to be the Levi-Civita connection, ${{\{\}}_{\mathbf{g}}}^{\lambda}{}_{\mu\nu}$.\\
In metric-affine gravity these two assumptions are relaxed; consequently, the metric-affine connection ${\tilde{\Gamma}}^{\lambda}{}_{\mu\nu}$ and the metric are independent dynamical variables.
The metric-affine connection can be decomposed into two distinct contributions: torsion, associated with the asymmetry of the connection $\tilde{\Gamma}^{\lambda}{}_{\mu\nu} \not= \tilde{\Gamma}^{\lambda}{}_{\nu\mu} $, and non-metricity, which measure the violation of metric-compability of ${\tilde{\Gamma}}^{\lambda}{}_{\mu\nu}$. Precisely, the torsion tensor is defined as the antisymmetric part of the connection
\begin{align}
    T^{\lambda}{}_{\mu\nu}\equiv{\tilde{\Gamma}}^{\lambda}{}_{\mu\nu}-\tilde{\Gamma}^{\lambda}{}_{\nu\mu},\label{eq:torsion}
\end{align}
and the non-metricity can be defined as the following tensor
\begin{align}
    Q_{\rho\mu\nu}\equiv {\tilde{\nabla}}_{\rho}g_{\mu\nu},
\end{align}
where $\tilde{\nabla}$ is the covariant derivative defined from the connection ${\tilde{\Gamma}}^{\lambda}{}_{\mu\nu}$.\\
Both torsion and non-metricity are encoded in the distortion tensor $C^{\lambda}{}_{\mu\nu}$, defined as the difference between the metric-affine connection ${\tilde{\Gamma}}^{\lambda}{}_{\mu\nu}$ and the Levi-Civita connection
\begin{align}
{\tilde{\Gamma}}^{\lambda}{}_{\mu\nu} \equiv {{\{\}}_{\mathbf{g}}}^{\lambda}{}_{\mu\nu} + C^{\lambda}{}_{\mu\nu}. \label{eq:disto}
\end{align}
Since the general connection $\tilde{\Gamma}$ is asymmetric in the last two indices, a convention is needed for the covariant derivative of a tensor. In this work we adopt the following convention
\begin{align}
    {\tilde{\nabla}}_{\mu}V_{\nu} \equiv \partial_{\mu}V_{\nu} - {\tilde{\Gamma}}^{\sigma}{}_{\nu\mu}V_{\sigma}, \quad  {\tilde{\nabla}}_{\mu}V^{\nu} \equiv \partial_{\mu}V^{\nu} + {\tilde{\Gamma}}^{\nu}{}_{\sigma\mu}V^{\sigma}.\label{eq:covder}
\end{align}
From now on we assume metricity condition, i.e. $Q_{\rho\mu\nu}=0$. Under this assumption, the torsionless components of the distorsion tensor vanish and $C^{\lambda}{}_{\mu\nu}$ reduces to the so-called contorsion tensor $K^{\lambda}{}_{\mu\nu}$, as follows
\begin{align}
C^{\lambda}{}_{\mu\nu}=\frac{1}{2}\Big(T^{\lambda}{}_{\mu\nu}-T_{\nu}{}^{\lambda}{}_{\mu}-T_{\mu}{}^{\lambda}{}_{\nu}\Big) \equiv K^{\lambda}{}_{\mu\nu}. \label{eq:contorsion}
\end{align}
The torsion tensor can be decomposed in three irreducible components: a trace vector $T_{\mu}$, a pseudovector $S_{\mu}$ and the reduced torsion tensor $q_{\mu\nu\rho}$ cf. ref.\,\cite{Karananas:2021zkl}
\begin{align}
    T_{\mu}=T^{\nu}{}_{\mu\nu},\quad S^{\mu} = \epsilon^{\nu\rho\sigma\mu}T_{\nu\rho\sigma},\quad q_{\mu\nu\rho}=\frac{2}{3}\big(T_{\mu\nu\rho} - T_{[\nu}g_{\rho]\mu}-T_{[\nu\rho]\mu} \big), \label{eq:irrepscomp}
\end{align}
where we use square brackets to denote antisymmetrization, $T_{[\nu\rho]\mu}\equiv \frac{1}{2}\left(T_{\nu\rho\mu}-T_{\rho\nu\mu}\right)$.\\
The tensor $q_{\mu\nu\rho}$ satisfies the constraints
\begin{align}
    q^{\nu}{}_{\mu\nu} = 0, \quad \epsilon^{\mu\nu\rho\sigma}q_{\mu\nu\rho}=0,
\end{align}
reducing its components from 24 to 16. $\epsilon^{\mu\nu\rho\sigma}$ is the usual Levi-Civita tensor, i.e. $\epsilon^{\mu\nu\rho\sigma}\equiv\frac{1}{\sqrt{-g}}[\mu,\nu,\rho,\sigma]$. From the definition of $q_{\mu\nu\rho}$ in eq.\,\eqref{eq:irrepscomp}, it is straightforward to check that the reduced torsion tensor satisfies the cyclic identity
\begin{align}
q_{\mu\nu\rho}+q_{\nu\rho\mu}+q_{\rho\mu\nu}=0, \label{eq:cyclic}
\end{align}
which proves useful in calculations involving $q_{\mu\nu\rho}$.\\
With the definitions in \eqref{eq:irrepscomp}, the torsion tensor can be expressed as follows
\begin{align}
    T_{\mu\nu\rho} = \frac{1}{3}\Big(T_{\nu}g_{\rho\mu}-T_{\rho}g_{\nu\mu}\Big) - \frac{1}{6}\epsilon_{\mu\nu\rho\sigma}S^{\sigma} + q_{\mu\nu\rho}. \label{eq:irreps}
\end{align}
In the presence of torsion, the additional degrees of freedom in \eqref{eq:irrepscomp} lead to the following generalization of the curvature tensor\,\cite{Shapiro:2001rz}
\begin{align}
\big[{\tilde{\nabla}}_{\mu},{\tilde{\nabla}}_{\nu}\big]P^{\lambda}=T^{\rho}{}_{\mu\nu}{\tilde{\nabla}}_{\rho}P^{\lambda}+P^{\sigma}{\tilde{R}}^{\lambda}{}_{\sigma\mu\nu}. \label{eq:commutator}
\end{align}
Adopting the convention defined in \eqref{eq:covder} in the above commutator, the Riemann tensor in the presence of torsion reads
\begin{align}
\tilde{R}^{\lambda}{}_{\sigma\mu\nu} = \partial_{\mu}{\tilde{\Gamma}}^{\lambda}{}_{\sigma\nu}-\partial_{\nu}{\tilde{\Gamma}}^{\lambda}{}_{\sigma\mu}+{\tilde{\Gamma}}^{\lambda}{}_{\gamma\mu}{\tilde{\Gamma}}^{\gamma}{}_{\sigma\nu}-{\tilde{\Gamma}}^{\lambda}{}_{\gamma\nu}{\tilde{\Gamma}}^{\gamma}{}_{\sigma\mu}.
\end{align}
Using \eqref{eq:disto}, $\tilde{R}^{\lambda}{}_{\sigma\mu\nu}$ can be written in terms of the torsion-less Riemann tensor $R^{\lambda}{}_{\sigma\mu\nu}$ and torsionless covariant derivatives of the contorsion tensor
\begin{align}
    \tilde{R}^{\lambda}{}_{\sigma\mu\nu} = R^{\lambda}{}_{\sigma\mu\nu} + \nabla_{\mu}C^{\lambda}{}_{\sigma\nu}- \nabla_{\nu}C^{\lambda}{}_{\sigma\mu} + C^{\lambda}{}_{\gamma\mu}C^{\gamma}{}_{\sigma\nu} - C^{\lambda}{}_{\gamma\nu}C^{\gamma}{}_{\sigma\mu}. \label{eq:riemann}
\end{align}
From \eqref{eq:riemann}, the Ricci tensor reads
\begin{align}
    \tilde{R}_{\sigma\nu}=\tilde{R}^{\mu}{}_{\sigma\mu\nu}= R_{\sigma\nu}+\nabla_{\mu}C^{\mu}{}_{\sigma\nu}-\nabla_{\nu}C^{\mu}{}_{\sigma\mu}+C^{\mu}{}_{\gamma\mu}C^{\gamma}{}_{\sigma\nu}-C^{\mu}{}_{\gamma\nu}C^{\gamma}{}_{\sigma\mu}.
\end{align}
The Ricci scalar then becomes
\begin{align}
    \tilde{R} &= g^{\sigma\nu}\tilde{R}_{\sigma\nu}= g^{\sigma\nu}g^{\lambda\mu}\tilde{R}_{\lambda\sigma\mu\nu}=R+2\nabla_{\mu}C^{\mu\nu}{}_{\nu}-C_{\nu\mu}{}^{\mu}C^{\nu\gamma}{}_{\gamma}+C_{\nu\gamma\mu}C^{\nu\mu\gamma}. \label{eq:ricciscalarcont}
\end{align}
Since it will be useful to express the Ricci scalar in terms of the irreducible components of torsion in \eqref{eq:irrepscomp}, we first rewrite \eqref{eq:ricciscalarcont} in terms of the full torsion tensor using \eqref{eq:contorsion}, finding
\begin{align}
    \tilde{R} &=\color{black} R-2\nabla_{\mu}T^{\nu\mu}{}_{\nu}-T_{\mu\nu}{}^{\mu}T^{\gamma\nu}{}_{\gamma}+\frac{1}{2}T_{\mu\nu\gamma}T^{\gamma\nu\mu}+\frac{1}{4}T_{\mu\nu\gamma}T^{\mu\nu\gamma}. \label{eq:riccitor}
\end{align}
Note that the quadratic in torsion terms appearing in \eqref{eq:riccitor} correspond exactly to the three linearly independent, parity-even quadratic invariants that can be constructed from the torsion tensor. Following the notation of ref.\,\cite{Diakonov:2011fs}, we write
\begin{align}
    K_{1}\equiv T_{\gamma\nu\mu}T^{\gamma\nu\mu}, \quad K_{2}\equiv T_{\mu\nu}{}^{\mu}T^{\gamma\nu}{}_{\gamma},\quad K_{3}\equiv T_{\gamma\nu\mu}T^{\mu\nu\gamma}.
\end{align}
Using eq.\,\eqref{eq:irreps}, it is possible to express $K_{1}$, $K_{2}$ and $K_{3}$ in terms of the irreducible components of torsion
\begin{align}
    &K_{1} = T_{\gamma\nu\mu}T^{\gamma\nu\mu}=\frac{2}{3}T_{\mu}T^{\mu}-\frac{1}{6}S_{\mu}S^{\mu}+q_{\gamma\nu\mu}q^{\gamma\nu\mu},\label{eq:k1} \\ 
    &K_{2} = T_{\mu\nu}{}^{\mu}T^{\gamma\nu}{}_{\gamma}=T_{\mu}T^{\mu},\label{eq:k2} \\ 
    &K_{3} = T_{\gamma\nu\mu}T^{\mu\nu\gamma}=\frac{1}{3}T_{\mu}T^{\mu}+\frac{1}{6}S_{\mu}S^{\mu}+q_{\gamma\nu\mu}q^{\mu\nu\gamma}. \label{eq:k3}
\end{align}
It is important to note that the quadratic invariants in $q_{\mu\nu\gamma}$ appearing in $K_{1}$ and $K_{3}$ are not independent.
Indeed, using the definition of the reduced torsion tensor in \eqref{eq:irrepscomp} and its cyclic property \eqref{eq:cyclic}, the quadratic term $q_{\gamma\nu\mu}q^{\mu\nu\gamma}$ appearing in $K_{3}$ can be rewritten as follows
\begin{align}
    q_{\gamma\nu\mu}q^{\mu\nu\gamma}=q^{\mu\nu\gamma}q_{\mu\nu\gamma}-q^{\mu\nu\gamma}q_{\nu\mu\gamma} = 2q^{[\mu\nu]\gamma}q_{[\mu\nu]\gamma}=\frac{1}{3}\Big(T_{\mu\nu\rho}T^{\mu\nu\rho}-T_{\mu\nu\rho}T^{\nu\rho\mu} - T_{\mu}T^{\mu} \Big). \label{eq:cyclicasym}
\end{align}
Substituting this expression for $q_{\gamma\nu\mu}q^{\mu\nu\gamma}$ into the quadratic invariant $K_{3}$, we obtain
\begin{align}
    T_{\gamma\nu\mu}T^{\mu\nu\gamma}=\frac{1}{2}T_{\mu\nu\gamma}T^{\mu\nu\gamma}+\frac{1}{4}S_{\mu}S^{\mu} \label{eq:trel}
\end{align}
Now, substituting the expressions for $K_{2}$ and $K_{3}$ in \eqref{eq:k3} into \eqref{eq:trel}, we finally obtain
\begin{align}
    q_{\gamma\nu\mu}q^{\mu\nu\gamma}=\frac{1}{2}q_{\gamma\nu\mu}q^{\gamma\nu\mu}, \label{eq:qrel}
\end{align}
showing explicitly that there are only three linear independent parity even quadratic terms that can be built using the irreducible components of torsion.\\
Finally, using \eqref{eq:k3} and \eqref{eq:qrel}, the Ricci scalar can be written as follows
\begin{align}
    \tilde{R} = R - 2\nabla_{\mu}T^{\mu} - \frac{2}{3} T_{\mu} T^{\mu} +\frac{1}{24}S_{\mu}S^{\mu} + \frac{1}{2} q_{\gamma\nu\mu}q^{\gamma\nu\mu}. \label{eq:finalricciscalar}
\end{align}
In metric-affine gravity, it is possible to build another (pseudo)scalar, the Holst invariant, defined as
\begin{align}
    \tilde{R}'&= \frac{1}{\sqrt{-g}}[\mu,\nu,\rho,\sigma]{\tilde{R}}_{\mu\nu\rho\sigma} = \epsilon^{\mu\nu\rho\sigma}{\tilde{R}}_{\mu\nu\rho\sigma} \label{eq:holstdef}\,.
\end{align}
Substituting the expression for the Riemann tensor in \eqref{eq:riemann} into the definition \eqref{eq:holstdef}, we find
\begin{align}
\tilde{R}'&=2\epsilon^{\mu\nu\rho\sigma}\Big(\nabla_{\rho}C_{\mu\nu\sigma}+C_{\mu\gamma\rho}C^{\gamma}{}_{\nu\sigma}\Big).
\end{align}
Using \eqref{eq:contorsion}, it is possible to express the Holst invariant in terms of torsion, obtaining
\begin{align}
\tilde{R}'
&=\epsilon^{\mu\nu\rho\sigma}\Big[\Big(\nabla_{\rho}T_{\mu\nu\sigma}\Big) -\frac{1}{2}T^{\gamma}{}_{\nu\sigma}T_{\gamma\mu\rho}\Big].
\end{align}
Following again the notation of \cite{Diakonov:2011fs}, we define
\begin{align}
    K_{4}\equiv \frac{1}{2}\epsilon^{\mu\nu\rho\sigma}T^{\gamma}{}_{\nu\sigma}T_{\gamma\mu\rho},
\end{align}
which is a parity-odd quadratic in torsion invariant\footnote{It is possible to write another parity-odd quadratic in torsion invariant \cite{Diakonov:2011fs}, namely $K_{5}\equiv \epsilon^{\mu\nu\rho\sigma}T_{\nu\sigma}{}^{\gamma}T_{\mu\rho\gamma}$, which does not appear in $\tilde{R}'$.} arising in $\tilde{R}'$.\\
Using \eqref{eq:irrepscomp}, we can express $K_{4}$ in terms of the irreducible components of torsion as
\begin{align}
    K_{4} = \frac{2}{3}S_{\mu}T^{\mu}+\frac{1}{2}\epsilon^{\mu\nu\rho\sigma}q^{\gamma}{}_{\nu\sigma}q_{\gamma\mu\rho}.
\end{align}
Applying \eqref{eq:irrepscomp} once again, $\tilde{R}'$ can be written in terms of the irreducible components of the torsion as follows
\begin{align}
    \tilde{R}'&=-\nabla_{\mu}S^{\mu}-\frac{2}{3}S_{\mu}T^{\mu}-\frac{1}{2}\epsilon^{\mu\nu\rho\sigma}q^{\gamma}{}_{\nu\sigma}q_{\gamma\mu\rho}. \label{eq:holstirreps}
\end{align}

In this work, we focus on metric-affine models of the form
\begin{align}
    S= \frac{\MPl^2}{2}\int d^4x \sqrt{-g} \Big[\tilde{R} + f({\tilde{R}}^{'})\Big]. \label{eq:frprime}
\end{align}
The metric-affine model proposed in \cite{Salvio:2025izr} is a concrete example of \eqref{eq:frprime}.\\
Similarly to the metric $f(R)$ case studied in section \ref{sec:phenoanalyis}, assuming $f''(\tilde{R'})\neq 0$, a Legendre transformation allows us to recast actions of the form \eqref{eq:frprime} into a dynamically equivalent model in which the Lagrangian is linear in both the curvature scalars \eqref{eq:finalricciscalar} \eqref{eq:holstirreps}, with the addition of an auxiliary pseudo-scalar degree of freedom $z$. We write
\begin{align}
    S&= \frac{\MPl^2}{2}\int d^4x \sqrt{-g} \Big[\tilde{R} + f(\tilde{R'})\Big]= \frac{\MPl^2}{2}\int d^4x \sqrt{-g} \Big[\tilde{R} + f(z) + f'(z)(\tilde{R'}-z)\Big]\nn\\
    &=\MPl^2\int d^4x \sqrt{-g} \Big\{\frac{\tilde{R}}{2} + \frac{f'(z)}{2}\tilde{R'} -\frac{1}{2}[zf'(z)-f(z)]\Big\}. \label{eq:maS}
\end{align}
Indeed, solving the equation of motion for $z$ gives: $z=\tilde{R'}$, showing that $z$ is a pseudo-scalar, often referred to as pseudo-scalaron. Note that, the pseuso-scalaron is a field degree of freedom with a geometric origin; precisely, it is a manifestation of torsion.\\
Following the discussion of ref.\,\cite{Pradisi:2022nmh}, we clarify which degrees of freedom associated with the torsion tensor in actions of the form \eqref{eq:frprime} are effectively dynamical.\\
We can rewrite \eqref{eq:maS} as follows \,\cite{Pradisi:2022nmh}
\begin{align}
    S = \int d^4x \sqrt{-g} \Big(\MPl^2T^{\mu\nu\rho\sigma}\tilde{R}_{\mu\nu\rho\sigma}-V(z)\Big), \label{eq:nodindist}
\end{align}
where
\begin{align}
    T^{\mu\nu\rho\sigma}= \frac{1}{2}\Big[g^{\mu\rho}g^{\nu\sigma}+f'(z)\epsilon^{\mu\nu\rho\sigma}\Big], \quad V(z) = \frac{\MPl^2}{2}[zf'(z)-f(z)].
\end{align}
Note that both $V(z)$ and $T^{\mu\nu\rho\sigma}$ are independent on the torsion tensor.\\
In \eqref{eq:nodindist}, it is manifest that all the dependence on torsion tensor is contained in the metric-affine Riemann tensor, which involves torsion-less covariant derivatives of the contorsion tensor, as shown in \eqref{eq:riemann}. Therefore, when evaluating equation of motion for torsion, one can first integrate by parts the first term in \eqref{eq:nodindist}, moving the derivatives onto the metric and on the pseudo-scalaron $z$. This implies that the equations of motion for the torsion tensor are algebraic, i.e., the torsion tensor in the dynamically equivalent model \eqref{eq:maS} is not dynamical.\\
In this way, we reduced the metric-affine model with a generally dynamical torsion in \eqref{eq:frprime} to a dynamically equivalent theory \eqref{eq:maS} containing non-dynamical torsion and a pseudo-scalaron. Whether the pseudo-scalaron itself is dynamical depends on the specific form of the functional $f({\tilde{R'}})$. Specifically, the pseudo-scalaron is dynamical if the action in eq.\,\eqref{eq:frprime} cannot be cast in the following form\,\cite{Pradisi:2022nmh}
\begin{align}
 S= \frac{\MPl^2}{2}\int d^4x \sqrt{-g}\,  \left[a(\gamma)\tilde{R} + b(\gamma)\tilde{R'}+f\left(a(\gamma)\tilde{R} + b(\gamma)\tilde{R'}\right)\right], \label{eq:FalseMA}
\end{align}
where $a(\gamma)$ and $b(\gamma)$ are generic functions and $\gamma$ denotes a set of fields independent of torsion. Indeed one can explicitly check that the above action  features non-dynamical torsion, and thus is a metric-theory\footnote{The action in eq.\,\eqref{eq:FalseMA} can be rewritten in the form of eq.\,\eqref{eq:nodindist}, where $T^{\mu\nu\rho\sigma}$ takes the following form: $T^{\mu\nu\rho\sigma}= \frac{1}{2}\Big[1+f'(z)\Big]\cdot\Big[a(\gamma)g^{\mu\rho}g^{\nu\sigma}+b(\gamma)\epsilon^{\mu\nu\rho\sigma}\Big]$ and $z=a(\gamma)\tilde{R}+\beta(\gamma)\tilde{R'}$. Then, using the expressions of the Ricci scalar and the Holst invariant in eqs.\,\eqref{eq:finalricciscalar} and \eqref{eq:holstirreps}, one can integrate out the non-dynamical torsion and show that the resulting kinetic term for $z$ can be precisely canceled by performing a conformal transformation, reducing the theory to a manifestly metric one which propagates only $g_{\mu\nu}$. Note that metric-affine $f(\tilde{R})$ models are a particular case of the action \eqref{eq:FalseMA}; therefore, they are metric theories.
}.
\\
We can thus integrate out the non dynamical torsion tensor in \eqref{eq:maS}. To do so, we use the expression of the Ricci scalar and the Holst invariant in terms of the irreducible components of torsion, given in \eqref{eq:finalricciscalar} and \eqref{eq:holstirreps}, respectively. Indeed, varying the action with respect to the torsion tensor $T^{\mu\nu\rho}$ is equivalent to varying it separately with respect to $T^{\mu}$, $S^{\mu}$ and $q^{\mu\nu\rho}$, and this latter procedure is considerably more practical.\\
Using the expressions \eqref{eq:finalricciscalar} and \eqref{eq:holstirreps}, after integrating by parts while neglecting total derivatives, the action \eqref{eq:maS} can be written as
\begin{align}
    S&=\int d^4 x \sqrt{-g}\Bigg\{\frac{\MPl^2}{2}\Big[R-\frac{2}{3}T_{\mu}T^{\mu}+\frac{1}{24}S_{\mu}S^{\mu}+\frac{1}{2}q_{\gamma\nu\mu}q^{\gamma\nu\mu} \nn\\
    &\quad- f^{'}(z)\Big(\frac{2}{3}S_{\mu}T^{\mu}+\frac{1}{2}\epsilon^{\mu\nu\rho\sigma}q^{\gamma}{}_{\nu\sigma}q_{\gamma\mu\rho} \Big) +f^{''}(z)S^{\mu} \partial_{\mu}z\Big] - V(z)\Bigg\}. \label{eq:qua}
\end{align}
Varying \eqref{eq:qua} with respect to $T^{\mu}$, $S^{\mu}$ and $q^{\mu\nu\rho}$ yields respectively $T^{\mu}=-\frac{1}{2}f^{'}(z)S^{\mu}$, $S^{\mu}=8f^{'}(z)T^{\mu}-12f^{''}(z)\partial^{\mu}z$ and $q^{\mu\nu\rho}=0$. Substituting the on-shell values of $T^{\mu}$ and $S^{\mu}$ back into \eqref{eq:qua}, we obtain
\begin{align}
     S&=\int d^4 x \sqrt{-g}\Bigg\{\frac{\MPl^2}{2}R-\frac{3\MPl^2f^{\prime\prime}(z)^2}{1+4f^{\prime}(z)^2}\partial_{\mu}z\partial^{\mu}z - V(z) \Bigg \}.\label{eq:qua1}
\end{align}
Unlike the metric $f(R)$ case studied in section\,\ref{sec:phenoanalyis}, eq.\,\eqref{eq:qua1} shows that metric-affine actions of the form \eqref{eq:frprime} are dynamically equivalent to the Einstein Hilbert action supplemented by a pseudo-scalar field in the Jordan frame, i.e. without requiring any conformal transformation. The only remark is that the connection associated with the model \eqref{eq:qua1} is ${\tilde{\Gamma}}^{\lambda}{}_{\mu\nu}$, which is not conformally equivalent to the Levi-Civita connection whenever the metric-affine theory features dynamical torsion. In the case of the model \eqref{eq:frprime}, the condition $f^{''}(z)\neq0$ implies that the action cannot be recast in the form of eq.\,\eqref{eq:FalseMA}, implying that the torsion is dynamical.
In order to write \eqref{eq:qua1} in a canonical form, we define
\begin{align}
     \varphi \equiv f^{'}(z),\label{eq:eqfieldred}
\end{align}
and introduce the field transformation:
\begin{align}
    \varphi = \frac{1}{2}\textrm{sinh}\Bigg\{\frac{\phi}{\MPl\sqrt{3/2}} + \textrm{sinh}^{-1}[2f^{'}(0)] \Bigg\}. \label{eq:eqfieldred1}
\end{align}
Using the above field redefinitions, we obtain
\begin{align}
     S&=\int d^4 x \sqrt{-g}\Big[\frac{\MPl^2}{2}R-\frac{1}{2}\partial_{\mu}\phi\partial^{\mu}\phi - V(\phi) \Big]\label{eq:finale}.
\end{align}
The explicit form of the pseudo-scalar potential $V(\phi)$ depends on the specific form of the functional $f(\tilde{R'})$. In this work, we concentrate on the metric-affine model proposed in ref.\,\cite{Salvio:2025izr}
\begin{align}
f(\tilde{R'})=\frac{2}{\MPl^2}\Big(\beta\tilde{R'}+c{\tilde{R'}}^2\Big),
\label{eq:Salviofunc}
\end{align}
where~$c$ and $\tilde{\beta}\equiv \beta/\MPl^2$ are dimensionless parameters.\\
In the case of \eqref{eq:Salviofunc} it is possible to solve \eqref{eq:eqfieldred} for $z$ and, using \eqref{eq:eqfieldred1}, obtain $z(\phi)$. Given that for \eqref{eq:Salviofunc} $V(z)=cz^2$, the pseudo-scalaron potential $V(\phi)$ is given by
\begin{align}
V(\phi) = \frac{\MPl^4}{4c}\left\{
\tilde{\beta} - 
\frac{1}{4}\sinh\left[
\frac{\phi}{\MPl\sqrt{3/2}}
+\tanh^{-1}\left(
\frac{4\tilde{\beta}}{
\sqrt{16\tilde{\beta}^2 + 1}
}
\right)
\right]
\right\}^2
\label{eq:VphiApp}.
\end{align}

In the spirit of Section \ref{sec:MetricAffineGravity}, we derive the explicit expression for the torsion tensor associated with metric-affine models of the form \eqref{eq:frprime}.\\
Substituting the on-shell values of the irreducible torsion components $T^{\mu}$ and $S^{\mu}$ into \eqref{eq:irreps}, we find
\begin{align}
    T_{\mu\nu\rho}=\frac{2f''(z)f'(z)}{1+4f'(z)^2}\Big(g_{\rho\mu}\partial_{\nu}z-g_{\nu\mu}\partial_{\rho}z \Big)+\frac{2f''(z)}{1+4f'(z)^2}\epsilon_{\mu\nu\rho\sigma}\partial^{\sigma}z
    \label{eq:torsiononshell}
\end{align}
The above expression shows that the pseudo-scalaron $z$ does not merely play the role of an additional degree of freedom, as the scalaron does in metric $f(R)$ theories. Rather, it also generates a deviation from the Levi-Civita connection of torsional origin.\\
For the specific case of the metric-affine model \eqref{eq:Salviofunc}, if $\tilde{\beta}$ and $c$ are treated as independent, it is manifest from the action \eqref{eq:qua1} and from eq.\,\eqref{eq:torsiononshell} that, in the limit $|\tilde{\beta}| \to \infty$, the pseudo-scalaron becomes non dynamical, the torsion tensor (and hence the contorsion tensor) vanishes, and the connection reduces to the Levi-Civita one
\begin{align}
|\tilde{\beta}|\to\infty\quad \Rightarrow\quad K^{\lambda}{}_{\mu\nu}\to0\quad \Rightarrow\quad  {\tilde{\Gamma}}^{\lambda}{}_{\mu\nu}\to{{\{\}}_{\mathbf{g}}}^{\lambda}{}_{\mu\nu}. \label{eq:Metriclimit}
\end{align}
In the above limit, the metric-affine model \eqref{eq:Salviofunc} reduces to a metric theory. More precisely, integrating out the non dynamical pseudo-scalaron $z$ from the action \eqref{eq:qua1} yields the Einstein-Hilbert action. It is worth noting that, in this setup, $|\tilde{\beta}| \to \infty$ is a torsionless limit that implies $M\to\infty$, as follows from eq.\,\eqref{eq:SaMass}. This is consistent with the fact that $z$, being a manifestation of torsion, must not propagate in a metric limit.\\ 
Therefore, the limit \eqref{eq:Metriclimit} shows that, in the large-$|\tilde{\beta}|$ regime, the metric-affine model \eqref{eq:Salviofunc} reduces to be equivalent to the metric $f(R)$ model in eq.\,\eqref{eq:resummedfR}, since both reduce to General Relativity, as discussed in section \ref{sec:MetricAffineGravity}. However, in an inflationary context, the ratio $c/\tilde{\beta}^2$ is fixed to a large value to reproduce the observed scalar power spectrum. As shown in eq.\,\eqref{eq:cOkApprox}, this prevents $M$ from diverging and yields $M\lesssim O(10^{-5})\MPl$.
It then follows from the action \eqref{eq:qua1} and from  \eqref{eq:torsiononshell} that the pseudo-scalaron remains dynamical and the torsion tensor does not vanish in the large-$|\tilde{\beta}|$ regime. As a result, the model \eqref{eq:Salviofunc} does not reduce to a metric theory.

\section{Metric-affine generalization of Starobinsky}
\label{app:C}
Consider the following metric-affine modification of the Starobinsky model, obtained by supplementing its metric-affine counterpart with a purely torsional operator
\begin{align}
    S=\frac{\MPl^2}{2}\int d^4x\sqrt{-g}\Big(\tilde{R}+\frac{\tilde{R}^2}{6\bar{M}^2} +2\tilde{\beta}\tilde{R'} \Big),
    \label{eq:MAStaro}
\end{align}
where $\bar{M}^2$ is given by
\begin{align}
\bar{M}^2\equiv \MPl^2\frac{\tilde{\beta}^2}{3c}.
\end{align}
A Legendre transformation allows to recast the action in eq.\,\eqref{eq:MAStaro} into a dynamically equivalent model in which the Lagrangian is linear in $\tilde{R}$, with the addition of an auxiliary scalar field $\sigma$ 
\begin{align}
    S=\int d^4x\sqrt{-g}\left\{\frac{\MPl^2}{2}\Big[\xi'(\sigma)\tilde{R}+2\tilde{\beta}\tilde{R'}\Big] -v(\sigma) \right\}, 
    \label{eq:MAStaroscalaron0}
\end{align}
where the functional $\xi(\sigma)$ and the scalar potential $v(\sigma)$ are defined as follows
\begin{align}
\xi(\sigma)\equiv\sigma+\frac{\sigma^2}{6\bar{M}^2},\quad v(\sigma)\equiv\frac{\MPl^2}{2}\Big(\sigma\xi'(\sigma)-\xi(\sigma) \Big),
\end{align}
and we assumed $\xi''(\tilde{R})\neq0$. Indeed, solving the equation of motion for $\sigma$ gives: $\sigma=\tilde{R}$, showing that $\sigma$ is a scalar of geometric origin. Unlike the pseudo-scalaron case treated in app.\,\ref{app:MetricAffine}, which vanishes in absence of torsion, $\sigma$ does not vanish when torsion is absent.\\
Following the discussion of app.\,\ref{app:MetricAffine}, it is possible to integrate out the non dynamical components of the torsion tensor in eq.\,\eqref{eq:MAStaro} by putting on shell the torsion tensor appearing in eq.\,\eqref{eq:MAStaroscalaron0}. Using eqs.\,\eqref{eq:finalricciscalar} and \eqref{eq:holstirreps}, after integrating by parts while neglecting total derivatives, the action \eqref{eq:MAStaroscalaron0} can be written as
\begin{align}
    S=&\int d^4x\sqrt{-g}\Bigg\{\frac{\MPl^2}{2}\Big[\xi'(\sigma)R+\xi'(\sigma)\Big(\frac{1}{24}S_{\mu}S^{\mu}-\frac{2}{3}T_{\mu}T^{\mu}+\frac{1}{2}q_{\gamma\nu\mu}q^{\gamma\nu\mu}\Big)\nn\\&+2\xi''(\sigma)T^{\mu}\partial_{\mu}\sigma- \frac{4}{3}\tilde{\beta}S_{\mu}T^{\mu}-\tilde{\beta}\epsilon^{\mu\nu\rho\sigma}q^{\gamma}{}_{\nu\sigma}q_{\gamma\mu\rho}\Big]-v(\sigma)\Bigg\}
    \label{eq:MAStaroirreps}
\end{align}
Contrary to the metric-affine case considered in ref.\,\cite{Salvio:2025izr}, and similarly to standard metric $f(R)$ theories, the Ricci scalar $R$ is here multiplied by the factor $\xi'(\sigma)$. In order to bring the action \eqref{eq:MAStaroirreps} into an Einstein-Hilbert-like form, we perform a conformal transformation to the Einstein frame, defined as follows
\begin{align}
\bar{g}_{\mu\nu}\equiv \Omega^2 g_{\mu\nu}\,,~~~~~~~
\textrm{with}~~~\Omega^2 \equiv \xi^{\prime}(\sigma)\,. \label{eq:confTrMA}
\end{align}
The action \eqref{eq:MAStaroirreps}, when expressed in the Einstein frame, takes the form (we drop the symbol $\,\bar{}\,$ from all metric-dependent functions)
\begin{align}
	S_{E}=&\int d^4x\sqrt{-g}\Bigg\{\frac{\MPl^2}{2}\Big[R-\frac{3}{2}\frac{\xi''(\sigma)^2}{\xi'(\sigma)^2}\partial_{\mu}\sigma\partial^{\mu}\sigma+\frac{1}{24}S_{\mu}S^{\mu}-\frac{2}{3}T_{\mu}T^{\mu}+\frac{1}{2}\xi'(\sigma)^2q_{\gamma\nu\mu}q^{\gamma\nu\mu} \nn\\&+2\frac{\xi''(\sigma)}{\xi'(\sigma)}T^{\mu}\partial_{\mu}\sigma- \frac{4}{3}\frac{\tilde{\beta}}{\xi'(\sigma)}S_{\mu}T^{\mu}-\tilde{\beta}\xi'(\sigma)\epsilon^{\mu\nu\rho\sigma}q^{\gamma}{}_{\nu\sigma}q_{\gamma\mu\rho}\Big]-V(\sigma)\Bigg\}\,,
	\label{eq:MAStaroEframe}
\end{align}
where $V(\sigma)\equiv\frac{v(\sigma)}{\xi'(\sigma)^2}$.\\
Varying eq.\,\eqref{eq:MAStaroEframe} with respect to $T^{\mu}$, $S^{\mu}$ and $q^{\mu\nu\rho}$ gives respectively $S_{\mu}=\frac{16\tilde{\beta}}{\xi'(\sigma)}T_{\mu}$, $T_{\mu}=\frac{3}{2}\frac{\xi''(\sigma)}{\xi'(\sigma)}\partial_{\mu}\sigma-\frac{\tilde{\beta}}{\xi'(\sigma)}S_{\mu}$ and $q^{\mu\nu\rho}=0$. Substituting the on-shell values of $S^{\mu}$ and $T^{\mu}$ back into eq.\,\eqref{eq:MAStaroEframe}, we obtain
\begin{align}
    S_{E}=\int d^4 x\sqrt{-g}\Bigg\{\frac{\MPl^2}{2}\Bigg[R-\frac{3}{2}\frac{\xi''(\sigma)^2}{\xi'(\sigma)^2}\Bigg(\frac{1}{1+\frac{\xi'(\sigma)^2}{16\tilde{\beta}^2}}\Bigg)\partial_{\mu}\sigma\partial^{\mu}\sigma\Bigg] -V(\sigma) \Bigg\}
    \label{eq:MAStaroscalaron}
\end{align}
In the Einstein frame, the action \eqref{eq:MAStaroscalaron} takes the form of a metric scalar–tensor theory. The only remark is that the connection associated with \eqref{eq:MAStaroscalaron} is the conformally transformed $\tilde{\Gamma}^{\lambda}{}_{\mu\nu}$, rather than the conformally transformed Levi-Civita connection, whenever the metric-affine theory features dynamical torsion. In the case of the model \eqref{eq:MAStaro}, the condition $\xi^{''}(\sigma)\neq0$ implies that the action cannot be recast in the form of eq.\,\eqref{eq:FalseMA}, implying that $\sigma$ is dynamical. Comparing \eqref{eq:MAStaroscalaron} with its metric $f(R)$ counterpart in \eqref{eq:EinsteinFrameFofR}, one can explicitly see that the coefficient of the kinetic term contains the usual metric contribution arising from the conformal transformation, together with an additional torsional contribution originating from the non-dynamical components of the torsion tensor.\\
In order to write the action \eqref{eq:MAStaroscalaron} in a canonical form, we define
\begin{align}
    \varphi\equiv\xi'(\sigma),
\end{align}
and introduce the field transformation
\begin{align}
    \phi\equiv\MPl\sqrt{\frac{3}{2}}\Big[\text{sinh}^{-1}\Big(\frac{4\tilde{\beta}}{\varphi}\Big)-\text{sinh}^{-1}\Big(4\tilde{\beta}\Big)\Big].
    \label{eq:MAFieldTra}
\end{align}
Using the above field redefinition, we obtain
\begin{align}
     S_{E}&=\int d^4 x \sqrt{-g}\Big[\frac{\MPl^2}{2}R-\frac{1}{2}\partial_{\mu}\phi\partial^{\mu}\phi - V(\phi) \Big].\label{eq:MAStarocanonical}
\end{align}
where $V(\phi)$ exactly reproduces the scalar potential in eq.\,\eqref{eq:VphiApp} proposed in ref.\,\cite{Salvio:2025izr}. 
This analysis shows that the inflationary potential \eqref{eq:VphiApp} can be reproduced by the metric-affine counterpart of the Starobinsky model supplemented with the purely torsional operator $\tilde{\beta}\tilde{R'}$.\\
It is important to note that the mass parameter $\bar{M}^2$ appearing in eq.\,\eqref{eq:MAStaro} does not correspond to the mass of the scalaron $\sigma$. The coefficient $\bar{M}^2$ multiplying the curvature-squared term coincides with the scalaron mass only in metric $f(R)$ theories, such as the one in \eqref{eq:resummedfR}. This distinction originates from the different field redefinitions required in the metric and metric-affine formulations, namely \eqref{eq:FieldTra} and \eqref{eq:MAFieldTra}, respectively, which must be properly taken into account when computing the second derivatives of the scalar potential.\\ 
The action in eq.\,\eqref{eq:MAStaro} reproduces, in the Einstein frame, the scalar potential \eqref{eq:VphiApp} and thus describes the following scalaron mass
\begin{align}
M^2\equiv\frac{\MPl^2(1+16\tilde{\beta}^2)}{48c}. \label{eq:Salviomass}
\end{align}
Substituting the on-shell values of the irreducible torsion components $T^{\mu}$ and $S^{\mu}$ into eq.\,\eqref{eq:irreps}, it is possible to write the on-shell expression of the torsion tensor
\begin{align}
    T_{\mu\nu\rho}=\frac{1}{2}\frac{\xi''(\sigma)\xi'(\sigma)}{\xi'(\sigma)^2+16\tilde{\beta}^2}\Big(g_{\rho\mu}\partial_{\nu}\sigma-g_{\nu\mu}\partial_{\rho}\sigma \Big)-\frac{4\tilde{\beta}\xi''(\sigma)}{\xi'(\sigma)^2+16\tilde{\beta}^2}\epsilon_{\mu\nu\rho\sigma}\partial^{\sigma}\sigma.
    \label{eq:MAStaroTorsion}
\end{align}
As for the metric-affine model discussed in app.\,\ref{app:MetricAffine}, the above expression shows that the scalaron $\sigma$ does not merely play the role of an additional degree of freedom but it also acts as a source of torsion generating a deviation from Levi-Civita connection of torsional origin.\\
It is meaningful to investigate the torsionless limit associated with the large-$|\tilde{\beta}|$ regime of the model in eq.\,\eqref{eq:MAStaro}. In a framework where $\tilde{\beta}$ and $c$ are treated as independent parameters, $|\tilde{\beta}|\to\infty$ implies $M\to \infty$ and thus $\xi''(\sigma)\to 0$. In this case, the limit should be taken directly at the level of the action \eqref{eq:MAStaro}. One then finds $S\sim\frac{\MPl^2}{2}\int d^4x\sqrt{-g}\Big(\tilde{R}+2\tilde{\beta}\tilde{R'} \Big)$, which can be casted in the form \eqref{eq:FalseMA}, implying that the torsion is non-dynamical and the resulting theory is metric. More precisely, integrating out the non dynamical torsion yields the Einstein-Hilbert action. In the inflationary scenario, instead, the ratio $c/\tilde{\beta}^2$ is fixed to a large value in order to reproduce the observed scalar power spectrum, keeping $M\simeq\bar{M}\lesssim O(10^{-5})\MPl$ and thus ensuring $\xi''(\sigma)\neq0$. It is then possible to use the action in the form \eqref{eq:MAStaroscalaron0}, where the auxiliary scalaron $\sigma$ can be introduced.
It then follows directly from eq.\,\eqref{eq:MAStaroTorsion} that the torsion tensor vanishes in the limit $|\tilde{\beta}|\to\infty$. Hence, one finds
\begin{align}
    |\tilde{\beta}|\to\infty\quad \Rightarrow\quad\tilde{R}\to R\quad \text{and}\quad  {\tilde{\Gamma}}^{\lambda}{}_{\mu\nu}\to{{\{\}}_{\mathbf{g}}}^{\lambda}{}_{\mu\nu}. 
\end{align}
Furthermore, in the same limit, the relevant mixed torsional component $\tilde{\beta}S_{\mu}T^{\mu}$ of the purely torsional operator $\tilde{\beta}\tilde{R'}$ appearing in eq.\,\eqref{eq:MAStaroscalaron0} also vanishes. 
Consequently, the action \eqref{eq:MAStaroscalaron0}, or, equivalently, its dynamical equivalent form \eqref{eq:MAStaro}, reduces to the metric Starobinsky model with the scalaron mass fixed at the inflationary scale, yielding
\begin{align}
|\tilde{\beta}|\to\infty\quad\Rightarrow\quad S=\frac{\MPl^2}{2}\int d^4x\sqrt{-g}\Bigg[ {R}+\frac{{{R}}^2}{6\bar{M}^2}\Bigg]. \label{eq:limitStaro}
\end{align}
In the large-$|\tilde{\beta}|$ regime, as evident from \eqref{eq:limitStaro}, $\bar{M}$ is promoted to be the mass of the scalaron. Indeed, we have
\begin{align}
|\tilde{\beta}|\to\infty\quad\Rightarrow\quad M^2=\MPl^2\frac{\tilde{\beta}^2}{3c}\Big[1+O(\tilde{\beta}^{-2}) \Big]={\bar{M}}^2\Big[1+O(\tilde{\beta}^{-2}) \Big],
\end{align}
which coincides with the mass \eqref{eq:Salviomass} introduced in ref.\,\cite{Salvio:2025izr} in the large-$|\tilde{\beta}|$ limit, i.e. in the torsionless regime.\\
It is important to stress that, contrary to the case of the pseudo-scalaron treated in app.\,\ref{app:MetricAffine}, a dynamical $\sigma$ can in general be a manifestation of both dynamical torsion and the additional dynamical scalar associated with the metric. Taking the torsionless limit with $M\to\infty$ suppresses the propagation of $\sigma$, thereby eliminating both manifestations simultaneously. By contrast, unlike the pseudo-scalaron which can't be dynamical in a torsionless limit, one can also consider a torsionless limit in which $\sigma$ remains dynamical. Indeed, for $|\tilde{\beta}| \to \infty$ with $M$ kept finite, it is evident from the action \eqref{eq:MAStaroscalaron} that only the torsional contribution to the coefficient of the kinetic term of $\sigma$ vanishes, while the standard metric contribution arising from the conformal transformation survives, allowing $\sigma$ to remain dynamical and drive inflation. Therefore, the large-$|\tilde{\beta}|$ limit at fixed $M$ removes only the purely torsional contribution to the dynamics of $\sigma$, while leaving intact its metric degree of freedom.\\
\textcolor{black}{This analysis clarifies the physical origin of the nontrivial exact convergence between the inflationary potential associated with the model proposed in ref.\,\cite{Salvio:2025izr} and the Starobinksy potential, thereby also explaining the specific structure of the metric $f(R)$ in eq.\,\eqref{eq:resummedfR} and its inflationary predictions in the $(n_{s},r)$ plane of fig.~\ref{fig:MetricAffine}, presented in sec.\,\ref{sec:MetricAffineGravity}.} Indeed, the metric-affine model of ref.\,\cite{Salvio:2025izr} reproduces the same inflationary potential as the theory in eq.\,\eqref{eq:MAStaro}, which is the metric-affine counterpart of Starobinsky supplemented with an additional purely torsional operator. As a consequence, when the torsion tensor is switched off, the theory manifestly reduces to Starosbinky.

\newpage

\bibliography{draft}

\end{document}